\documentclass{article}

\usepackage{arxiv}

\usepackage[english]{babel}
\usepackage{amsmath}
\usepackage{bm}
\usepackage{amssymb}
\usepackage{amsthm}
\usepackage{graphicx}
\usepackage{ulem}
\usepackage[colorlinks=true, allcolors=blue]{hyperref}
\usepackage{algorithm}
\usepackage{setspace}
\let\Algorithm\algorithm
\renewcommand\algorithm[1][]{\Algorithm[#1]\setstretch{1.38}}

\usepackage{algpseudocode}
\usepackage{placeins} 
\usepackage{tikz} 
\usepackage{subfigure} 
\usepackage{caption}
\usepackage{booktabs}
\usepackage{siunitx} 

\definecolor{color1}{HTML}{0060AD} 
\definecolor{color2}{HTML}{FF4500} 
\definecolor{color3}{HTML}{FFA500} 
\definecolor{color4}{HTML}{006400} 
\definecolor{color5}{HTML}{9400D3} 
\definecolor{color6}{HTML}{800000} 
\definecolor{color7}{HTML}{000000} 
\definecolor{color8}{HTML}{0000FF} 
\definecolor{color9}{HTML}{FF0000} 
\definecolor{mycolor_blue}{RGB}{66,124,161}
\definecolor{mycolor_grey}{RGB}{198,198,198} 

\tikzstyle{line1} = [color=color7, thick]
\tikzstyle{line2} = [color=color2,densely dotted,thick]
\tikzstyle{line3} = [color=color1,densely dashed,thick]
\tikzstyle{line4} = [color=color5,dash dot,thick]
\tikzstyle{line5} = [color=color4,dash dot dot,thick]
\tikzstyle{line6} = [color=color6,thick]

\tikzstyle{mark1} = [color=color4, mark=x,mark size=2pt,mark options=solid, thick] 
\tikzstyle{mark2} = [color=color7, mark=o,mark size=2pt,mark options=solid, thick]
\tikzstyle{mark3} = [color=color2, mark=triangle,mark size=2pt,mark options=solid,thick]
\tikzstyle{mark4} = [color=color8,mark=square,mark size=2pt,mark options=solid,thick]
\tikzstyle{mark5} = [color=color3,mark=diamond, mark size=2pt,mark options=solid,thick]
\tikzstyle{mark6} = [color=color5, mark=star,mark size=2pt,mark options=solid,thick]
\tikzstyle{mark7} = [color=color7,mark=*,mark size=2pt,mark options=solid,thick]
\tikzstyle{mark8} = [color=color7,mark=triangle,mark size=2pt,mark options=solid,thick]

\title{Parameter-free shape optimization: various \\ shape updates for engineering applications}

\author{ \href{https://orcid.org/0000-0001-7015-8928}{\includegraphics[scale=0.06]{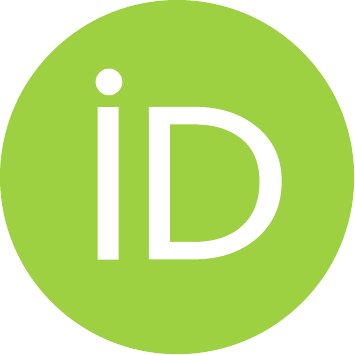}\hspace{1mm}Lars Radtke} \\
	Institute for Ship Structural Design and Analysis (M-10) \\
	Numerical Structural Analysis with Appl. in Ship Technology \\
	Hamburg University of Technology \\
	Hamburg, 21073 \\
	\texttt{lars.radtke@tuhh.de} \\
	\And
	\href{https://orcid.org/0000-0001-8033-9861}{\includegraphics[scale=0.06]{orcid.pdf}\hspace{1mm}Georgios Bletsos} \\
	Institute for Fluid Dynamics and Ship Theory (M-8) \\
	Hamburg University of Technology \\
	Hamburg, 21073 \\
	\texttt{george.bletsos@tuhh.de} \\
	\And
	\href{https://orcid.org/0000-0002-4229-1358}{\includegraphics[scale=0.06]{orcid.pdf}\hspace{1mm}Niklas K\"uhl} \\
	Institute for Fluid Dynamics and Ship Theory (M-8) \\
	Hamburg University of Technology \\
	Hamburg, 21073 \\
	\texttt{kuehl@hsva.de} \\
		\And
	\href{https://orcid.org/0000-0001-5847-5613}{\includegraphics[scale=0.06]{orcid.pdf}\hspace{1mm}Tim Suchan} \\
	Faculty for Mechanical and Civil Engineering \\
	Helmut Schmidt University \\
	Hamburg, 22008 \\
	\texttt{suchan@hsu-hh.de} \\
		\And
	\href{https://orcid.org/0000-0002-3454-1804}{\includegraphics[scale=0.06]{orcid.pdf}\hspace{1mm}Thomas Rung} \\
	Institute for Fluid Dynamics and Ship Theory (M-8) \\
	Hamburg University of Technology \\
	Hamburg, 21073 \\
	\texttt{thomas.rung@tuhh.de} \\
		\And
	\href{https://orcid.org/0000-0002-2162-3675}{\includegraphics[scale=0.06]{orcid.pdf}\hspace{1mm}Alexander D\"uster} \\
	Institute for Ship Structural Design and Analysis (M-10) \\
	Num. Struct. Analysis with Appl. in Ship Technology \\
	Hamburg University of Technology \\
	Hamburg, 21073 \\
	\texttt{alexander.duester@tuhh.de} \\
		\And
	\href{https://orcid.org/0000-0002-6673-9436}{\includegraphics[scale=0.06]{orcid.pdf}\hspace{1mm}Kathrin Welker} \\
	Institute of Numerical Mathematics and Optimization \\
	Technische Universit\"{a}t Bergakademie Freiberg \\
	Freiberg, 09599 \\
	\texttt{kathrin.welker@math.tu-freiberg.de} \\
}

\renewcommand{\undertitle}{Preprint submitted to \textit{CMAME}}
\renewcommand{\shorttitle}{Shape updates for engineering applications}

\hypersetup{
pdftitle={Parameter-free shape optimization: various shape updates for engineering applications},
pdfsubject={cs.CE, math.OC},
pdfauthor={Lars Radtke},
pdfkeywords={shape optimization, shape gradient, steepest descent, continuous adjoint method,  computational fluid dynamics},
}

\begin{document}
\maketitle

\begin{abstract}
	In the last decade, parameter-free approaches to shape optimization problems have matured to a state where they provide a versatile tool for complex engineering applications. 
However, sensitivity distributions obtained from shape derivatives in this context cannot be directly used as a shape update in gradient-based optimization strategies.
Instead, an auxiliary problem has to be solved to obtain a gradient from the sensitivity.
While several choices for these auxiliary problems were investigated mathematically, the complexity of the concepts behind their derivation has often prevented their application in engineering.
This work aims at an explanation of several approaches to compute shape updates from an engineering perspective. 
We introduce the corresponding auxiliary problems in a formal way and compare the choices by means of numerical examples.
To this end, a test case and exemplary applications from computational fluid dynamics are considered.

\end{abstract}


\section{Introduction}
\label{sec:introduction}

Shape optimization is a broad topic with many applications and a large variety of methods.
We focus on optimization methods designed to solve optimization problems that are constrained by partial differential equations (PDE).
These arise, for example, in many fields of engineering such as fluid mechanics \cite{soto2004,othmer2008continuous,lohner2003adjoint}, structural mechanics \cite{allaire2004,upadhyay2021} and acoustics \cite{schmidt2016,kapellos2019unsteady}.

In order to solve computationally a PDE constraint of an optimization problem, the domain under investigation needs to be discretized, i.e., a computational mesh is required.
In this paper, we are particularly concerned with boundary-fitted meshes and methods, where shape updates are realized through updates of the mesh. 
In the context of boundary-fitted meshes, solution methods for  shape optimization problems may be loosely divided into parameterized and parameter-free approaches.
With parameterized, we denote methods that apply a finite dimensional description of the geometry, which is prescribed beforehand and is part of the derivation process of suitable shape updates, see e.g. \cite{papoutsis2019}.
With parameter-free, we denote methods that are derived on the continuous level independently of a parameterization.
Of course, in an application scenario, also parameter-free approaches finally discretize the shape using the mesh needed for the solution of the PDE.

In general, optimization methods for PDE-constrained problems aim at the minimization (or maximization, respectively) of an objective functional that depends on the solution (also called the state) of the PDE, e.g. the compliance of an elastic structure \cite{allaire2004} or dissipated power in a viscous flow \cite{othmer2008continuous}. 
Since a maximization problem can be expressed as a minimization problem by considering the negative objective functional, we only consider minimization problems in this paper.
An in-depth introduction is given in \cite{hinze2008optimization}.
In this paper, we are concerned with iterative methods that generate shape updates such that the objective functional is reduced.
In order to determine suitable shape updates, the so-called shape derivative of the objective functional is utilized.
Typically, adjoint methods are used to compute shape derivatives, when the number of design variables is high.
This is the case in particular for parameter-free shape optimization approaches, where shapes are not explicitly parameterized, e.g. by splines and after a final discretization, the number of design variables typically corresponds to the number of nodes in the mesh that is used to solve the constraining PDE.
Adjoint methods are favorable in this scenario, because their computational cost to obtain the shape derivative is independent of the number of design variables.
Only a single additional problem, the adjoint problem, needs to be derived and solved to obtain the shape derivative.
For a general introduction to the adjoint method, we refer to \cite{errico1997adjoint,giles2000introduction}.
In the continuous adjoint method, the shape derivative is usually obtained as an integral expression over the design boundary identified with the shape and gives rise to a scalar distribution over the boundary, the sensitivity distribution, which is expressed in terms of the solution of the adjoint problem.
As an alternative to the continuous adjoint method, the discrete adjoint method may be employed. 
It directly provides sensitivities at discrete points, likely nodes of the computational mesh. 
A summary of the continuous and the discrete adjoint approach is given in~\cite{giannakoglou2008}.

Especially in combination with continuous adjoint approaches, it is not common to use the derived expression for the sensitivity directly as a shape update within the optimization loop.
Instead, sensitivities are usually \textit{smoothed} or \textit{filtered} \cite{bletzinger2014consistent}.
A focus of this work lies on the explanation of several approaches to achieve this in such a way that they can be readily applied in the context of engineering applications. 
To this end, we concentrate on questions like \textit{How to apply an approach?} and \textit{What are the benefits and costs?} rather than \textit{How can approaches of this type be derived?}

Nevertheless, we would like to point out that there is a large amount of literature concerned with the mathematical foundation of shape optimization. 
For a deeper introduction, one may consult standard text books such as \cite{delfour2011shapes,sokolowski1992}. 
More recently, an in-depth overview on state-of-the-art concepts has been given in \cite{allaire2021} including many references.
We include Sobolev gradients into our studies, which can be seen as a well-established concept that is applied in many studies to obtain a so-called descent direction (which leads to the shape update) from a shape derivative, see e.g. \cite{kroger2015cad,bletsos2021adjoint} for engineering and \cite{Schulz2015a,Welker2016,welker2021} for mathematical studies. 
We also look at more recently-developed approaches like the Steklov-Poincar\'e approach developed in \cite{schulz2015} and further investigated in \cite{schulz2016,welker2021} and the $p$-harmonic descent approach, which was proposed in \cite{deckelnick2021novel} and further investigated in \cite{mueller2021novel}.
In addition, we address discrete filtering approaches as used e.g. in \cite{stavropoulou2014plane, bletzinger2014consistent} into our studies.

The considered shape updates have to perform well in terms of mesh distortion. 
Over the course of the optimization algorithm, the mesh has to be updated several times, including the position of the nodes in the domain interior. 
The deterioration of mesh quality especially if large steps are taken in a given direction is a severe issue that is the subject of several works, see e.g. \cite{onyshkevych2021mesh,stavropoulou2014plane} and plays a major role in the present study as well.
Using an illustrative example and an application from computational fluid dynamics (CFD), the different approaches are compared and investigated.
However, we do not extensively discuss the derivation of the respective adjoint problem or the numerical solution of the primal and the adjoint problem but refer to the available literature on this topic, see e.g.~\cite{mohammadi2010applied,othmer2008continuous,heners2017adjoint,schmidt2013three,soto2004,vassberg2006aerodynamic, vassberg2006aerodynamic_II}.
Instead, we focus on an investigation of the performances of the different approaches to compute a suitable shape update from a given sensitivity.

The remainder of this paper is structured as follows. 
In Sec.~\ref{sec:shape_spaces_metrics_gradients}, we explain the shape optimization approaches from a mathematical perspective and provide some glimpses on the mathematical concepts behind the approaches. 
This includes an introduction to the concept of shape spaces, and the definition of metrics on tangent spaces that lead to the well-known Hilbertian approaches or Sobolev gradients. 
These concepts are then applied in Sec.~\ref{sec:parameter_free_shape_optimization} to formulate shape updates that reduce an objective functional.
In Sec.~\ref{sec:illustrative_example}, we apply the various approaches to obtain shape updates in the scope of an illustrative example, which is not constrained by a PDE.
This outlines the different properties of the approaches, e.g. their convergence behavior under mesh refinement.
In Sec.~\ref{sec:cfd_application} a PDE-constrained optimization problem is considered. 
In particular, the energy dissipation for a laminar flow around a two-dimensional obstacle and in a three-dimensional duct is minimized. 
The different approaches to compute a shape update are investigated and compared in terms of applicability in the sense of  being able to yield good mesh qualities and efficiency in the sense of yielding fast convergence.

\section{Shape spaces, metrics and gradients}
\label{sec:shape_spaces_metrics_gradients}

This section focuses on the mathematical background behind parameter-free shape optimization and aims at introducing the required terminology and definitions for Sec.~\ref{sec:parameter_free_shape_optimization}, which is aimed more at straightforward application.
However, we will reference back to the mathematical section several times, since some information in Sec.~\ref{sec:parameter_free_shape_optimization} may be difficult to understand without the mathematical background. 
In general, we follow the explanations in~\cite{doCarmo1992,Absil2008}, to which we also refer for further reading, and for application to shape optimization, we refer to~\cite{Ring2012,Welker2016,allaire2021}.

\subsection{Definition of shapes} \label{sec:definition_of_shapes}
\begin{figure}%
	\centering%
	\subfigure[]{%
		\includegraphics[width=0.24\textwidth]{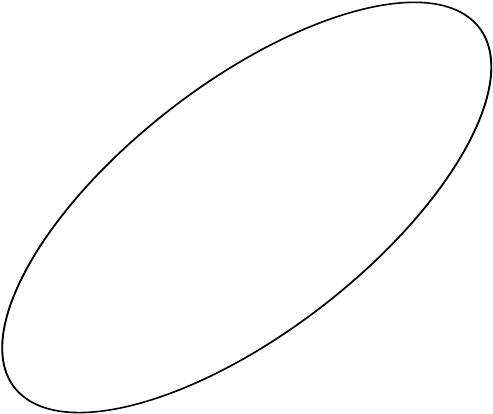}
		\label{fig:InfinitelySmoothShape}%
	}%
	\hfill%
	\subfigure[]{%
		\includegraphics[width=0.24\textwidth]{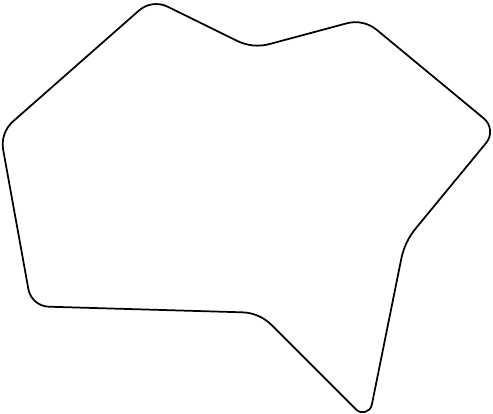}
		\label{fig:C1ContinuousShape}%
	}%
	\hfill%
	\subfigure[]{%
		\includegraphics[width=0.24\textwidth]{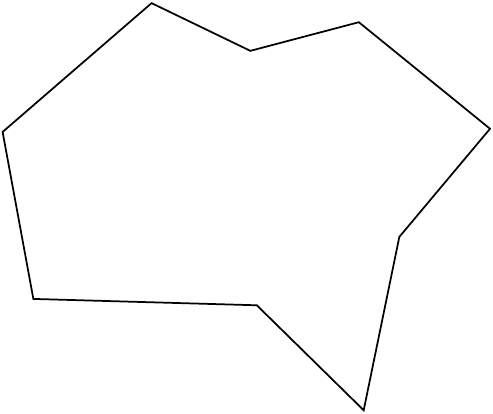}		
		\label{fig:LipschitzContinuousShape}%
	}%
	\hfill%
	\subfigure[]{%
		\includegraphics[width=0.24\textwidth]{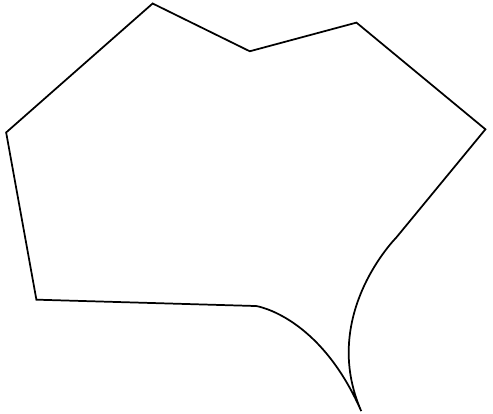}
		\label{fig:NotLipschitzContinuousShape}%
	}%
	\caption{\label{fig:sketchShapes} Sketch of shapes in $\mathbb{R}^2$ from classes of different smoothness. a) Infinitely smooth ($C^\infty$). b) Continuously differentiable ($C^1$). c) Lipschitz-continuous and $C^0$. d) Non-Lipschitz-continuous but $C^0$.}
\end{figure}

To enable a theoretical investigation of gradient descent algorithms, we first need to define what we describe as a shape. 
There are multiple options, e.g. the usage of landmark vectors~\cite{Cootes1995,Hafner2000,Kendall1984,Perperidis2005,Soehn2005}, plane curves \cite{Michor2006,Michor2007,Michor2007a,Mio2006} or surfaces \cite{Bauer2011a,Bauer2012,Kilian2007,Kurtek2010,Michor2005} in higher dimensions, boundary contours of objects \cite{Fuchs2009,Ling2007,Rumpf2009}, multiphase objects \cite{Wirth2010}, characteristic functions of measurable sets \cite{Zolesio2007} and morphologies of images \cite{Droske2007}. 
For our investigations in a two-dimensional setting, we will describe the shape as a plane curve embedded in the surrounding two-dimensional space, the so-called \textit{hold-all domain} $D\subset \mathbb{R}^2$ similar to~\cite{Geiersbach2021a}, and for three-dimensional models, we use a two-dimensional surface embedded in the surrounding three-dimensional space $D\subset \mathbb{R}^3$. 
Additionally, we need the definition of a Lipschitz shape, which is a curve embedded in $\mathbb{R}^2$ or a surface embedded in $\mathbb{R}^3$ that can be described by (a graph of) a Lipschitz-continuous function. 
Furthermore, we define a Lipschitz domain as a domain that has a Lipschitz shape as boundary. 
The concept of smoothness of shapes in two dimensions is sketched in Fig.~\ref{fig:sketchShapes}. 

\subsection{The concept of shape spaces}
The definition of a shape space, i.e. a space of all possible shapes, is required for theoretical investigations of shape optimization. Since we focus on gradient descent algorithms, the possibility to use these algorithms requires the existence of gradients. Gradients are trivially computed in Euclidean space (e.g.~$\mathbb{R}^d$, $d\in\mathbb{N}$), however shape spaces usually do not have a vector space structure. Determining what type of structure a shape space inherits is usually a challenging task and therefore exceeds this paper, however it is common that a shape space does not have a vector space structure. Instead, the next-best option is to aim for a manifold structure with an associated Riemannian metric, a so-called \textit{Riemannian manifold}.

A finite-dimensional manifold is a topological space and additionally fulfills the three conditions.
\begin{enumerate}
	\item It locally can be described by an Euclidean space.
	\item It can be described completely by countably many subsets (second axiom of countability).
	\item Different points in the space have different neighborhoods (Hausdorff space).
\end{enumerate}
If the subsets, so-called \textit{charts}, are compatible, i.e. there are differentiable transitions between charts, then the manifold is a differentiable manifold and allows the definition of tangent spaces and directions, which are paramount for further analysis in the field of shape optimization. 
The tangent space at a point on the manifold is a space tangential to the manifold and describes all directions in which the point could move. 
It is of the same dimension as the manifold. 
If the transition between charts is infinitely smooth, then we call the manifold a smooth manifold. 

Extending the previous definition of a finite-dimensional manifold into infinite dimensions while dropping the second axiom of countability and Hausdorff  yields infinite-dimensional manifolds. A brief introduction and overview about concepts for infinite-dimensional manifolds is given in \cite[Section 2.3]{Welker2016} and the references therein.

In case a manifold structure cannot be established for the shape space in question, an alternative option is a diffeological space structure. These describe a generalization of manifolds, i.e. any previously-mentioned manifold is also a diffeological space. Here, the subsets to completely parametrize the space are called \textit{plots}. As explained in~\cite{IglesiasZemmour2013}, these plots do not necessarily have to be of the same dimension as the underlying diffeological space, and the mappings between plots do not necessarily have to be reversible.
In contrast to shape spaces as  Riemannian manifolds, research for diffeological spaces as shape spaces has just begun, see e.g.~\cite{Goldammer2020,welker2021}. Therefore, for the following section we will focus on Riemannian manifolds first, and then briefly consider diffeological spaces.

\subsection{Metrics on shape spaces} \label{sec:metrics_on_shape_spaces}
In order to define distances and angles on the shape space a metric on the shape space is required.
Distances between iterates (in our setting, shapes) are necessary, e.g. to state convergence properties or to formulate appropriate stopping criteria of optimization algorithms.
For all points $m$ on the manifold~$M$, a Riemannian metric defines a positive definite inner product $g_m ( \cdot,\cdot )$ on the tangent space~$T_m(M)$ at each $m \in M$.\footnote{If the inner product is not positive definite but at least non-degenerate as defined in e.g.~\cite[Def.~8.6]{Kwak1997}, then we call the metric a pseudo-Riemannian metric.} This yields a family of inner products such that we have a positive definite inner product available at any point of the manifold. Additionally, it also defines a norm on the tangent space at $m$ as $\|\cdot \|_{g_m}=\sqrt{g_m ( \cdot,\cdot )}$. If such a Riemannian metric exists, then we call the differentiable manifold a Riemannian manifold, often denoted as $(M, g)$.

Different types of metrics on shape spaces can be identified, e.g. inner metrics~\cite{Bauer2011a,Bauer2012,Michor2007}, outer metrics~\cite{Beg2005,Bookstein1997,Glaunes2008,Kendall1984,Michor2007}, metamorphosis metrics~\cite{Holm2009,Trouve2005}, the Wasserstein or Monge-Kantorovic metric for probability measures~\cite{Ambrosio2004,Benamou2000,Benamou2002}, the Weil-Peterson metric~\cite{Kushnarev2009,Sharon2006}, current metrics~\cite{Durrleman2009,Durrleman2008,Vaillant2005} and metrics based on elastic deformations~\cite{Fuchs2009,Rumpf2015}.

Additional to the Riemannian metric, we also need a definition of distance to obtain a metric in the classical sense. Following~\cite{Absil2008,Ring2012,Welker2016}, to obtain an expression for distances on the manifold, we first define the length of a differentiable curve~$\gamma$ on the manifold starting at $m$ using the Riemannian metric~$g_m ( \cdot,\cdot )$ as
\begin{equation}
	L(\gamma) = \int_0^1 \sqrt{g_m (\dot{\gamma}(t), \dot{\gamma}(t))} \,\mathrm{d} t
\end{equation}
and then define the distance function $d(m_1,m_2)$ as the infimum of any curve length which starts at $m_1$ and ends at $m_2$, i.e.
\begin{align}
d(m_1,m_2) = \inf_{\gamma} L(\gamma), \quad \text{ with } \gamma(0)=m_1 \text{ and } \gamma(1)=m_2.
\label{eq:distanceRiemannianManifold}
\end{align}
This distance function is called the \textit{Riemannian distance} or \textit{geodesic distance}, since the so-called \textit{geodesic} describes the shortest distance between two points on the manifold. For more details about geodesics, we refer to~\cite{Lang1999}. 

If one were able to obtain the geodesic, then a local mapping from the tangent space to the manifold would already be available: the so-called \textit{exponential map}. However, finding the exponential map requires the solution of a second-order ordinary differential equation. This is often prohibitively expensive or inaccurate using numerical schemes. The exponential map is a specific retraction (cf. e.g.~\cite{Absil2008,Ring2012,Welker2016}), but different retractions can also be used to locally map an element of the tangent space back to the manifold. A retraction is a mapping from $T_m(M) \rightarrow M$ which fulfills the following two conditions.
\begin{enumerate}
	\item The zero-element of the tangent space at $m$ gets mapped to $m$ itself, i.e. $\mathcal{R}_m(0) = m$.
	\item The tangent vector $\dot{\gamma}(t)$ of a curve $\gamma: t\mapsto \mathcal{R}_m(t \,\xi)$ starting at $m$ satisfies $\dot{\gamma}(0) = \xi$. Figuratively speaking, this means that a movement along the curve~$\gamma$ is described by a movement in the direction~$\bm{\xi}$ while being constrained to the manifold~$M$.
\end{enumerate}

\begin{figure}[tbp]%
	\centering%
	\includegraphics[width=0.3\textwidth]{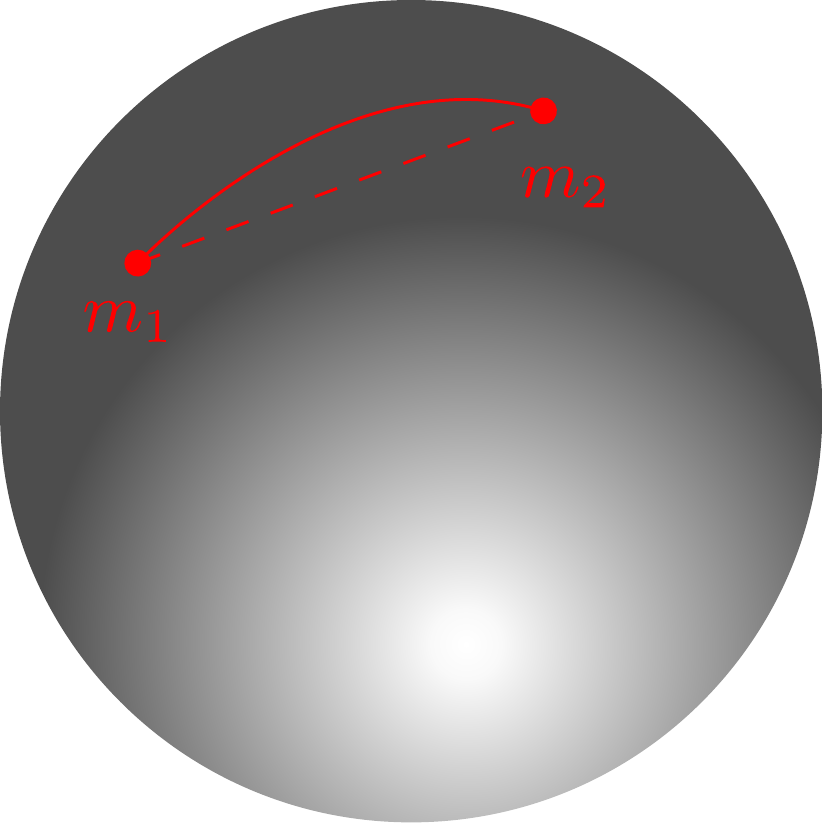}
	\caption{Illustration of two points on a sphere (a manifold), connected by the straight connection through the sphere (leaving the manifold) and a curve on the sphere.}%
	\label{fig:sphere_with_points}%
\end{figure}

\paragraph{Example}
To illustrate the previous point, we would like to introduce a relatively simple example. Let us assume we have a sphere without interior (a two-dimensional surface) embedded in $\mathbb{R}^3$ as illustrated in Fig.~\ref{fig:sphere_with_points}. 
This sphere represents a manifold~$M$. 
Additionally, let us take two arbitrary points $m_1$ and $m_2$ on the sphere. 
The shortest distance of these two points \textit{while remaining on the sphere} is not trivial to compute. 
If one were to use that the sphere is embedded in $\mathbb{R}^3$ then the shortest distance of these two points can be computed by subtracting the position vector of both points and is depicted by the red dashed line.
However, this path does not stay on the sphere, but instead goes through it.
In consideration of the above concepts, the shortest distance between two points on the manifold is  given by the geodesic, indicated by a solid red line.
Similarly, obtaining the shortest distance along earth's surface suffers from the same issue.
Here, using the straight path through the earth is not an option (for obvious reasons). In a local vicinity around point~$m_1$ it is sufficient to move on the tangential space~$T_{m_1}(M)$ at point~$m_1$ and project back to the manifold using the exponential map to calculate the shortest distance to point~$m_2$. However, at larger distances, this may not be a valid approximation anymore.

\bigskip

Several difficulties arise when trying to transfer the previous concepts to infinite-dimensional manifolds.
As described in~\cite{geiersbach2021b}, most Riemannian metrics are only weak, i.e. lack an invertible mapping between tangent and cotangent spaces, which is required for inner products.\footnote{We do not go into more detail about this issue, the interested reader is referred to \cite{Bauer2014} for more information on this topic.} 
Further, the geodesic may not exist or is not unique, or the distance between two different elements of the infinite-dimensional manifold may be $0$ (the so-called \textit{vanishing geodesic distance phenomenon}). Thus, even though a family of inner products is a Riemannian metric on a finite-dimensional differentiable manifold, it may not be a Riemannian metric on an infinite-dimensional manifold. 
Due to these challenges, infinite-dimensional manifolds as shape spaces are still subject of ongoing research.

Metrics for diffeological spaces have been researched to a lesser extent. 
However most concepts can be transferred, and in \cite{Goldammer2020} a Riemannian metric is defined for a diffeological space, which yields a Riemannian diffeological space. Additionally, the Riemannian gradient and a steepest descent method on diffeological spaces are defined, assuming a Riemannian metric is available. To enable usage of diffeological spaces in an engineering context, further research is required in this field. 

\subsection{Riemannian shape gradients} \label{sec:Riemannian_shape_gradients}

The previous sections were kept relatively general and tried to explain the concept of manifolds and metrics on manifolds. 
Now we focus specifically on shape optimization based on Riemannian manifolds. 
Following~\cite{Welker2016}, we introduce an objective functional which is dependent on a shape\footnote{We use the description of a shape as an element of the manifold and as a $d-1$-dimensional subset of the hold-all domain $D\subset \mathbb{R}^d$ interchangeably.}~$\Gamma \in M$, where $M$ denotes the shape space, in this case a Riemannian manifold.
In shape optimization, it is often also called \textit{shape functional} and reads $J\colon M \rightarrow \mathbb{R},\, \Gamma \mapsto J(\Gamma)$.
Furthermore, we denote the perturbation of the shape~$\Gamma$ as $\Gamma_t = F_t (\Gamma) = \{ F_t(\bm{x}): \bm{x}\in \Gamma \}$ with $t \geq 0$. The two most common approaches for $F_t$ are the velocity method and the perturbation of identity. The velocity method or speed method requires the solution of an initial value problem as described in~\cite{sokolowski1992}, while the perturbation of identity is defined by $F_t(\bm{x}) = \bm{x} + t \, \bm{v}^\Gamma(\bm{x})$, $\bm{x} \in \Gamma$, with a sufficiently smooth vector field $\bm{v}^\Gamma$ on $\Gamma$. 
We focus on the perturbation of identity for this publication. 
Reciting Sec.~\ref{sec:definition_of_shapes} a shape is described as a plane curve in two or as a surface in three dimensional surrounding space here, which means they are always embedded in the hold-all domain $D$.

To minimize the shape functional, i.e. $\min_{\Gamma \in M} J(\Gamma)$, we are interested in performing an optimization based on gradients. In general, the concept of a gradient can be generalized to Riemannian (shape) manifolds, but some differences between a standard gradient descent method and a gradient descent method on Riemannian manifolds exist. For comparison, we show a gradient descent method on $\mathbb{R}^d$, $d \in \mathbb{N}$ and on Riemannian manifolds in Algorithms~\ref{alg:gradient_descent_Rn} and \ref{alg:gradient_descent_manifold}, respectively, for which we introduce the required elements in the following.

\renewcommand\algorithm[1][]{\Algorithm[#1]\setstretch{1.38}}
\begin{figure}
\begin{minipage}[t]{0.47\linewidth}
\begin{algorithm}[H]
\caption{Steepest (gradient) descent algorithm \\ on Riemannian manifold $(M, g)$}
\label{alg:gradient_descent_manifold}
\begin{algorithmic}[1]
	\Require \raggedright shape functional~$J$, initial value $\Gamma^0 \in M$, ~~~~~~ $\epsilon > 0$, retraction $\mathcal{R}$ on $(M,g)$
	\For{$i=0,1,...$}
		\State Compute $J(\Gamma^i)$
		\State Compute shape gradient $\nabla J(\Gamma^i)$ from
$$
			g_{\Gamma^i}(\nabla J(\Gamma^i), \bm{v}^\Gamma) = (J_*)_{\Gamma^i} (\bm{v}^\Gamma) \quad \forall \bm{v}^\Gamma \in T_{\Gamma^i}(M) \phantom{\frac{\partial J}{\partial \bm{x}}}
			\phantom{\left.  \frac{\partial J}{\partial \bm{x}} \right|_{\bm{x}^i}}
$$
		\State Compute $\| \nabla J(\Gamma^i) \|_{g_{\Gamma^i}}$
		\If{$\| \nabla J(\Gamma^i) \|_{g_{\Gamma^i}} \leq \epsilon$}
			\State \textbf{break}
		\EndIf
		\State Compute direction $\bm{\theta}^i = -\frac{\nabla J(\Gamma^i)}{\| \nabla J(\Gamma^i) \|_{g_{\Gamma^i}}}$
		\State Determine step size $\alpha^i$
		\State Set $\Gamma^{i+1} = \mathcal{R}_{\Gamma^i} ( \alpha^i \, \bm{\theta}^i )$ 
	\EndFor
    \end{algorithmic}
\end{algorithm}
\end{minipage}
\hfill
\begin{minipage}[t]{0.47\linewidth}
\begin{algorithm}[H]
	\caption{Steepest (gradient) descent algorithm \\ in Euclidean space $(\mathbb{R}^d,\|\cdot\|_2)$}
	\label{alg:gradient_descent_Rn}
	\begin{algorithmic}[1]
		\Require \raggedright differentiable~function~$J$, \phantom{$\Gamma^0$} initial~value~$\bm{x}^0~\in~\mathbb{R}^d$, $\epsilon>0$
		\For{$i=0,1,...$}
			\State Compute $J(\bm{x}^i)$
			\State Compute gradient $\nabla J(\bm{x}^i)$ from
$$
			\nabla J(\bm{x}^i) =\left.  \frac{\partial J}{\partial \bm{x}} \right|_{\bm{x}^i}
$$
			\State Compute $\| \nabla J(\bm{x}^i) \|_2$
			\If{$\| \nabla J(\bm{x}^i) \|_2 \leq \epsilon$}
				\State \textbf{break}
			\EndIf
			\State Compute direction $\bm{\theta}^i = -\frac{\nabla J(\bm{x}^i)}{\| \nabla J(\bm{x}^i) \|_2}$ \phantom{$\frac{2}{\|1\|_{g_{\Gamma^i}}}$}
			\State Determine step size $\alpha^i$
			\State Set $\bm{x}^{i+1} = \bm{x}^i + \alpha^i \, \bm{\theta}^i$
		\EndFor
	\end{algorithmic}
\end{algorithm}
\end{minipage}
\end{figure}

On Euclidean spaces, an analytic or numerical differentiation suffices to calculate gradients. In contrast, we consider a Riemannian manifold $(M,g)$ now, where the \textit{pushforward} is required in order to determine the Riemannian (shape) gradient of $J$.
We use the definition of the pushforward from~\cite[p.~28]{Lang1999} and~\cite[p.~56]{Lee2012}, which has been adapted to shape optimization in e.g.~\cite{Geiersbach2021a}. The pushforward~$(J_*)_\Gamma$ describes a mapping between the tangent spaces $T_\Gamma(M)$ and $T_{J(\Gamma)} (\mathbb{R})$.
Using the pushforward, the Riemannian (shape) gradient~$\nabla J(\Gamma)$ of a (shape) differentiable function~$J$ at $\Gamma \in M$ is then defined as
\begin{align}
	\label{eq:shape_gradient_boundary_chap2_push_forward}
	g_\Gamma(\nabla J(\Gamma), \bm{v}^\Gamma) = (J_*)_\Gamma (\bm{v}^\Gamma)\quad \forall\, \bm{v}^\Gamma\in T_\Gamma M.
\end{align}
Further details about the pushforward can be found in e.g.~\cite{Kriegl1997,Lang1999}.

As is obvious from the computation of the gradient in Algorithm~\ref{alg:gradient_descent_manifold} in line~4 $\rightarrow$ Eq.~\eqref{eq:shape_gradient_boundary_chap2_push_forward}, the Riemannian shape gradient lives on the tangent space at $\Gamma$, which (in contrast to the gradient for Euclidean space) is not directly compatible with the shape  $\Gamma$. 
A movement on this tangent space will lead to leaving the manifold, unless a projection back to the manifold is performed by the usage of a retraction as in line~10 of the algorithm and previously described in Sec.~\ref{sec:metrics_on_shape_spaces}.

In practical applications the pushforward is often replaced by the so-called \textit{shape derivative}. A shape update direction~$\bm{u}^\Gamma$ of a (shape) differentiable function~$J$ at $\Gamma \in M$ is computed by solving
\begin{align}
	\label{eq:shape_gradient_boundary_chap2}
g_\Gamma (\bm{u}^\Gamma, \bm{v}^\Gamma) = J'(\Gamma)(\bm{v}^\Gamma) \quad \forall\, \bm{v}^\Gamma\in T_\Gamma M.
\end{align}
The term $J'(\Gamma)(\bm{v}^\Gamma)$ describes the shape derivative of $J$ at $\Gamma$ in the direction of $\bm{v}^\Gamma$. The shape derivative is defined by the so-called \textit{Eulerian derivative}. The Eulerian derivative of a functional~$J$ at $\Gamma$ in a sufficiently smooth direction $\bm{v}^\Gamma$ is given by
\begin{align}
\label{eq:shape_derivative}
DJ(\Gamma)(\bm{v}^\Gamma) =  J'(\Gamma)(\bm{v}^\Gamma) = \lim_{t \to 0^+} \frac{J(\Gamma_t) - J(\Gamma)}{t}.
\end{align}
If the Eulerian derivative exists for all directions $\bm{v}^\Gamma$ and if the mapping $\bm{v}^\Gamma \mapsto J'(\Gamma)(\bm{v}^\Gamma)$ is linear and continuous, then we call the expression $J'(\Gamma)(\bm{v}^\Gamma)$ the \textit{shape derivative} of $J$ at $\Gamma$ in the direction $\bm{v}^\Gamma$.

In general, a shape derivative depends only on the displacement of the shape $\Gamma$ in the direction of its local normal $\bm{n}$ such that it can be expressed as
\begin{align}
\label{eq:shape_derivative_boundary_chap2}
J'(\Gamma)(\bm{v}^\Gamma) 
= 
\int_\Gamma \bm{v}^\Gamma \cdot \bm{n} \, s(\bm{x}) \, \mathrm{d}\Gamma,
\end{align}
the so-called \textit{Hadamard form} or \textit{strong formulation}, where $s$ is called \textit{sensitivity distribution} here.
The existence of such a scalar distribution $s$ is the outcome of the well-known Hadamard theorem, see e.g. \cite{hadamard1968memoire,sokolowski1992,delfour2011shapes}. It should be noted that a weak formulation\footnote{If the objective functional is defined over the surrounding domain then the weak formulation is also an integral over the domain; if it is defined over $\Gamma$ then the weak formulation is an integral over $\Gamma$, however not in Hadamard form. Using the weak formulation reduces the analytical effort for the derivation of shape derivatives. If the objective functional is a domain integral then using the weak formulation requires an integration over the surrounding domain instead of over $\Gamma$. Further details as well as additional advantages and drawbacks can be found e.g. in~\cite{sokolowski1992,schulz2015,Welker2016,welker2021}. } of the shape derivative is derived as an intermediate result, however in this publication only strong formulations as in Eq.~\eqref{eq:shape_derivative_boundary_chap2} will be considered. 

\subsection{Examples of shape spaces and their use for shape optimization}
\label{sec:examples_of_shape_spaces}

Now we shift our focus towards specific spaces which have been used as shape spaces, and metrics on these shape spaces. 
In this publication, we concentrate on the class of inner metrics, i.e. metrics defined on the shape itself, see Sec.~\ref{sec:metrics_on_shape_spaces}.

\paragraph{The shape space~$\mathcal{B}_e$}

Among the most common is the shape space often denoted by $\mathcal{B}_e$ from~\cite{Michor2006}. We avoid a mathematical definition here and instead describe it as the following: The shape space $\mathcal{B}_e$ contains all shapes which stem from embeddings of the unit circle into the hold-all domain excluding reparametrizations. This space only contains infinitely-smooth shapes (see Fig.~\ref{fig:InfinitelySmoothShape}). It has been shown in~\cite{Michor2006} that this shape space is an infinite-dimensional Riemannian manifold, which means we can use the previously-described concepts to attain Riemannian shape gradients for the gradient descent algorithm in Algorithm~\ref{alg:gradient_descent_manifold} on~$\mathcal{B}_e$, but two open questions still have to be addressed: \textit{Which Riemannian metric can (or should) we choose as $g$?} and \textit{Which method do we use to convert a direction on the tangential space into movement on the manifold?} The latter question has been answered in~\cite{Schulz2018,Geiersbach2021a}, where a possible retraction on~$\mathcal{B}_e$ is described as
\begin{align}
	\mathcal{R}_{\Gamma^i}: T_{\Gamma^i}(M) \rightarrow M, \bm{v}^\Gamma \mapsto \mathcal{R}_{\Gamma^i}(\bm{v}^\Gamma)=\Gamma^i+\bm{v}^\Gamma,
\end{align}
i.e. all $\bm{x} \in \Gamma^i$ are displaced to $\bm{x}+\bm{v}^\Gamma(\bm{x})$ $\forall \bm{x} \in \Gamma^i$.
Due to its simplicity of application this is what will be used throughout this paper.

The former question is not so easily-answered. Multiple types of Riemannian metrics could be chosen in order to compute the Riemannian shape gradient, each with its advantages and drawbacks. To introduce the three different classes of Riemannian metrics, we first introduce an option which does not represent a Riemannian metric on~$\mathcal{B}_e$.

As has been proven in~\cite{Michor2006}, the standard $L^2$ metric on $T_\Gamma(\mathcal{B}_e)$ defined as
\begin{align}
	\label{eq:L2innerproduct}
	g_\Gamma: T_{\Gamma}(\mathcal{B}_e) \times T_{\Gamma}(\mathcal{B}_e), (\bm{u}^\Gamma, \bm{v}^\Gamma) \mapsto \int_\Gamma \bm{u}^\Gamma \cdot \bm{v}^\Gamma \,\mathrm{d}\Gamma
\end{align}
is \textit{not} a Riemannian metric on $\mathcal{B}_e$ because it suffers from the vanishing geodesic distance phenomenon. This means that the whole theory for Riemannian manifolds cannot be used, i.e. it is not guaranteed that the computed ``gradient'' w.r.t. the $L^2$ metric is a steepest descent direction.

Based on the $L^2$ metric not being a Riemannian metric on $\mathcal{B}_e$, alternative options have been proposed which do not suffer from the vanishing geodesic distance phenomenon. As described in~\cite{Welker2016}, three groups of $L^2$-metric-based Riemannian metrics can be identified.
\begin{enumerate}
	\item \textit{Almost local metrics} include weights into the $L^2$ metric (cf.~\cite{Bauer2010,Bauer2012,Michor2007}).
	\item \textit{Sobolev metrics} include derivatives into the $L^2$ metric (cf.~\cite{Bauer2011a,Michor2007}).
	\item \textit{Weighted Sobolev metrics} include both weights and derivatives into the the $L^2$ metric (cf.~\cite{Bauer2012}).
\end{enumerate}
The first group of Riemannian metrics can be summarized as
\begin{align}
	g_\Gamma: T_{\Gamma}(\mathcal{B}_e) \times T_{\Gamma}(\mathcal{B}_e), (\bm{u}^\Gamma, \bm{v}^\Gamma) \mapsto \int_\Gamma \Phi \, \bm{u}^\Gamma \cdot \bm{v}^\Gamma \,\mathrm{d}\Gamma
\end{align}
with an arbitrary function~$\Phi$. As described in~\cite{Michor2007}, this function could be dependent e.g. on the length of the two-dimensional shape to varying degrees, the curvature of the shape, or both.

According to~\cite{Michor2007}, the more common approach falls into the second group. In this group, higher derivatives are used to avoid the vanishing geodesic distance phenomenon. To so-called \textit{Sobolev metric} exists up to arbitrarily high order. Commonly-used (cf. e.g.~\cite{Schulz2015a}) is the first-order Sobolev metric
\begin{align}
\label{eq:first_order_sobolev_weak}
	g_\Gamma: T_{\Gamma}(\mathcal{B}_e) \times T_{\Gamma}(\mathcal{B}_e), (\bm{u}^\Gamma, \bm{v}^\Gamma) \mapsto \int_\Gamma \bm{u}^\Gamma \cdot \bm{v}^\Gamma + A \, \nabla_\Gamma \bm{u}^\Gamma \cdot \nabla_\Gamma \bm{v}^\Gamma \,\mathrm{d}\Gamma
\end{align}
with the arc length derivative $\nabla_\Gamma$ and a metric parameter~$A>0$. An equivalent metric can be obtained by partial integration and reads 
\begin{align}
\label{eq:first_order_sobolev_strong}
	g_\Gamma(\bm{u}^\Gamma, \bm{v}^\Gamma):=	\int_\Gamma \bm{u}^\Gamma \cdot \bm{v}^\Gamma - A \, \Delta_\Gamma \bm{u}^\Gamma \cdot \bm{v}^\Gamma \,\mathrm{d}\Gamma,
\end{align}
where $\Delta_\Gamma$ represents the Laplace-Beltrami operator. Therefore, the first-order Sobolev metric is also sometimes called the \textit{Laplace-Beltrami approach}.

The third group combines the previous two, thus a first-order weighted Sobolev metric is given by
\begin{align}
	g_\Gamma: T_{\Gamma}(\mathcal{B}_e) \times T_{\Gamma}(\mathcal{B}_e), (\bm{u}^\Gamma, \bm{v}^\Gamma) \mapsto \int_\Gamma \Phi \left( \bm{u}^\Gamma \cdot \bm{v}^\Gamma + A \, \nabla_\Gamma \bm{u}^\Gamma \cdot \nabla_\Gamma \bm{v}^\Gamma \right) \,\mathrm{d}\Gamma,
\end{align}
or equivalently, 
\begin{align*}
	g_\Gamma(\bm{u}^\Gamma, \bm{v}^\Gamma):=	\int_\Gamma \Phi \left( \bm{u}^\Gamma \cdot \bm{v}^\Gamma - A \, \Delta_\Gamma \bm{u}^\Gamma \cdot \bm{v}^\Gamma \right) \,\mathrm{d}\Gamma.
\end{align*}

As already described in Algorithm~\ref{alg:gradient_descent_manifold}, the solution of a PDE to obtain the Riemannian shape gradient cannot be avoided.
In most cases, the PDE cannot be solved analytically. 
Instead, a discretizetion has to be used to numerically solve the PDE.
However, the discretized domain $\Omega \subseteq D$ in which the shape~$\Gamma$ is embedded will not move along with the shape itself, which causes a quick deterioration of the computational mesh. Therefore, the Riemannian shape gradient has to be extended into the surrounding domain. The Laplace equation $\Delta \bm{u} = \bm{0}$ is commonly used for this, with the Riemannian shape gradient as a Dirichlet boundary condition on $\Gamma$. Then, we call $\bm{u}$ the \textit{extension of the Riemannian shape gradient into the domain~$\Omega$}, i.e. $\bm{u}^\Gamma$ denotes the restriction of $\bm{u}$ to $\Gamma$.

An alternative approach on $\mathcal{B}_e$ that avoids the use of Sobolev metrics has been introduced in~\cite{schulz2016} and is named  \textit{Steklov-Poincar\'e approach}, where one uses a member of the family of \textit{Steklov-Poincar\'e metrics} $g_s(\cdot, \cdot)$ to calculate the shape update. The name stems from the Poincar{\'e}-Steklov operator, which is an operator to transform a Neumann- to a Dirichlet boundary condition. Its inverse is then used to transform the Dirichlet boundary condition on $\Gamma$ to a Neumann boundary condition. More specifically, the resulting Neumann boundary condition gives a deformation equivalent to a Dirichlet boundary condition. Let $V(\Omega)$ be an appropriate function space with an inner product defined on the domain~$\Omega$. Then, using the Neumann solution operator $E_N(\bm{u}^\Gamma) = \bm{u}$, where $\bm{u}$ is the solution of the variational problem $a(\bm{u},\bm{v}) = \int_{\Gamma} \bm{u}^\Gamma \cdot \bm{v}^\Gamma \, \mathrm{d} \Gamma$ $\forall \bm{v} \in V(\Omega)$, we can combine the Steklov-Poincar\'e metric $g_s$, the shape derivative $J'(\Gamma)(\bm{v})$, and the symmetric and coercive bilinear form $a(\cdot,\cdot)$ defined on the domain $\Omega$ to determine the extension of the Riemannian shape gradient w.r.t. the Steklov-Poincar\'e metric into the domain, which we denote by $\bm{u} \in V(\Omega)$, as
\begin{align}
	g_s(\bm{u}^\Gamma, \bm{v}^\Gamma) = J'(\Gamma)(\bm{v}^\Gamma) = a(\bm{u},\bm{v}) \quad \forall \, \bm{v} \in V(\Omega).
	\label{eq:SteklovPoincareBilinearForm}
\end{align}
For further details we refer the interested reader to~\cite{Welker2016}. Different choices for the bilinear form $a(\cdot,\cdot)$ yield different Steklov-Poincar\'e metrics, which motivates the expression of the family of Steklov-Poincar\'e metrics. Common choices for the bilinear form are 
\begin{align}
\label{eq:bilinear_form_sp}
	a(\bm{u},\bm{v})=\int_\Omega \nabla \bm{u} \cdot \nabla \bm{v} \, \mathrm{d} \Omega \quad \text{or} \quad a(\bm{u},\bm{v}) = \int_\Omega \nabla \bm{u} \cdot \mathcal{D} \, \nabla \bm{v} \, \mathrm{d}\Omega,
\end{align}
where $\mathcal{D}$ could represent the material tensor of linear elasticity. The extension of the Riemannian shape gradient $\bm{u}$ w.r.t. the Steklov-Poincar\'e metric $g_s$ is directly obtained and can immediately be used to update the mesh in all of $\Omega$, which avoids the solution of an additional PDE on $\Gamma$.
Additionally, the weak formulation of the shape derivative can be used in equation~\eqref{eq:SteklovPoincareBilinearForm} to simplify the analytical derivation, as already described in Sec.~\ref{sec:Riemannian_shape_gradients}.

\paragraph{The shape space~$\mathcal{B}^\frac{1}{2}$}
An alternative to the shape space $\mathcal{B}_e$ has been introduced in~\cite{welker2021}. It is denoted as $\mathcal{B}^\frac{1}{2}(\Gamma^0)$ and it is shown that this shape space is a diffeological space. This shape space contains all shapes which arise from admissible transformations of an initial shape $\Gamma^0$, where $\Gamma^0$ is at least Lipschitz-continuous.
This is a much weaker requirement on the smoothness of admissible shapes (compared to to the infinitely-smooth shapes in $\mathcal{B}_e$). An overview of shapes with different smoothness has already been given in Fig.~\ref{fig:sketchShapes}. 
Opposed to optimization on Riemannian manifolds, optimization on diffeological spaces is not yet a well-established topic.
Therefore, the main objective for formulating optimization algorithms on a shape space, i.e. the generalization of concepts like the definition of a gradient, a distance measure and optimality conditions, is not yet reached for the novel space $\mathcal{B}^{\frac{1}{2}}(\Gamma^0)$. However, the necessary objects for the steepest descent method on a diffeological space are established and the corresponding algorithm is formulated in~\cite{Goldammer2020}.
It is nevertheless worth to mention that various numerical experiments, e.g.~\cite{schulz2015,schulz2016computational,Siebenborn2017,Blauth2021}, have shown that shape updates obtained from the Steklov-Poincar\'e metric can also be applied to problems involving non-smooth shapes.
However, questions about the vanishing geodesic distance, a proper retraction and the dependency of the space on the initial shape $\Gamma^0$ remain open.

\paragraph{The largest-possible space of bi-Lipschitz transformations $W^{1,\infty}(\Omega, \mathbb{R}^d)$}
On fi\-nite-di\-men\-sio\-nal manifolds, the direction of steepest descent\footnote{The source gives the direction of steepest ascent, but the direction of steepest descent is defined accordingly.} can be described by two equivalent formulations, see~\cite{Absil2008}, and reads
\begin{align}
	-\frac{\nabla J(\Gamma)}{\| \nabla J(\Gamma) \|_{g_\Gamma}} = \operatornamewithlimits{arg\,min}_{\bm{u}^\Gamma\in T_\Gamma(M): \| \bm{u}^\Gamma \|_{g_\Gamma} = 1} J'(\Gamma)(\bm{u}^\Gamma).
	\label{eqn:GradientSteepestDescentDirection}
\end{align}
Instead of solving for the shape gradient $\nabla J(\Gamma)$, another option to obtain a shape update direction is to solve the optimization problem on the right-hand side of equation~\eqref{eqn:GradientSteepestDescentDirection}, but this usually is prohibitively expensive.
Introduced in~\cite{Ishii2005} and applied in shape optimization in~\cite{deckelnick2021novel} as the \textit{$W^{1,\infty}$ approach}, it is proposed to approximate the solution to the minimization problem~\eqref{eqn:GradientSteepestDescentDirection} by solving
\begin{align}
	\min_{\bm{u} \in W^{1,p}(\Omega, \mathbb{R}^d)} \int_\Omega \frac{1}{p} \left| \nabla \bm{u} \right|^p \,\mathrm{d} \Omega + J'(\Gamma)(\bm{u}^\Gamma)
\end{align}
while taking $p\to\infty$ with $p>2$, see~\cite{Evans1982}. Due to the equivalence to the extension equation as described in~\cite{Evans1982,Ishii2005,mueller2021novel} in weak formulation
\begin{align}
\label{eq:phd_weak_form}
	\underbrace{\int_\Omega \left| \nabla \bm{u} \right|^{p-2} \left( \nabla \bm{u} \cdot \nabla \bm{v} \right) \mathrm{d} \Omega}_{a(\bm{u},\bm{v})} = J'(\Gamma)(\bm{v}^\Gamma) \quad \forall \bm{v} \in W^{1,p}(\Omega, \mathbb{R}^d),
\end{align}
this PDE can be solved numerically with iteratively increasing $p$. In a similar fashion to the Steklov-Poincar\'e approach, we can equate the weak form of the extension equation~$a(\bm{u},\bm{v})$ to the shape derivative~$J'(\Gamma)(\bm{v}^\Gamma)$ in strong or weak formulation to obtain the shape update direction. In~\cite{mueller2021novel}, this approach is called the \textit{$p$-harmonic descent approach}. The Sobolev space for the extension of the shape update direction $W^{1,\infty}(\Omega, \mathbb{R}^d)$ is motivated as the largest-possible space of bi-Lipschitz shape updates. However, it is not yet clear which additional assumptions are needed in order to guarantee that a Lipschitz shape update preserves Lipschitz continuity in this manner, see \cite[Sec. 3.2]{welker2021} and \cite[Sec. 4.1]{hofmann2007} for further details on this topic. Moreover, a theoretical investigation of the underlying shape space that results in shape update directions from the space $W^{1,\infty}(\Omega, \mathbb{R}^d)$ is still required. Since neither a manifold structure has been established which would motivate the minimization over the tangent space in equation~\eqref{eqn:GradientSteepestDescentDirection}, nor has it been shown that $g_s$ is possibly a Riemannian metric for this manifold\footnote{There is no inner product defined on $W^{1,p}(\Omega, \mathbb{R}^d)$ unless $p=2$ and $a(\bm{u},\bm{v})$ does not fulfill the condition of linearity in the arguments unless $p=2$ to classify as a bilinear form. A bilinear form is required for Eq.~\eqref{eq:SteklovPoincareBilinearForm} to hold.}, it is not guaranteed that equation~\eqref{eq:SteklovPoincareBilinearForm} yields a steepest descent direction in this scenario.

If we assume $W^{1,\infty}(\Omega, \mathbb{R}^d)$ to be the largest possible space for $\bm{u}$ that yields shape updates conserving Lipschitz continuity, then only $W^{1,\infty}(\Omega, \mathbb{R}^d)$ itself or subspaces of $W^{1,\infty}(\Omega, \mathbb{R}^d)$ yield shape updates conserving Lipschitz continuity. 
For example, when working with the Sobolev metrics of higher order and an extension which does not lose regularity, one needs to choose the order~$p$ high enough such that the corresponding solution from the Hilbert space $H^p(\Omega, \mathbb{R}^d)$ is also an element of $W^{1,\infty}(\Omega, \mathbb{R}^d)$.
The Sobolev embedding theorem yields that this is only the case for $p \geq \frac{d}{2}+1$.
Therefore, one would need to choose at least $p=2$ in two dimensions and $p=3$ in three dimensions. 
However, this requirement is usually not fulfilled in practice due to the demanding requirement of solving nonlinear PDEs for the shape update direction.
Further, already the shape gradient w.r.t. the first-order Sobolev metric is sufficient to meet the above requirement under certain conditions, as described in~\cite[Sec. 2.2.2]{Welker2016}.

After introducing the necessary concepts to formulate shape updates from a theoretical perspective, we will now reiterate these concepts in the next section with a focus on applicability.

\FloatBarrier

\section{Parameter-free shape optimization in engineering}
\label{sec:parameter_free_shape_optimization}
\begin{figure}[t]
\centering
\includegraphics[width=0.8\textwidth]{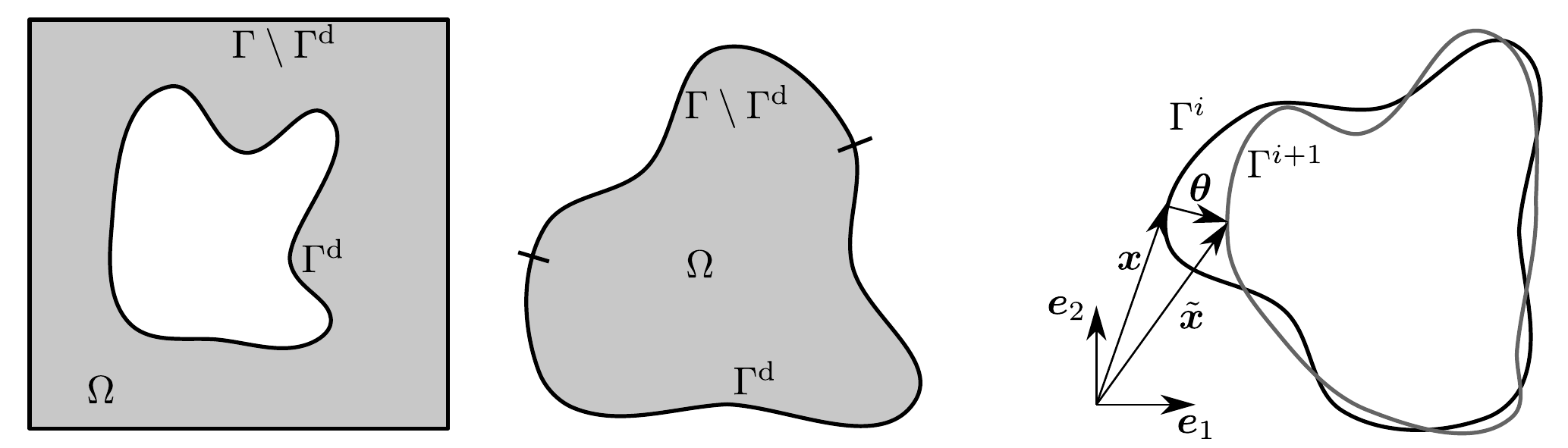}
\caption{\label{fig:domain_examples} Examples for computational domains and their boundaries (left) and domain transformation (right).}
\end{figure}
In an engineering application, the shape $\Gamma$ to be optimized may be associated with a computational domain $\Omega$ in different ways as illustrated in Fig.~\ref{fig:domain_examples}.
Independently of this setting the main goal of an optimization algorithm is not only to compute updated shapes $\Gamma^{i+1}$ from a given shape $\Gamma^i$ such that $J(\Gamma^{i+1}) < J(\Gamma^i)$ but also to compute updated domains $\Omega^{i+1}$ that preserve the quality of a given discretization of $\Omega^{i}$.
Similar to the updated shape according to the perturbation of identity, the updated domain is computed as
\begin{align}
 \label{eq:domain_transformation}
        \Omega^{i+1} = \left\{ \tilde{\bm{x}}: \tilde{\bm{x}} = \bm{x} + \alpha \, \bm{\theta}(\bm{x}) \quad \forall \bm{x} \in \Omega^i \right\},
\end{align}
which is applied in a discrete sense, e.g. by a corresponding displacement of all nodes by $\alpha \, \bm{\theta}$.
Summarizing the elaborations in the previous section, a gradient descent algorithm that achieves a desired reduction of the objective functions involves four steps that compute
\begin{enumerate}
\item the objective function $J(\Gamma^i)$ and its shape derivative $J'(\Gamma^i)(\bm{v}^\Gamma)$,
\item the shape update direction $\bm{\theta}^\Gamma$ (the negative shape gradient $-\bm{u}^\Gamma$),
\item the domain update direction $\bm{\theta}$ (the  extension of the negative shape gradient $-\bm{u}$),
\item a step size $\alpha$ and an updated domain $\Omega^{i+1}$.
\end{enumerate}
We introduce $\bm{\theta}^\Gamma$ and $\bm{\theta}$ here in a general way as \textit{shape update direction} and \textit{domain update direction}, respectively, because not all approaches yield an actual \textit{shape gradient} according to its definition in Eq.~\eqref{eq:shape_gradient_boundary_chap2}.
In the remainder of this section, we focus on Step 2 -- 4 starting with a description of several approaches to compute $\bm{\theta}^\Gamma$ in a simplified way that allows for a direct application.
Some approaches combine Steps 2 and 3 and directly yield the domain update direction $\bm{\theta}$.
For all other approaches, the extension is computed separately as explained at the end of this section, which includes an explanation of the step size control.

We do not give details about Step 1 (the computation of the shape derivative $J'(\Gamma_i)$) and refer to the literature cited in Sec.~\ref{sec:introduction} about the derivation of adjoint problems in order to compute $J'(\Gamma)$ in an efficient way independently of the number of design variables.
However, we assume that the objective function is given as
\begin{align}
    J(\Gamma) = \int_\Omega j_\Omega \, \mathrm{d}\Omega + \int_\Gamma j_\Gamma \, \mathrm{d}\Gamma,
\end{align}
which is the case for all problems considered in this work and arises in many engineering applications as well.
Further, we assume that the shape derivative is given in the strong formulation (see Eq.~\eqref{eq:shape_derivative_boundary_chap2}).
The main input for Step 2 is accordingly the sensitivity distribution $s$.

\subsection{Shape and domain update approaches}
\label{sec:practical_descent_directions}
Before collecting several approaches for the computation of a shape update direction $\bm{\theta}^\Gamma$ from a sensitivity $s$ we would like to give some general remarks about why the computed directions are reasonable candidates for a shape update that yields a reduction of $J$. 
To this end, the definition of the shape derivative in Eq.~\eqref{eq:shape_derivative} can be used to obtain a first-order approximation
\begin{align}
\label{eq:tailor_series_ds}
    J(\Gamma^{i+1}) \approx J(\Gamma^i) + \alpha \, J'(\Gamma^i)(\bm{\theta}^\Gamma).
\end{align}
Using the expression of the shape derivative from Eq.~\eqref{eq:shape_derivative_boundary_chap2} and setting $\bm{\theta}^\Gamma = -\bm{n} \, s$, one obtains
\begin{align}
\label{eq:inequality_sensitivity}
J(\Gamma^{i+1}) \approx J(\Gamma^i) - \alpha \int_\Gamma s^2 \, \mathrm{d}\Gamma \lessapprox J(\Gamma^i),
\end{align}
which formally shows that a decrease of the objective function can be expected at least for small $\alpha$.
However, several problems arise when trying to use $\bm{\theta}^\Gamma=-\bm{n}\,s$ in practice and in theory, when used for further mathematical investigations as detailed in Sec.~\ref{sec:shape_spaces_metrics_gradients}.
An obvious practical problem is that neither $\bm{n}$ nor $s$ can be assumed to be smooth enough such that their product and the subsequent extension result in a valid displacement field $\bm{\theta}$ that can be applied according to Eq.~\eqref{eq:domain_transformation}.
All approaches considered here overcome this problem by providing a shape update direction $\bm{\theta}^\Gamma$, which is smoother than $\bm{n} \, s$. 
Several approaches make use of the Riemannian shape gradient $\bm{u}^\Gamma$ as defined in Eq.~\eqref{eq:shape_gradient_boundary_chap2} for this purpose.
A corresponding first-order approximation reads
\begin{align}
\label{eq:tailor_series_rsg}
    J(\Gamma^{i+1}) \approx J(\Gamma^i) + \alpha \, g_\Gamma(\bm{u}^\Gamma, \bm{\theta}^\Gamma).
\end{align}
Setting $\bm{\theta}^\Gamma = -\bm{u}^\Gamma$, one obtains
\begin{align}
\label{eq:inequality_riemannian}
J(\Gamma^{i+1}) \approx J(\Gamma^i) - \alpha \, g_\Gamma(\bm{\theta}^\Gamma, \bm{\theta}^\Gamma),
\end{align}
which shows that also these approaches yield a decrease in the objective function provided that $\alpha$ is small.

\subsubsection{Discrete filtering approaches}
\label{sec:dicrete_filtering}
Several authors successfully apply discrete filtering techniques to obtain a smooth shape update, see e.g. \cite{stavropoulou2014plane, bletzinger2014consistent, kroger2015cad}. 
As the name suggests, they are formulated based on the underlying discretization, e.g. on the nodes or points $\bm{x}_n$ on $\Gamma$ and the sensitivity at these points $s_n=s(\bm{x}_n)$. 
The shape update direction at the nodes, i.e. the direction of the displacement to be applied there, is computed by
\begin{align}
\label{eq:filtered_nodal_value}
    \bm{\theta}^\Gamma_n = \bm{\theta}^\Gamma(\bm{x}_n) = - \sum_{j \in N_n} w_{n,j} \, s_j \, \bm{n}_j.
\end{align}
Therein, $w_{n,j}$ denotes the weight and $N_n$ is the set indices of nodes in the neighborhood of node $n$.
We introduce a particular choice for the neighborhoods $N_n$ and the weights $w_{n,j}$ in Sec.~\ref{sec:illustrative_example} and denote it as \textit{Filtered Sensitivity (FS)} approach.

The discrete nature of a filter according to Eq.~\eqref{eq:filtered_nodal_value} demands for a computation of a normal vector $\bm{n}_n$ at the nodal positions.
Since $\bm{n}(\bm{x}_n)$ is not defined, a special heuristic computation rule must be applied.
In the example considered in Sec.~\ref{sec:illustrative_example}, the nodes on $\Gamma$ are connected by linear edges, and we compute the normal vector $\bm{n}_n$ as the average of normal vectors $\bm{n}^{e_1}$ and $\bm{n}^{e_2}$ of the two adjacent edges,
\begin{align}
\label{eq:node_normal}
    \bm{n}_n = \frac{1}{2} \left( \bm{n}^{e_1} + \bm{n}^{e_2} \right).
\end{align}
An analogue computation rule is established for the three-dimensional problem considered in Sec.~\ref{sec:cfd_application}.
In this discrete setting, it also becomes possible to directly use the sensitivity and the normal vector as a shape update direction, even for non-smooth geometries. 
It is just a special case of \eqref{eq:filtered_nodal_value} using a neighborhood $N_n=\left\{n\right\}$ and weight $w_{n,n}=1$, which results in $\bm{\theta}^\Gamma_n = - \bm{n}_n \, s_n$.
The resulting approach is denoted here as the \textit{direct sensitivity (DS)} approach.

We would like to emphasize that the corresponding choice in the continuous setting $\bm{\theta}^\Gamma= - \bm{n} \, s$ that led to Eq.~\eqref{eq:inequality_sensitivity} cannot be applied for the piece-wise linear shapes that arise when working with computational meshes -- the normal vectors at the nodal points are simply not defined.
The same problem arises for any shape update in normal direction.
However, we include such methods in our study, because they are widely used in literature and can be successfully applied when combined with a special computation rule for the normal direction at singular points like Eq.~\eqref{eq:node_normal}.
It is noted that having computed $\bm{\theta}^\Gamma$ according to the FS or DS approach one needs to extend it into the domain to obtain $\bm{\theta}$ as described in Sec.~\ref{sec:mesh_morphing}.

Finally, we would like to point out that in an application scenario, also the continuously-derived shape update directions eventually make use of a discrete update of nodal positions (Sec.~\ref{sec:illustrative_example}) or cell centers (Sec.~\ref{sec:cfd_application}).
Accordingly, all approaches -- including those introduced in the following sections -- finally undergo an additional discrete filtering.

\subsubsection{Laplace-Beltrami approaches}
\label{sec:laplace_beltrami}
A commonly applied shape update is based on the first-order Sobolev metric (see Eq.~\eqref{eq:first_order_sobolev_weak}), which yields as an identification problem for the shape gradient:
\begin{align}
\label{eq:identification_problem}
    \text{Find }\bm{u}^\Gamma, \text{ s.t. } \quad
    \int_{\Gamma^\mathrm{d}}
    A \, \nabla_\Gamma \bm{u}^\Gamma \cdot \nabla_\Gamma \bm{v}^\Gamma
    +
    \bm{u}^\Gamma \cdot \bm{v}^\Gamma
    \, \mathrm{d}\Gamma^\mathrm{d}
    = 
    J'(\Omega)(\bm{v}^\Gamma)
    =
    \int_{\Gamma^\mathrm{d}}
    \bm{n} \cdot \bm{v}^\Gamma \, s
    \, \mathrm{d}\Gamma
    \quad \forall \bm{v}^\Gamma \in V(\Gamma^\mathrm{d}).
\end{align}
We denote the constitutive parameter $A$ as conductivity here.
A strong formulation involves the tangential Laplace-Beltrami operator $\Delta_\Gamma$ suggesting the name for this type of approach.
Formulated as a boundary value problem it reads
\begin{alignat}{2}
\label{eq:pde_lapalce_beltrami}
    \bm{u}^\Gamma - A \, \Delta_\Gamma \bm{u}^\Gamma &= \bm{n} \, s \quad && \text{ in } \Gamma^\mathrm{d}, \\
\label{eq:dirichlet_lapalce_beltrami}
    \bm{u}^\Gamma &= \bm{0} \quad && \text{ on } \partial \Gamma^\mathrm{d}.
\end{alignat}
This auxiliary problem yields $\bm{u}^\Gamma$ on $\Gamma^\mathrm{d}$, while on $\Gamma \setminus \Gamma^\mathrm{d}$ we set $\bm{u}^\Gamma = \bm{0}$. 
Means to extend $\bm{\theta}^\Gamma = - \bm{u}^\Gamma$ into the domain to obtain $\bm{\theta}$, respectively $\bm{u}$, are described in Sec.~\ref{sec:mesh_morphing}.
We denote this approach as \textit{Vector Laplace Beltrami (VLB)} in the following.
Due to the fact that $\Delta_\Gamma$ operates only in the tangential direction, the components of $s \, \bm{n}$ are mixed, such that $\bm{\theta}^\Gamma$ is not parallel to $\bm{n}$, see \cite{kroger2015cad,stavropoulou2014plane} for further details.

As an alternative, we consider a scalar variant of the VLB approach applied in \cite{heners2017adjoint} and call it \textit{Scalar Laplace Beltrami (SLB)} in the following.
A scalar field $\bar{u}$ is computed using the tangential Laplace Beltrami operator and the sensitivity $s$ as a right-hand side:
\begin{alignat}{2}
\label{eq:pde_lapalce_beltrami_scalar}
    \bar{u} - A \, \Delta_\Gamma \bar{u} &= s \quad && \text{ in } \Gamma^\mathrm{d}, \\
\label{eq:dirichlet_lapalce_beltrami_scalar}
    \bar{u} &= 0 \quad && \text{ on } \partial \Gamma^\mathrm{d}.
\end{alignat}
As a shape update direction $\bm{\theta}^\Gamma = -\bar{u} \, \bm{n}$ is taken.
As in the VLB case, some smoothness is gained in the sense that $\bar{u}$ is smoother than $s$.
However, this choice has the same shortcomings as any direction that is parallel to the normal direction.
It is further noted that the discrete filtering approach from Sec.~\ref{sec:dicrete_filtering} is equivalent to a finite-difference approximation of the VLB method, if the weights in Eq.~\eqref{eq:filtered_nodal_value} are chosen according to the bell-shaped Gaussian function, see \cite{bletzinger2014consistent,kroger2015cad}.

\subsubsection{Steklov-Poincar\'e approaches}
\label{sec:steklov_poincare}
As mentioned in Sec.~\ref{sec:shape_spaces_metrics_gradients}, these approaches combine the identification of $\bm{\theta}^\Gamma$ and the computation of its extension into the domain. 
This leads to an identification problem, similar to Eq.~\eqref{eq:identification_problem}, however, now using a function space $V(\Omega)$ defined over the domain $\Omega$ and a bilinear form $a(\cdot, \cdot)$ on $\Omega$ instead of an inner product $g(\cdot, \cdot)$ on $\Gamma$.
Choosing the second bilinear form from Eq.~\eqref{eq:bilinear_form_sp}, the identification problem for the shape gradient reads
\begin{align}
\label{eq:identification_problem_sp}
    \text{Find }\bm{u}, \text{ s.t. } \quad
    \int_\Omega 
    \nabla \bm{u} \cdot \mathcal{D} \, \nabla \bm{v} 
    \, \mathrm{d}\Omega
    = 
    J'(\Omega)(\bm{v})
    =
    \int_\Gamma 
    \bm{n} \cdot \bm{v}^\Gamma \, s
    \, \mathrm{d}\Gamma
    \quad \forall \bm{v} \in V(\Omega).
\end{align}
If $\mathcal{D}$ is chosen as the constitutive tensor of an isotropic material, Eq. \eqref{eq:identification_problem_sp} can be interpreted as a weak formulation of the balance of linear momentum.
In this linear elasticity context, $s \, \bm{n}$ plays the role of a surface traction.
Appropriately in this regard, the approach is also known as the traction method, see e.g. \cite{azegami1996domain, azegami2006smoothing}.
To complete the formulation, the constitutive tensor is expressed as
\begin{align}
\label{eq:linear_elasticity_tensor}
    \mathcal{D} = \lambda \, \mathcal{T} + 2 \, \mu \, \mathcal{S},
\end{align}
where $\mathcal{T}$ denotes the fourth order tensor that yields the trace ($\mathcal{T} \, \bm{A} = \mathrm{tr}\left(\bm{A}\right) \, \bm{I}$), $\mathcal{S}$ is the fourth order tensor that yields the symmetric part ($\mathcal{S}\,\bm{A} = \frac{1}{2} \left( \bm{A} + \bm{A}^\mathrm{T} \right)$) and $\lambda$ and $\mu$ are the Lam\'e constants.
Suitable choices for these parameters are problem-dependent and are usually chosen, such that the quality of the underlying mesh is preserved as good as possible.
Through integration by parts, a strong formulation of the identification problem can be obtained that further needs to be equipped with Dirichlet boundary conditions to arrive at
\begin{alignat}{2}
\label{eq:pde_steklov_poincare}
    \mathrm{div}\left(\mathcal{D}\, \nabla\bm{u}\right) &= \bm{0} \quad &&\text{ in } \Omega, \\
\label{eq:neumann_steklov_poincare}    
    \mathcal{D}\, \nabla\bm{u} \, \bm{n} &= \bm{n}\, s \quad &&\text{ on } \Gamma^\mathrm{d}, \\
\label{eq:dirichlet_steklov_poincare}
    \bm{u} &= \bm{0} \quad &&\text{ on } \Gamma \setminus \Gamma^\mathrm{d}.
\end{alignat}
We will refer to this choice as \textit{Steklov-Poincar\'e structural mechanics (SP-SM)} in the following.
An advantage is the quality of the domain transformation that is brought along with it -- a domain that is perturbed like an elastic solid with a surface load will likely preserve the quality of the elements that its discretization is made of.
Of course, the displacement must be rather small, as no geometric or physical nonlinearities are considered.
Further, the approach makes it possible to use weak formulations of the shape derivative as mentioned in Sec.~\ref{sec:Riemannian_shape_gradients}.
To this end, the integrand in the shape derivative can then be interpreted as a volume load in the elasticity context and applied as a right-hand side in \eqref{eq:pde_steklov_poincare}.

Diverse alternatives exist that employ an effective simplification of the former.
In \cite{kuhl2021phd} the spatial cross coupling introduced by the elasticity theory is neglected and a spatially varying scalar conductivity is introduced.
The conductivity is identified with the inverse distance to the boundary such that
\begin{align}
\label{eq:constituive_wall_distance}
   \mathcal{D} = \frac{1}{w+\varepsilon} \, \mathcal{I},
\end{align}
where $\mathcal{I}$ denotes the fourth order identity tensor and $w$ refers to the distance to the boundary.
A small value $\varepsilon$ is introduced to circumvent singularities for points located on the wall.
In the sequel, we denote this variant as \textit{Steklov-Poincar\'e wall distance (SP-WD)}.
It is emphasized that it is now a diffusivity or heat transfer problem that is solved, instead of an elasticity problem.
More precisely, $d$ decoupled diffusivity or heat transfer problems are solved -- one for each component of $\bm{u} = \left[ u_1~u_2~u_3\right]$ -- since with \eqref{eq:constituive_wall_distance} the PDE \eqref{eq:pde_steklov_poincare} reduces to
\begin{alignat}{2}
    \nabla \cdot \left( \frac{1}{w+\varepsilon} \, \nabla u_i \right) &= 0 \quad &&\text{ in } \Omega \quad \text{ for } i = 1, 2, 3.
\end{alignat}

For completeness, we would like to refer to an alternative from \cite{onyshkevych2021mesh} that introduces a nonlinearity into the identification problem (\ref{eq:identification_problem_sp}).
Another choice for $\mathcal{D}$ employed in \cite{schulz2016computational,geiersbach2019} is $\mathcal{D} = 2 \, \mu \, \mathcal{S}$, where $\mu$ is set to a user-defined maximum value on $\Gamma^\mathrm{d}$ and a minimum value on the remaining part of the boundary. Values inside $\Omega$ are computed as the solution of a Laplace equation such that the given boundary values are smoothly interpolated.
However, we do not consider these choices in our investigations in Sections~\ref{sec:illustrative_example} and \ref{sec:cfd_application}.

\subsubsection{$p$-harmonic descent approach}
\label{sec:harmoic_descent}
As introduced at the end of Sec.~\ref{sec:examples_of_shape_spaces}, the \textit{$p$-harmonic descent approach} (PHD) yields another  identification problem for the domain update direction $\bm{\theta}^*$ as given in Eq.~\eqref{eq:phd_weak_form}. 
A minor reformulation yields
\begin{align}
    \label{eq:p_harmonic_descent}
    \int_\Omega
    \left( \nabla\bm{u} \cdot \nabla \bm{u} \right)^{\frac{p-2}{2}} 
    \left( \nabla \bm{u} \cdot \nabla \bm{v} \right) \, d \Omega = \alpha \, J'(\Omega)(\bm{v}) = \alpha \, \int_{\Gamma^\mathrm{d}} \bm{v} \cdot \bm{n} \, s \, \mathrm{d}\Gamma^\mathrm{d}.
\end{align}
A strong form of the problem reads
\begin{alignat}{2}
\label{eq:p_laplacian_pde}
    \mathrm{div}\left( 
    \left( \nabla \bm{u} \cdot \nabla \bm{u} \right)^{\frac{p-2}{2}} 
    \nabla \bm{u}
    \right) 
    &=\bm{0} \quad &&\mathrm{in} \ \Omega, \\
\label{eq:p_laplacian_neumann}
    \left( \nabla \bm{u} \cdot \nabla \bm{u} \right)^{\frac{p-2}{2}} 
    \nabla \bm{u} \, \bm{n}
    &= \alpha \, s \, \bm{n} \quad &&\mathrm{on} \  \Gamma^\mathrm{d}, \\
\label{eq:p_laplacian_dirichlet}
    \bm{u} 
    &= \bm{0} \quad &&\mathrm{on} \  \Gamma \backslash \Gamma^\mathrm{d}.
\end{alignat}
The domain update direction is then taken to be $\bm{\theta} = - \frac{1}{\alpha}\bm{u}$.
Due to the nonlinearity of \eqref{eq:p_laplacian_pde} we have introduced the scaling parameter $\alpha$ here.
In the scope of an optimization algorithm $\alpha$ represents a step size and may be determined by a step size control.
All other approaches introduced above establish a linear relation between $s$ and $\bm{\theta}$ such that the scaling can be done independently of the solution of the auxiliary problem.
For the PHD approach, Problem (\ref{eq:p_laplacian_pde} -- \ref{eq:p_laplacian_dirichlet}) may need to be solved repeatedly in order to find the desired step size.

The main practical advantage of this choice is the parameter $p$, which allows to get arbitrarily close to the case of bi-Lipschitz transformations~$W^{1,\infty}(\Omega, \mathbb{R}^d)$.
Sharp corners can therefore be resolved arbitrarily close as discussed in Sec.~\ref{sec:shape_spaces_metrics_gradients} and demonstrated in \cite{mueller2021novel, deckelnick2021novel}.
Another positive aspect demonstrated therein is that the PHD approach yields comparably good mesh qualities. 
Like the SP approaches the PHD approach further allows for a direct utilization of a weak formulation of the shape derivative.

\subsection{Mesh morphing and step size control}
\label{sec:mesh_morphing}
Several methods are commonly applied to extend shape update directions $\bm{\theta}^\Gamma$ obtained from the approaches DS, FS, VLB, and SLB into the domain. 
For example, interpolation methods like radial basis functions may be used, see e.g. \cite{heners2017adjoint}.
Another typical choice is the solution of a Laplace equation, with $\bm{\theta}$ as its state and $\bm{\theta}^\Gamma$ as a Dirichlet boundary condition on $\Gamma^\mathrm{d}$ for this purpose, see e.g. \cite{lohner1996improved}.
We follow a similar methodology and base our extension on the general PDE introduced for the Steklov-Poincar\'e approach.
The boundary value problem to be solved when applied in this context reads
\begin{alignat}{2}
    \label{equ:domain_transport}
    \mathrm{div}\left ( \mathcal{D} \, \nabla \bm{\theta} \right)
    &= \bm{0} \quad && \text{ in } \Omega, \\
    \label{equ:domain_transport_design}
    \bm{\theta} &= \bm{\theta}^\Gamma \quad &&\text{ on } \Gamma^\mathrm{d}, \\
    \label{equ:domain_transport_fixed}
    \bm{\theta} &= \bm{0} \quad && \text{ on } \Gamma \, \backslash \, \Gamma^\mathrm{d}.
\end{alignat}
As a constitutive relation, we choose again linear elasticity (see Eq.~\eqref{eq:linear_elasticity_tensor}) or component-wise heat transfer (see Eq.~\eqref{eq:constituive_wall_distance}).
Once a deformation field is available in the entire domain, its discrete representation can be updated according to Eq.~\eqref{eq:domain_transformation}.
It is recalled here that the domain update direction $\bm{\theta}$ can be computed independently of the step size $\alpha$ for all approaches except for the PHD approach, where it has a nonlinear dependence on $\alpha$, see Sec.~\ref{sec:harmoic_descent}.

In order to compare different shape updates, we apply a step size control.
We follow two different methods to obtain a suitable step size $\alpha$ for the optimization.
\begin{enumerate}
    \item We perform a line search, where $\alpha$ is determined by a divide and conquer approach such that $J(\Omega^{i+1})$ is minimized. By construction, the algorithm approaches the optimal value from below and leads to the smallest $\alpha > 0$ that yields such a local minimum. If the mesh quality falls below a certain threshold, the algorithm quits before a minimum is found and yields the largest $\alpha$, for which the mesh is still acceptable.
    For all considered examples and shape update directions, this involves repeated evaluations of $J$. For the PHD approach, it further involves repeated computations of $\bm{\theta}$.
    \item We prescribe the maximum displacement for the first shape update  $\theta^\mathrm{max} = \underset{\bm{x} \in \Omega_0}{\max} \Vert \alpha \, \bm{\theta}(\bm{x}) \Vert$. 
    This does not involve evaluations of $J$ but for the PHD approach it involves again repeated computations of $\bm{\theta}$. 
    For all other methods, we simply set
    \begin{align}
    \alpha = \theta^\mathrm{max} \left( \underset{\bm{x} \in \Omega_0}{\max} \Vert \, \bm{\theta}(\bm{x}) \Vert \right)^{-1}.    
    \end{align}
\end{enumerate}
Because we aim at a comparison of the different approaches to compute a shape update rather than an optimal efficiency of the steepest descent algorithm, we do not make use of advanced step size control strategies such as Armijo backtracking. 

As mentioned in the previous section, the evaluation of the shape update direction depends on the application and the underlying numerical method.
In particular, the evaluation of the normal vector $\bm{n}$ is a delicate issue that may determine whether or not a method is applicable.
We include a detailed explanation of the methods used for this purpose in Sections~\ref{sec:illustrative_example}~and~\ref{sec:cfd_application}.

\section{Illustrative test case}
\label{sec:illustrative_example}
\begin{figure}
\centering
\includegraphics[width=\textwidth]{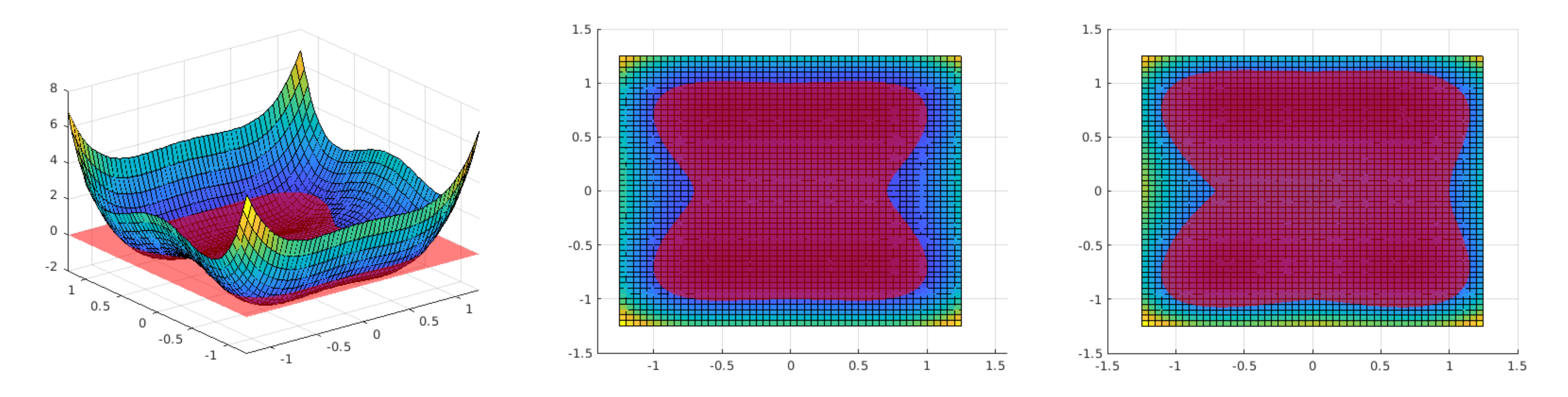}
\caption{\label{fig:levelset_function} Graph of the function $f$ described by~(\ref{eq:test_problem}) with $C_2 = 0$. Left and center: $C_1=0$. Right: $C_1=1$.}
\end{figure}

In order to investigate the different shape and domain updates, we consider the following unconstrained optimization problem.
\begin{align}
    \label{eq:test_problem}
    \min_{\Gamma \in M} J(\Gamma) = \int_\Omega f(\bm{x}) \, \mathrm{d}\Omega,
\end{align}
where
\begin{align}
f(\bm{x}) = f(x_1, x_2) = 2\,x_1^4 + x_2^4 - x_1^2 - 4 \, x_2^2 - 3 \, C_1 \, \left| \mathrm{max}(x_1,x_2)  \right| + \frac{1}{10} \, C_2 \left( \sin(50\,x_1) + \sin(50\,x_2) \right).
\end{align}
The graph of $f$ is shown in Fig.~\ref{fig:levelset_function}, including an indication of the curve, where $f=0$, i.e. the level-set of $f$.
Since inside this curve, $f \leq 0$ and outside $f>0$, the level-set is exactly the boundary of the minimizing domain. 
Through the term that is multiplied by $C_1$, a singularity is introduced --
if $C_1 \neq 0$, the optimal shape has two kinks, while it is smooth for $C_1 = 0$. 
Through the term that is multiplied by $C_2$, high-frequency content is introduced.
Applying the standard formula for the shape derivative (see e.g. \cite{allaire2021}), we obtain
\begin{align}
    J'(\Gamma)(\bm{v}^\Gamma) =  \int_\Gamma f \, \bm{v}^\Gamma \cdot \bm{n} \, \mathrm{d} \Gamma
\end{align}
such that $s=f$.

We start the optimization process from a smooth initial shape -- a disc with outer radius $R=1$ and inner radius $r=0.3$. The design boundary $\Gamma^\mathrm{d}$ corresponds to the outer boundary only, the center hole is fixed.
This ensures the applicability of the SP-SM approach, which can only be applied as described if at least rigid body motions are prevented by Dirichlet boundary conditions. 
This requires $\Gamma \backslash \Gamma^\mathrm{d} \neq \emptyset$ in the corresponding auxiliary problem (Eqs.~(\ref{eq:pde_steklov_poincare}--\ref{eq:neumann_steklov_poincare})).

We perform an iterative algorithm to solve the minimization problem by successively updating the shape (and the domain) using the various approaches introduced in Sec.~\ref{sec:parameter_free_shape_optimization}.
For a fair comparison of the different shape and domain updates the line search technique sketched in Section~\ref{sec:mesh_morphing} is used to find the step size $\alpha$ that minimizes $J(\Gamma^{i+1})$ for a given $\bm{\theta}$, i.e. the extension of $\bm{\theta^\Gamma}$ into the domain is taken into account when determining the step size $\alpha$.

\subsection{Discretization}
We discretize the initial domain using a triangulation and in a first step keep this mesh throughout the optimization. In a second step, re-meshing is performed every third optimiation iteration and additionally, whenever the line search method yields a step size smaller than $10^{-6}$.
The boundary is accordingly discretized by lines (triangle edges).
In order to practically apply the theoretically infeasible shape updates, which are parallel to the boundary normal field, the morphing of the mesh is done based on the nodes.
A smoothed normal vector is obtained at all boundary nodes by averaging the normal vectors of the two adjacent edges. 
The sensitivity $s$ is evaluated at the nodes as well and then used in combination with the respective auxiliary problem to obtain the domain update direction $\bm{\theta}$, respectively $\bm{\theta}^\Gamma$ at the nodes.
The evaluation of the integral in $J$ is based on values at the triangle centers.

The auxiliary problems for the choices from Section~\ref{sec:laplace_beltrami} (VLB, and SLB) are solved using finite differences. 
Given $\bm{u}^\Gamma$, the tangential divergence at a boundary node $j$ is approximated based on the adjacent boundary nodes by
\begin{align}
\Delta_\Gamma \bm{u}^\Gamma(\bm{x}_j) \approx 2 \, \frac{\bm{u}^\Gamma(\bm{x}_{j+1})-\bm{u}^\Gamma(\bm{x}_{j})}{ h_{j+1} \left( h_j + h_{j+1} \right) } - 2 \, \frac{\bm{u}^\Gamma(\bm{x}_{j})-\bm{u}^\Gamma(\bm{x}_{j-1})}{h_{j} \left( h_j + h_{j+1} \right) },
\end{align}
where $h_j = \Vert \bm{x}_{j} - \bm{x}_{j-1} \Vert$ denotes the distance between nodes $j$ and ${j-1}$.

The auxiliary problems for the choices from Sections~\ref{sec:steklov_poincare} and \ref{sec:harmoic_descent} (SP-SM, SP-TM, and PHD) are solved with the finite element method.
Isoparametric elements with linear shape functions based on the chosen triangulation are used.
Dirichlet boundary conditions are prescribed by elimination of the corresponding degrees of freedom. 

The auxiliary problem (\ref{equ:domain_transport}-\ref{equ:domain_transport_fixed}) needed in combination with all choices from Section \ref{sec:practical_descent_directions} that provide only $\bm{\theta}^\Gamma$ (DS, FS, VLB, SLB) is solved using the same finite-element method.
All computations are done in MATLAB \cite{matlab2021}.
The code is available through \url{http://collaborating.tuhh.de/M-10/radtke/soul}.

\subsection{Results}
\begin{figure}
\centering
\includegraphics[width=\textwidth]{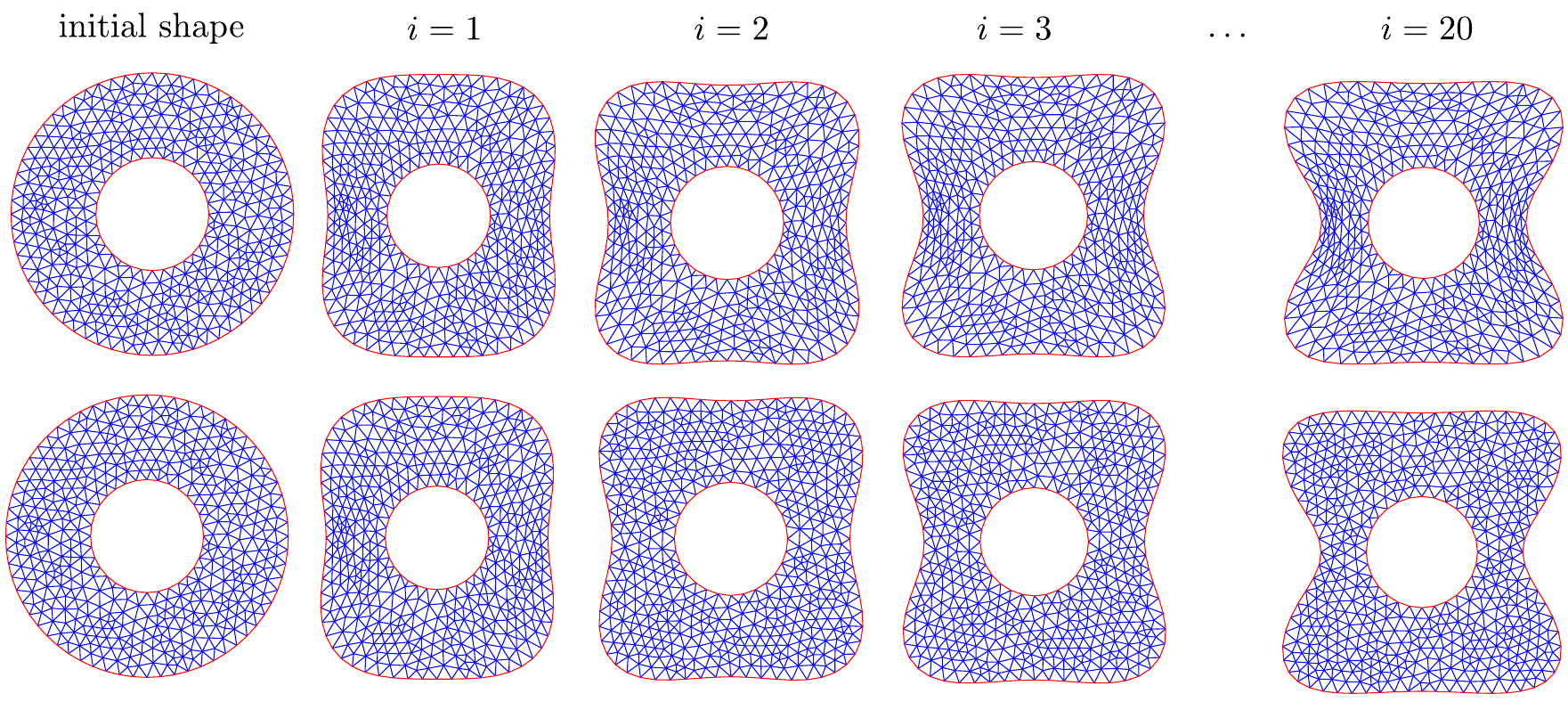}
\caption{\label{fig:levelset_shapes} Shapes encountered during the optimization iterations for different initial shapes using the VLB method and a coarse mesh ($h=0.1$). Top: no remeshing. Bottom: remeshing every second iteration.}
\end{figure}

Figure~\ref{fig:levelset_shapes} illustrates the optimization process with and without remeshing for a coarse discretization to give an overview. 
The mean edge length is set to $h=0.1$ for this case.
In the following, a finer mesh with $h=0.05$ is used if not stated differently. 
Preliminary investigations based on a solution with $h=0.01$ show that the approximation error when evaluating $J$ drops below $10^{-6}$ then.

\begin{figure}
\centering
\includegraphics[width=\textwidth]{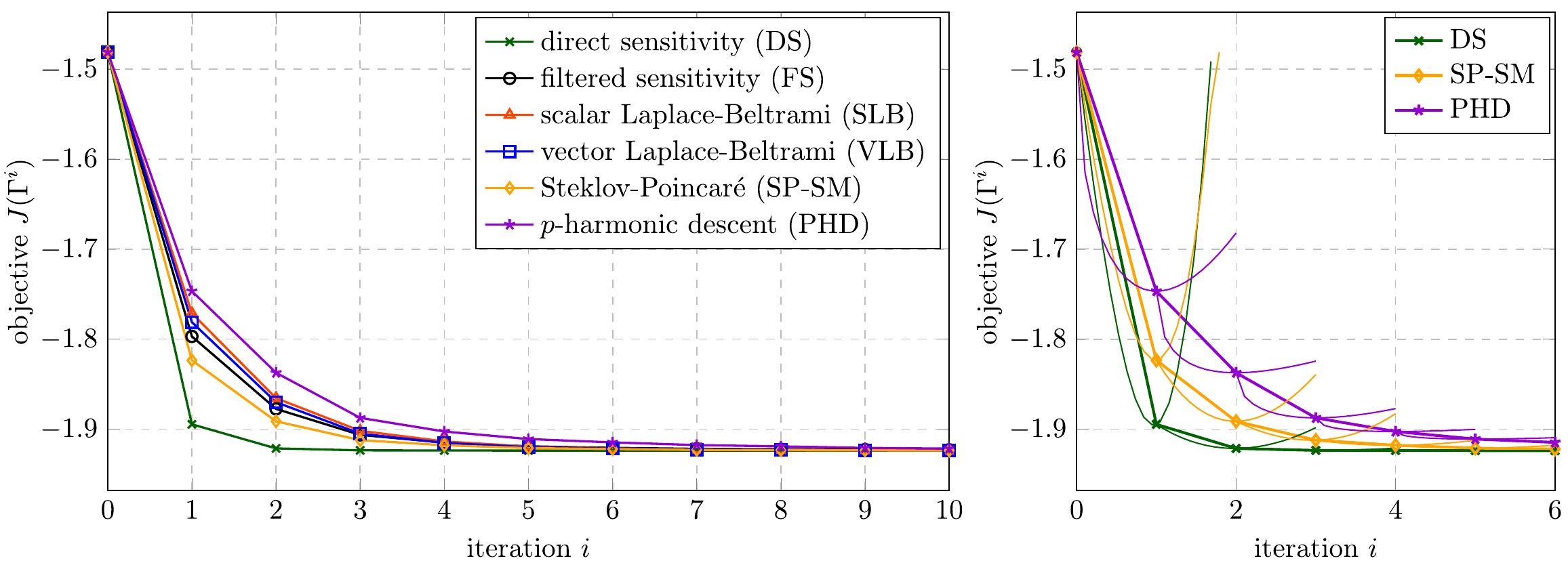}
\caption{\label{fig:levelset_convergence_smooth} Convergence of $J$ during the optimization iterations for different shape updates including values for (untaken) steps with sizes between $0$ and $2\,\alpha$.}
\end{figure}

To begin with, we consider the smooth case without high frequency content, i.e. $C_1=0$ and $C_2=0$.
Figure~\ref{fig:levelset_convergence_smooth} (left) shows the convergence of $J$ over the optimization iterations for the different approaches to compute the shape update.
For this particular example, the DS approach yields the fastest reduction of $J$, while the $PHD$ yields the slowest.
In order to ensure that the line search algorithm works correctly and does not terminate early due to mesh degeneration, a check was performed as shown in Figure~\ref{fig:levelset_convergence_smooth} (right).
The thin lines indicate the values of $J$ that correspond to steps with sizes from $0$ to $2\,\alpha$.
It can be seen that the line search iterations did not quit early but lead to the optimal step size at all times.

\begin{figure}
\centering
\includegraphics[width=0.9\textwidth]{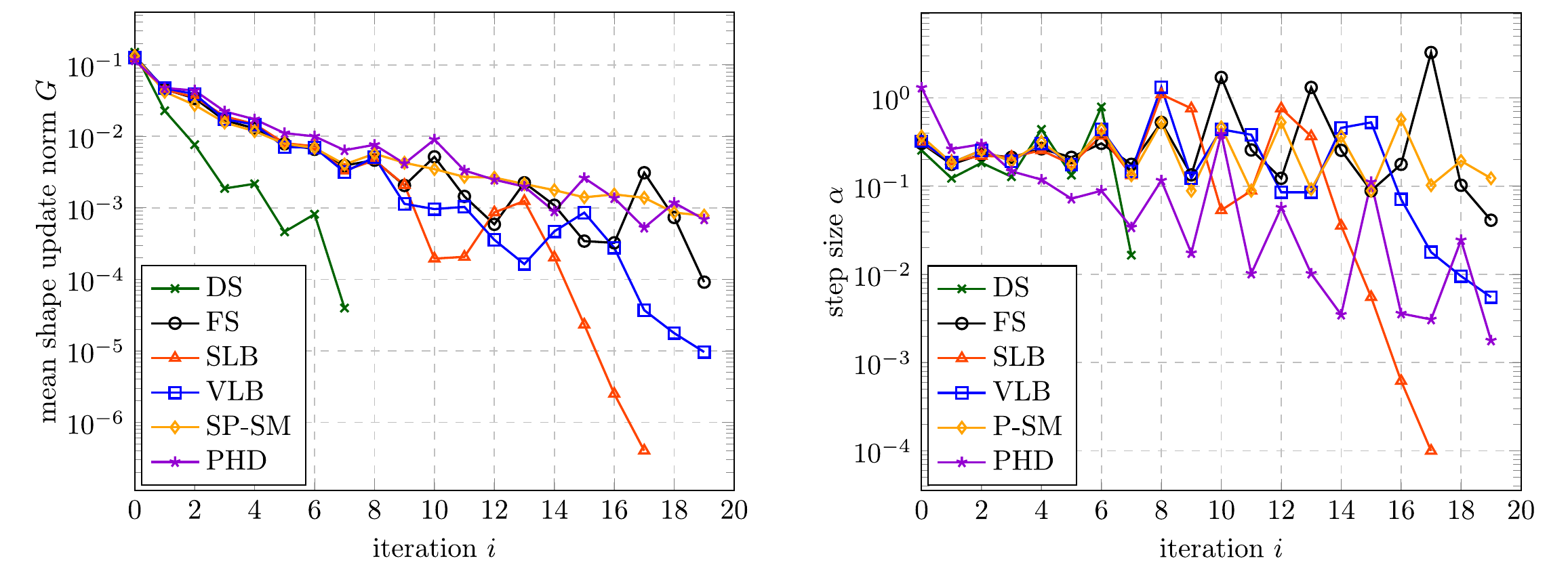}
\caption{\label{fig:levelset_gradient_and_stepsize_smooth} Left: Mean norm of the nodal boundary displacement. Right: Optimal step size.}
\end{figure}

The progression of the norm of the domain update direction and the step size is shown in Fig.~\ref{fig:levelset_gradient_and_stepsize_smooth}. 
More precisely, we plot there the mean norm of the displacement of all nodes on the boundary, i.e.
\begin{align}
    G = \frac{\alpha}{N^\mathrm{n}} \sum_{n=1}^{N^\mathrm{n}} \Vert \bm{\theta}_n \Vert_2,
\end{align}
where $N^\mathrm{n}$ is the total number of nodes on the boundary. 
As expected, $G$ converges to a small value, which yields no practical shape updates anymore after a certain number of iterations.

\subsubsection{Behavior under mesh refinement}
While we have ensured that the considered discretizations are fine enough to accurately compute the cost functional in a preliminary step, the effect of mesh refinement on the computed optimal shape shall be looked at more closely. 
To this end, the scenario $C_1=0$ and $C_2=0$ considered so far does not yield new insight.
All methods successfully converged to the same optimal shape as shown in Fig.~\ref{fig:levelset_shapes} and the convergence behavior was indistinguishable from that shown in Fig.~\ref{fig:levelset_convergence_smooth}.
This result was obtained with and without remeshing.

\begin{figure}
\centering
\includegraphics[width=\textwidth]{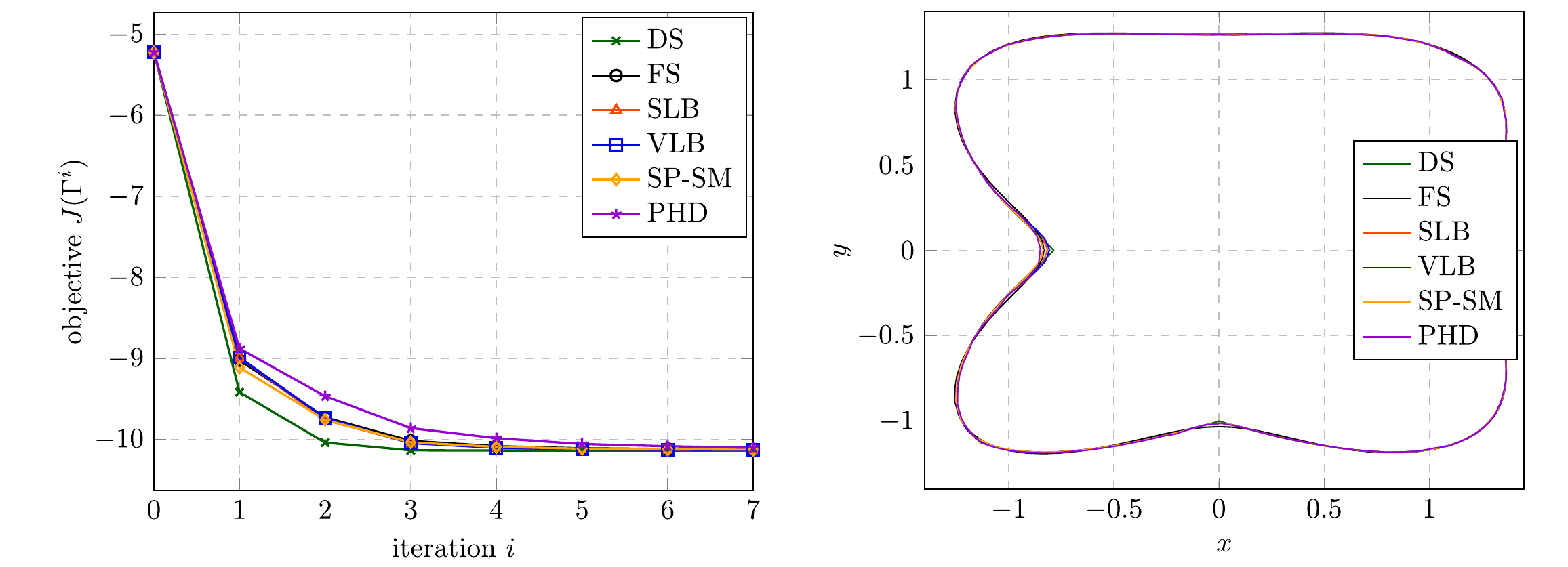}
\caption{\label{fig:levelset_convergence_anz} Results for $C_1=1$. Left: convergence of $J$ during the first optimization iterations for different shape updates. Right: Shapes  obtained after 20 iterations.}
\end{figure}
For the scenario  $C_1=1$ and $C_2=0$ with sharp corners (see Fig.~\ref{fig:levelset_function}), different behaviors were observed.
Figure~\ref{fig:levelset_convergence_anz} shows the convergence of the objective functional (left) and final shapes obtained with the different shape updates.
All shapes are approximately equal except in the region of the sharp corners on the $x$-axis close to $x=-1$ and on the $y$-axis close to $y=-1$.
\begin{figure}
\centering
\includegraphics[width=\textwidth]{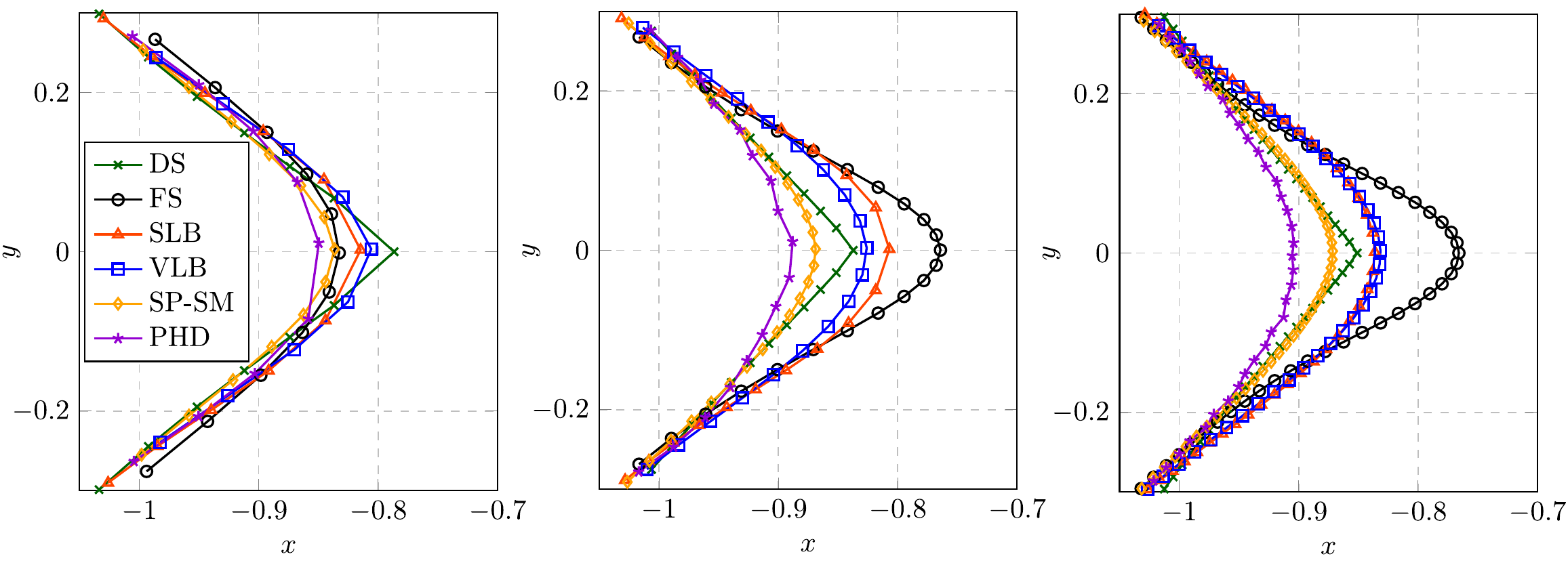}
\caption{\label{fig:levelset_geometry_anz} Geometries obtained for $C_1=1$ and $C_2=0$. Left: results for $h=0.05$. Middle: results for $h=0.025$. Right: results for $h=0.0125$.}
\end{figure}
Figure~\ref{fig:levelset_geometry_anz} shows a zoom into the region of the first sharp corner for the final shapes obtained with different mesh densities.
It is observed that only the DS approach resolves the sharp corner while all other approaches yield smoother shapes.
For further mesh refinements the obtained shapes were indistinguishable from those shown in Figure~\ref{fig:levelset_geometry_anz} (right).

\begin{figure}
\centering
\includegraphics[width=\textwidth]{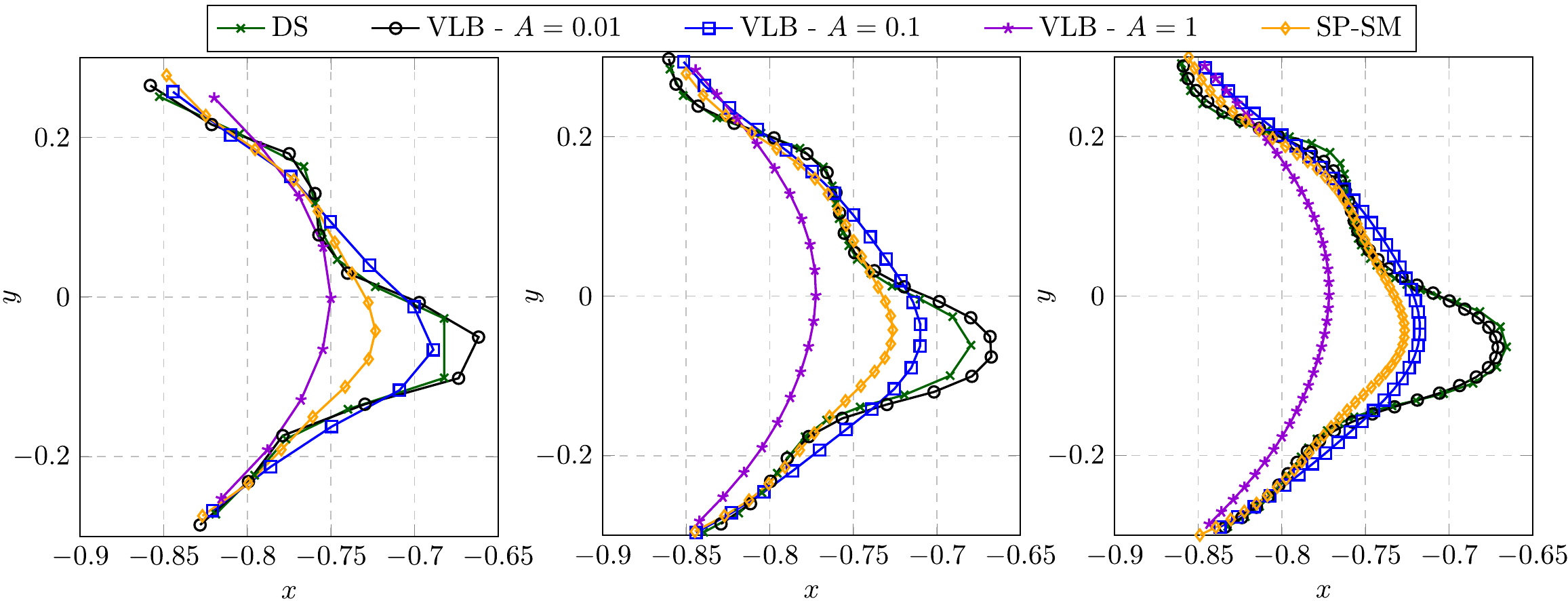}
\caption{\label{fig:levelset_convergence_bnz} Geometries obtained for $C_1=0$ and $C_2=1$. Left: results for $h=0.05$. Right: results for $h=0.025$.}
\end{figure}
Next we consider the scenario $C_1=0$ and $C_2=1$, which introduces high-frequency content into the optimal shape.
The high-frequency content may be interpreted in two different ways, when making an analogy to real world applications.
\begin{enumerate}
    \item It may represent a numerical artifact, arising due to the discretization of the primal and the adjoint problem (we do not want to find it in the predicted optimal shape then).
    \item It may represent physical fluctuations, e.g. due to a sensitivity that depends on a turbulent flow field (we do not want to find it in the predicted optimal shape then). 
    \item It may represent the actual and desired optimal shape (we want to find it in the predicted optimal shape).
\end{enumerate}
With this being said, no judgement about the suitability of the different approaches can be made. Depending on the interpretation, a convergence to a shape that includes the high-frequency content can be desired or not.

Fig.~\ref{fig:levelset_convergence_bnz} shows the shapes obtained with selected approaches when refining the mesh.
The approaches FS, SLB and PHD were excluded because they yield qualitatively the same results as the SP-SM approach, i.e. convergence to a smooth shape without high frequency content.
In order to illustrate the influence of the conductivity $A$, three variants are considered for the VLB approach. 
For a large conductivity of $A=1$, the obtained shape is even smoother than that obtained for the SP-SM approach, while $A=0.1$ (the value chosen so far in all studies) yields a similar shape. 
Reducing the conductivity to $A=0.01$, the obtained shape is similar to that obtained for the DS approach, which does resolve the high frequency content.

\subsubsection{Behavior for a non-smooth initial shape}
\begin{figure}
\centering
\includegraphics[width=0.35\textwidth]{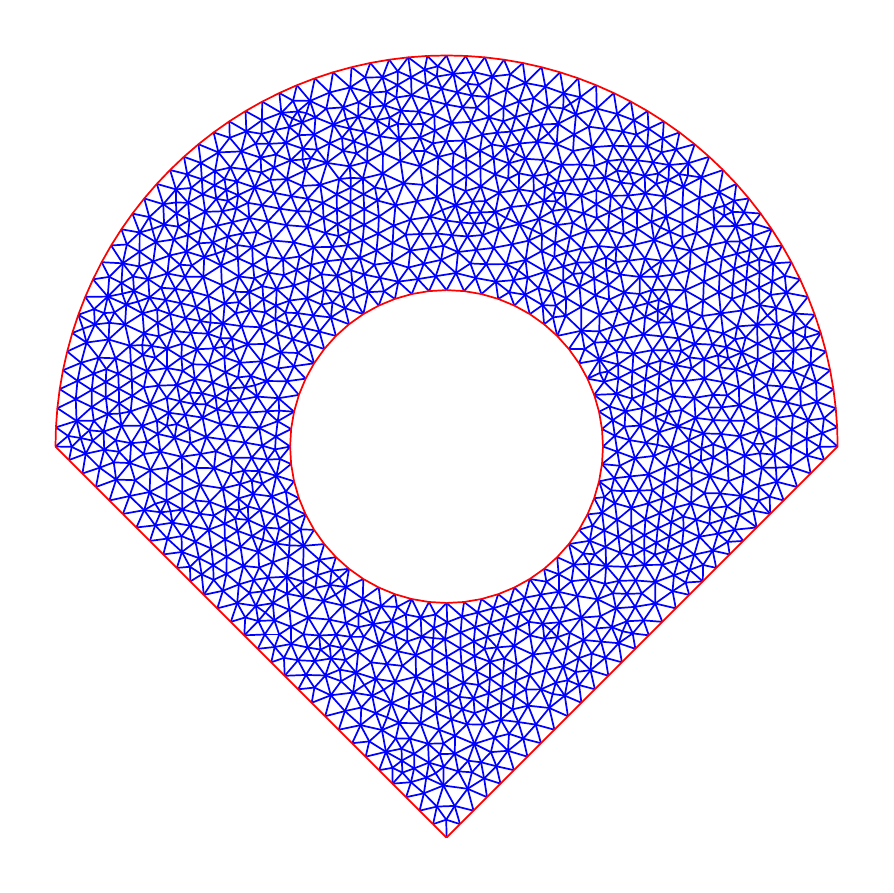}
\includegraphics[width=0.5\textwidth]{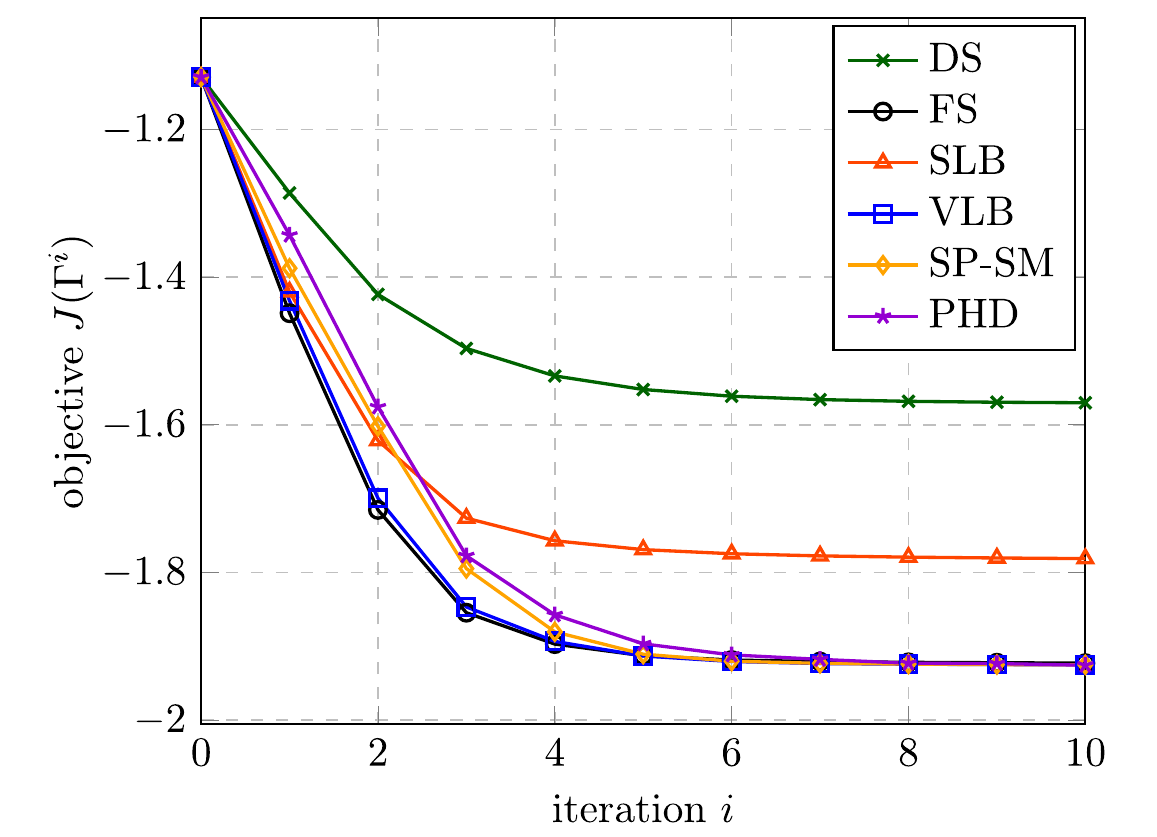}
\caption{\label{fig:levelset_convergence_diamond} Left: initial shape with sharp corners. Right: convergence of $J$ during the optimization iterations for different shape updates.}
\end{figure}
Finally, we test the robustness of the different shape updates by starting the optimization process from a non-smooth initial shape. A corresponding mesh is shown in Fig.~\ref{fig:levelset_convergence_diamond} (left).
The convergence behavior in Fig.~\ref{fig:levelset_convergence_diamond} (right) already indicates that not all approaches converged to the optimal shape.
Instead, the DS and the SLB approach yield different shapes with a much higher value of the objective functional.

\begin{figure}
\centering
\includegraphics[width=0.245\textwidth]{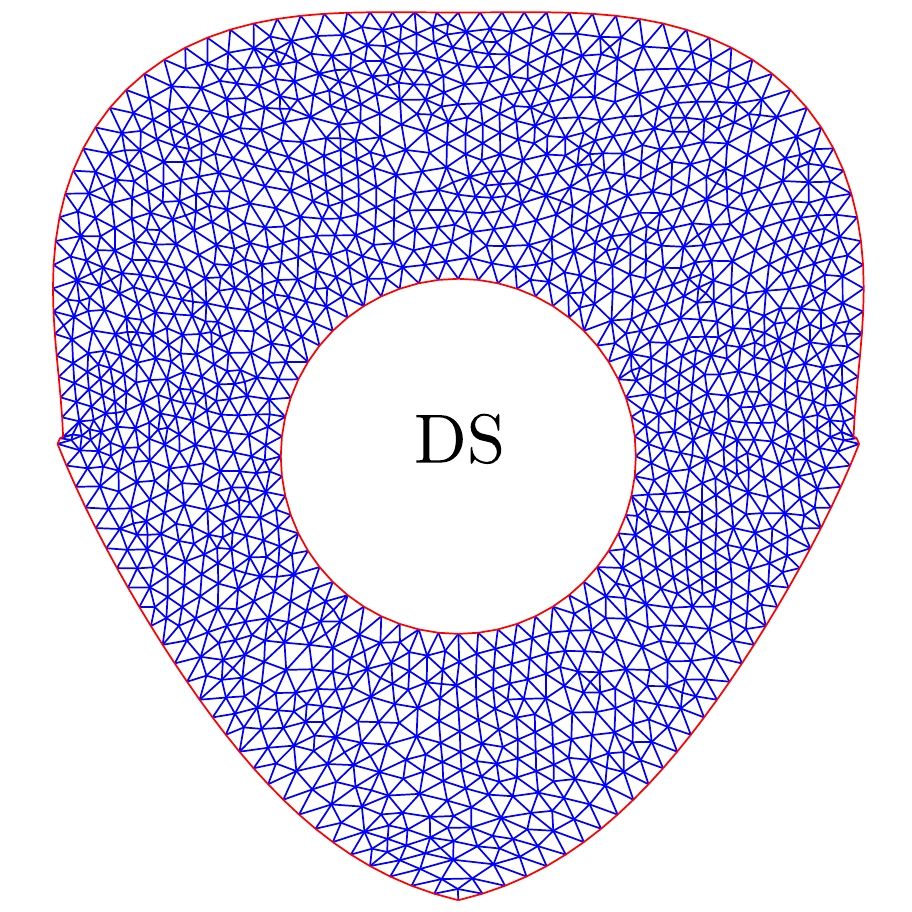}
\includegraphics[width=0.245\textwidth]{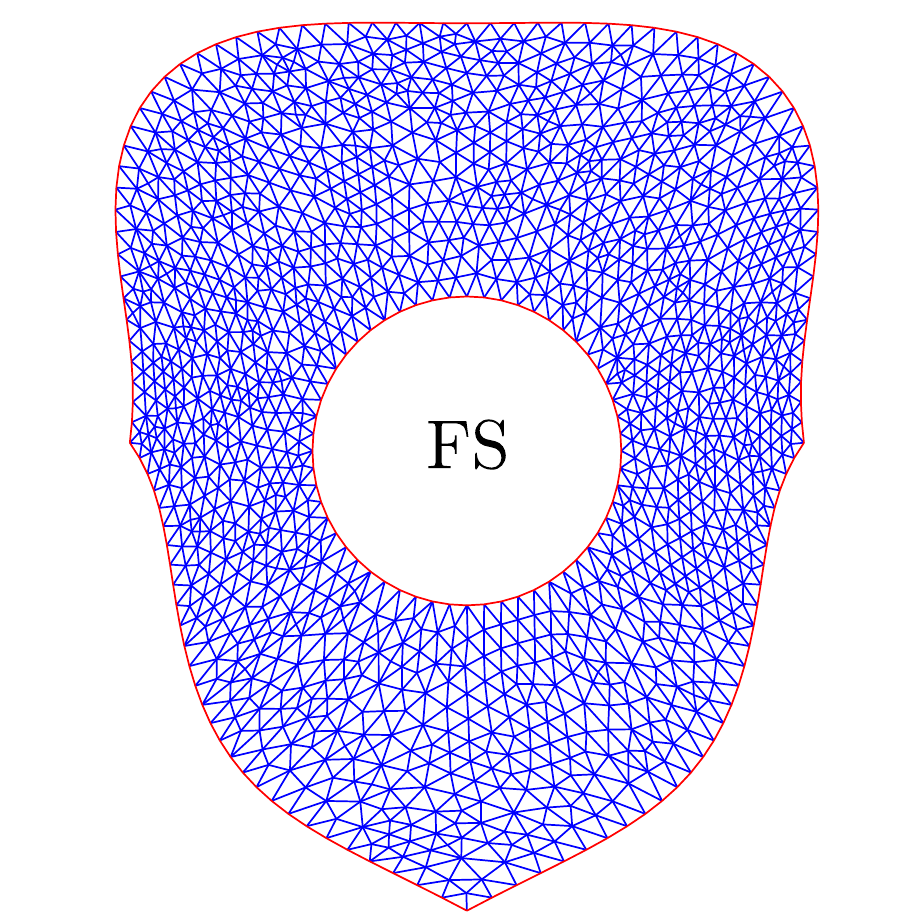}
\includegraphics[width=0.245\textwidth]{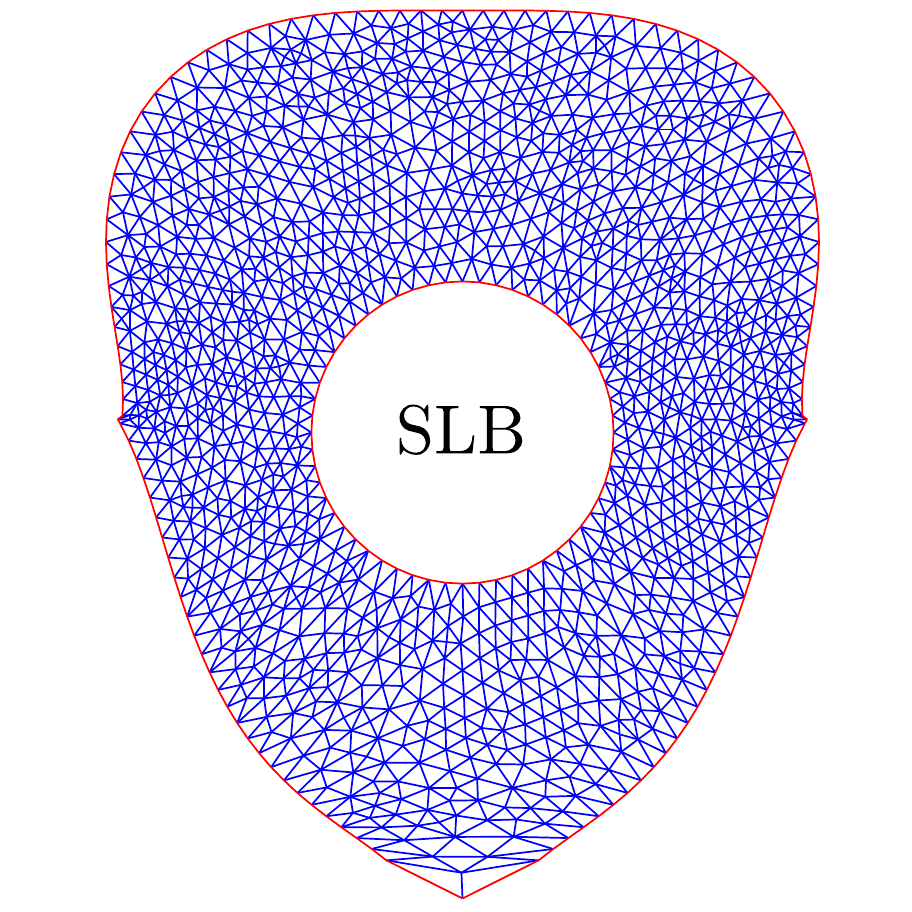}
\includegraphics[width=0.245\textwidth]{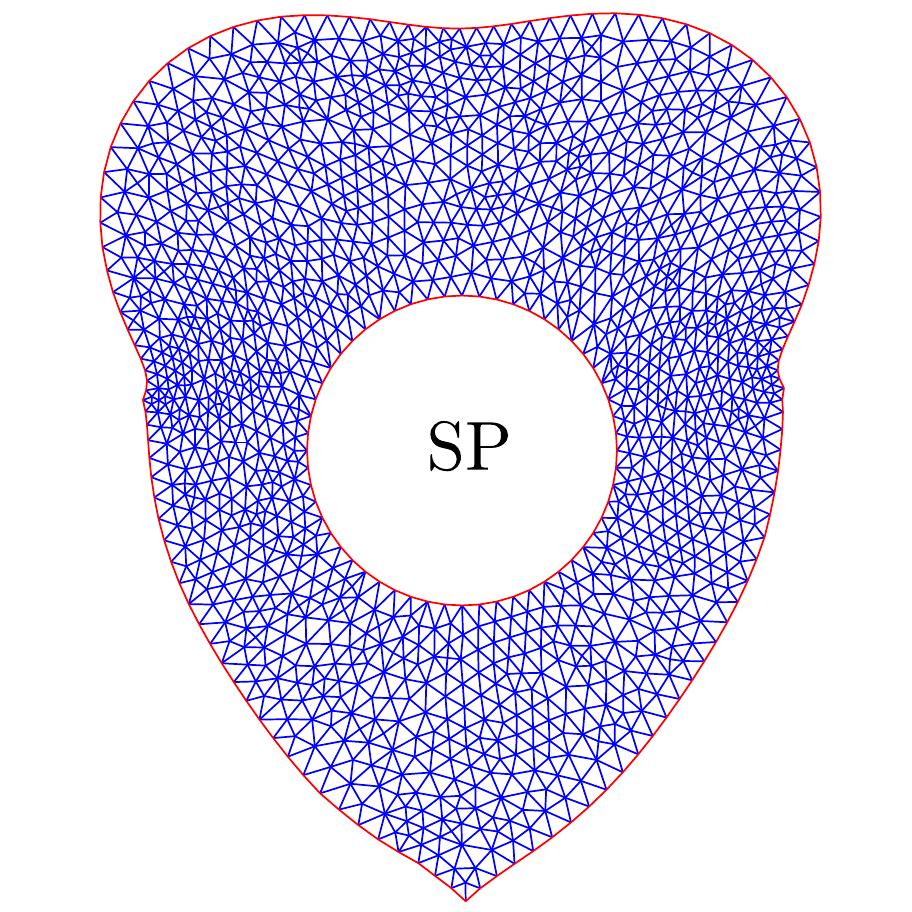}

\vspace{1em}

\includegraphics[width=0.245\textwidth]{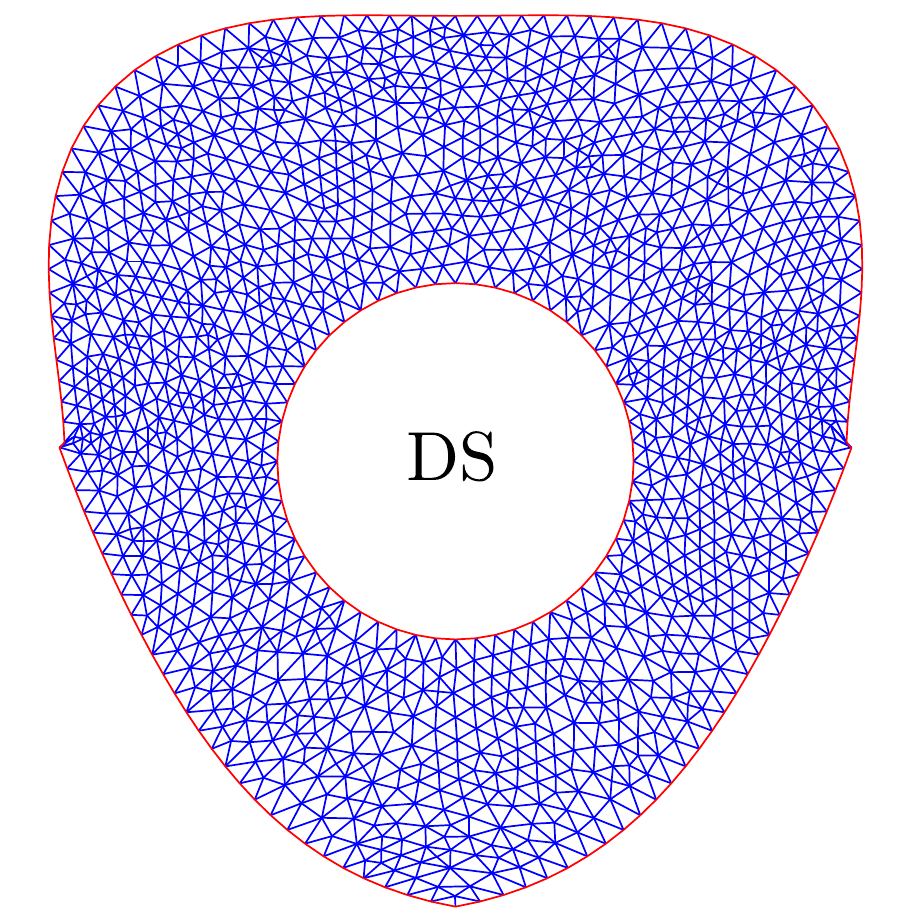}
\includegraphics[width=0.245\textwidth]{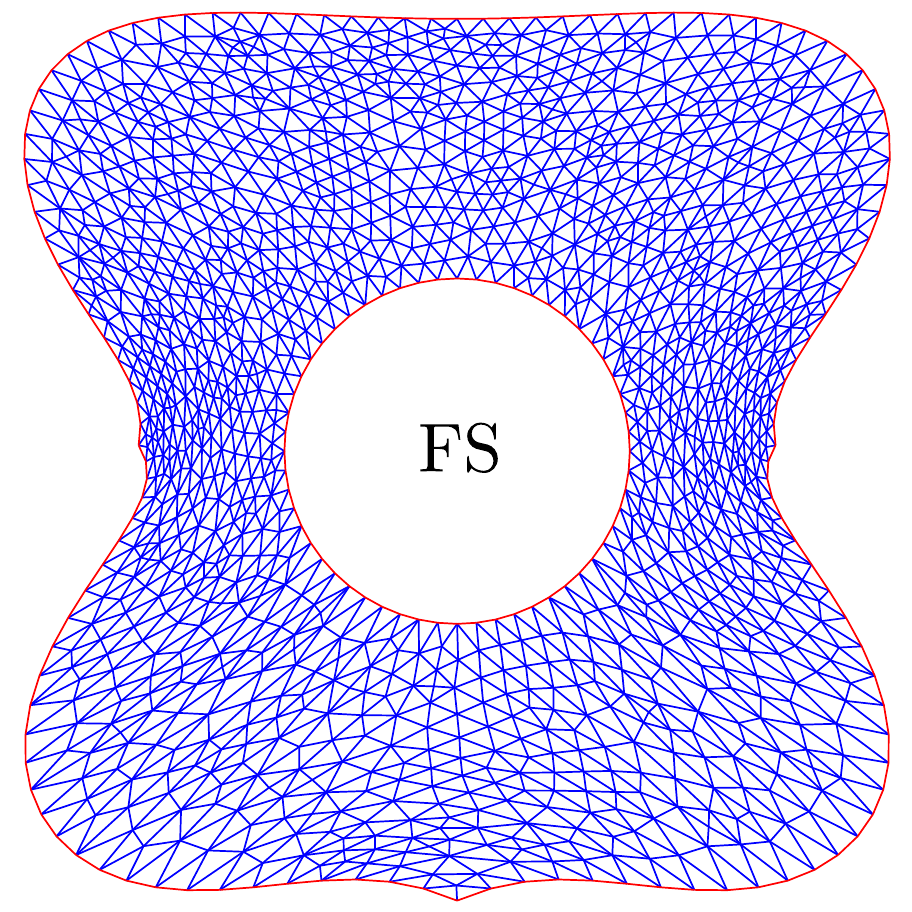}
\includegraphics[width=0.245\textwidth]{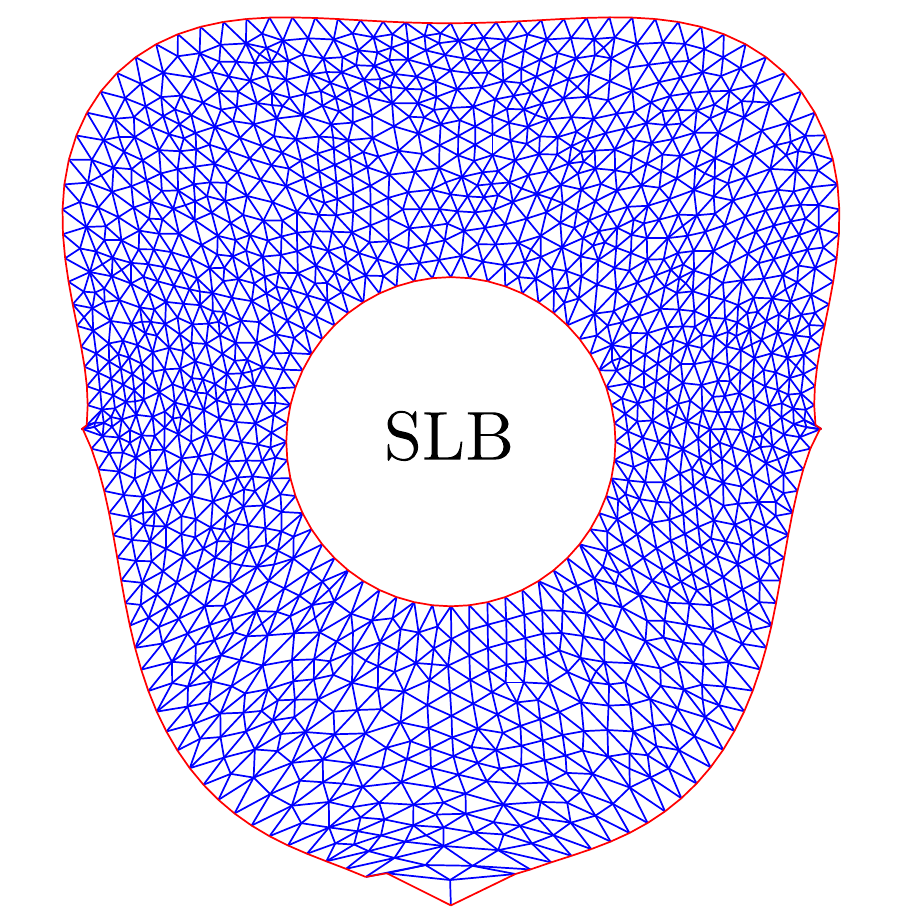}
\includegraphics[width=0.245\textwidth]{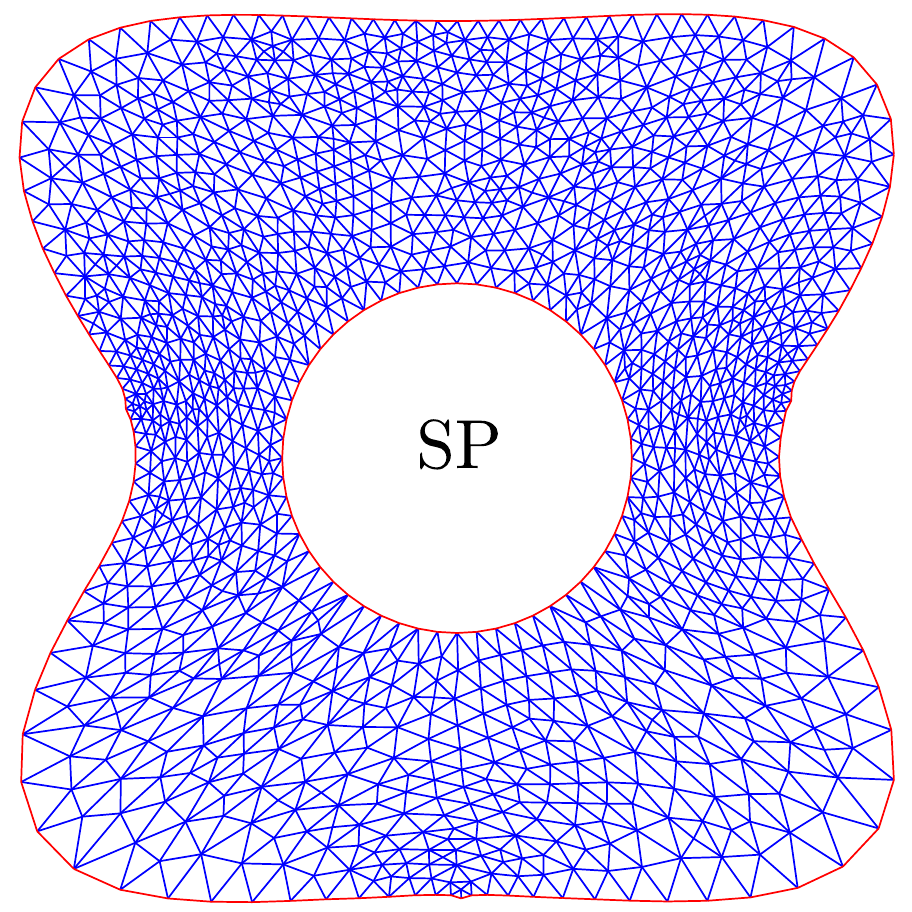}
\caption{\label{fig:levelset_geometry_diamond} Meshes encountered during selected optimizations based on a non-smooth initial shape without remeshing. Top: meshes after the first iteration. Bottom: meshes after 20 iterations.}
\end{figure}
Figure~\ref{fig:levelset_geometry_diamond} provides an explanation for the convergence behavior. 
After the first iteration, the DS and the SLB approach show a severe mesh distortion in those regions, where the initial shape had a sharp corner (see Fig.~\ref{fig:levelset_convergence_diamond} (left)).
In order to prevent at least self-penetration of the triangular elements, the step sizes become very small for the following iterations and after 9 (for DS) or 8 (for SLB) iterations, no step sizes larger that $10^{-6}$ could be found that reduce the objective functional.
Opposed to that, the FS and the SP approach yield shapes which are very close to the optimal shape.
Still, the initial corners are visible also for these approaches, not only due to the distorted internal mesh but also as a remaining corner in the shape, which is more pronounced for the FS approach.
The VLB and the PHD approach behave very similar to the SP approach and are therefore not shown here.

We would like to emphasize that even if different approaches yield approximately the same optimal shape, the intermediate shapes, i.e. the path taken during the optimization, may be fundamentally different as apparent in Fig.~\ref{fig:levelset_geometry_diamond}. 
This is to be kept in mind especially when comparing the outcome of optimizations with different shape updates that had to be terminated early. e.g. due to mesh degeneration, which is the case for several of the studies presented in the next section.

\FloatBarrier

\section{Exemplary applications}
\label{sec:cfd_application}

In this section we showcase CFD-based shape optimization applications on a 2D and 3D geometry, while considering the introduced shape update approaches. Emphasis is given to practical aspects and restrictions that arise during an optimization procedure. The investigated applications refer to steady, laminar internal and external flows. 
The optimization problems are constrained by the Navier-Stokes (NS) equations of an incompressible, Newtonian fluid with density $\rho$ and dynamic viscosity $\mu$, viz.
\begin{align}
    R^p &= - \mathrm{div}( \bm{u} ) = 0  \, ,  
    \label{eq:primal_continuity} 
    \\
    \bm{R}^{\bm{u}} &= \rho \, \nabla \bm{u} \, \bm{u} - \mathrm{div}(  2 \, \mu \, \bm{S} - p \, \bm{I} ) = \bm{0} \, ,
    \label{eq:primal_momentum}  
\end{align}
where, $\bm{u}$, $p$, $\bm{S} = 1/2 (\nabla \bm{u} + (\nabla \bm{u})^\mathrm{T})$ and $\bm{I}$ refer to the velocity, static pressure, strain-rate tensor and identity tensor, respectively.
The adjoint state of (\ref{eq:primal_continuity})-(\ref{eq:primal_momentum}) is defined by the adjoint fluid velocity $\hat{\bm{u}}$ and adjoint pressure $\hat{p}$ that follow from the solution of
\begin{align}
    R^{\hat{p}} &= - \mathrm{div}( \hat{\bm{u}} ) = 0,  
    \label{eq:adjoint_continuity} \\
    \bm{R}^{\hat{\bm{u}}} &=  \rho \left( (\nabla \bm{u})^\mathrm{T} \, \hat{\bm{u}} - \nabla \hat{\bm{u}} \, \bm{u} \right) - \mathrm{div}(  2 \, \mu \, \hat{\bm{S}} - \hat{p} \, \bm{I} ) = \bm{0} \, ,
    \label{eq:adjoint_momentum}  
\end{align}
where, $\hat{\bm{S}}=1/2 (\nabla \hat{\bm{u}} + (\nabla \hat{\bm{u}})^\mathrm{T})$ refers to the adjoint strain rate tensor.

The employed numerical procedure refers to an implicit, second-order accurate finite-volume method (FVM) using arbitrarily-shaped/structured polyhedral grids. 
The segregated algorithm uses a cell-centered, collocated storage arrangement for all transport properties, cf. \cite{rung2009challenges}. The primal and adjoint pressure-velocity coupling, which has been extensively verified and validated
\cite{stuck2013adjoint, kroger2018adjoint, kuehl2019decoupling, kuhl2021adjoint, bletsos2021adjoint}, follows the SIMPLE method, and possible parallelization is realized using a domain decomposition approach \cite{yakubov2013hybrid, yakubov2015experience}. Convective fluxes for the primal [adjoint] momentum are approximated using the Quadratic Upwind [Downwind] Interpolation of Convective Kinematics (QUICK) [QDICK] scheme \cite{stuck2013adjoint} and the self-adjoint diffusive fluxes follow a central difference approach.

The auxiliary problems of the various approaches to compute a shape update are solved numerically using the finite-volume strategies described in the previously-mentioned publications. 
Accordingly, $\bm{\theta}$ is computed at the cell centers $\bm{c}_c$ in a first step.
In a second step, it needs to be mapped to the nodal positions $\bm{x}_n$, which is done using an inverse distance weighting, also known as Shepard's interpolation \cite{shepard1968two}.
We use $\bm{\theta}_n$ to denote the value at a node
\begin{align}
\label{equ:mesh_update}
\bm{\theta}_n &= \frac{1}{N_n^\mathrm{c}} \sum_\mathrm{c \in C_n} 
\bm{\theta}(\bm{c}_c) \left(
1
-
\frac{||\bm{x}_n - \bm{c}_c||}{\sum_{d \in C_n} ||\bm{x}_n - \bm{c}_d||} \right). 
\end{align}
Therein, $C_{n}$ contains the $N_n^\mathrm{c}$ indices of all adjacent cells at node $n$.
After the update of the grid, geometric quantities are recalculated for each FV. Topological relationships remain unaltered and the simulation continues by restarting from the previous optimization step to evaluate the new objective functional value. Due to the employed iterative optimization algorithm and comparably small step sizes, field solutions of two consecutive shapes are usually nearby. Compared to a simulation from scratch, a speedup in total computational time of about one order of magnitude is realistic for the considered applications.

\subsection{Two-dimensional flow around a cylinder}\label{sec:2d_cylinder}
\begin{figure}
\centering
\subfigure[]{
\centering
\includegraphics[width=0.4\textwidth]{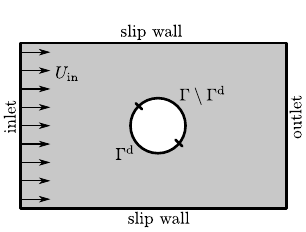}
}
\subfigure[]{
\centering
\includegraphics[width=0.45\textwidth]{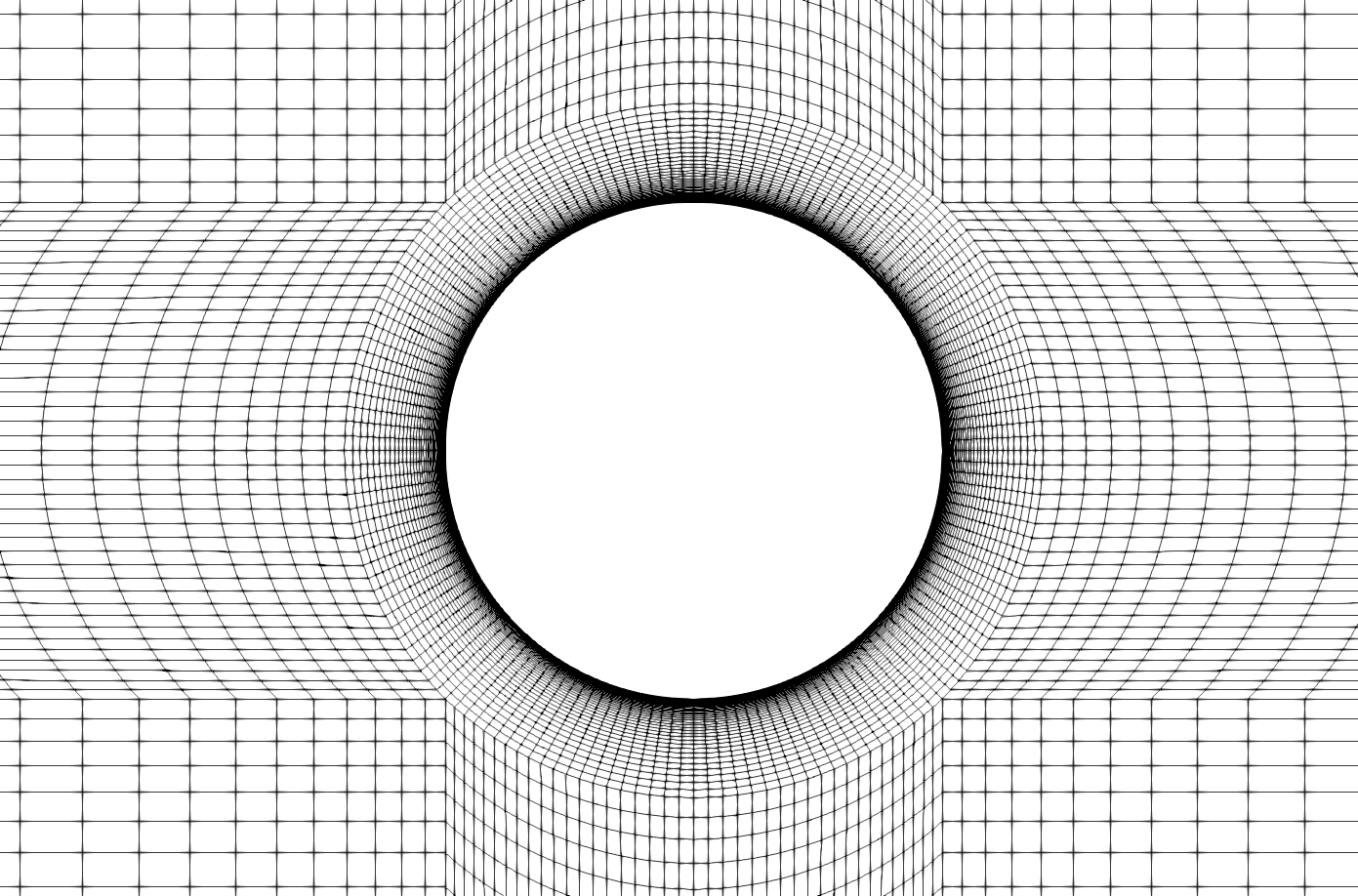}
}
\caption{Cylinder ($\mathrm{Re}_\mathrm{D} = 20$): (a) Sketch of the investigated 2D optimization problem where the dashed line denotes the section free for design ($\Gamma^\mathrm{d}$) and (b) detail of the employed numerical grid near the cylinder.}
\label{fig:cylinder_2d_sketch}
\end{figure}

We consider a benchmark problem which refers to a fluid flow around a cylinder, as schematically depicted in Fig. \ref{fig:cylinder_2d_sketch}(a).
This application targets to minimize the flow-induced drag of the cylinder by optimizing parts of its shape. The objective $J(\Gamma)$ and its shape derivative read
\begin{align}
    J(\Gamma) = \int_{\Gamma} \left( p \, \bm{I} - 2 \mu \bm{S} \right) \bm{n} \cdot \bm{e}_1 \mathrm{d} \Gamma
    \qquad \mathrm{and} \qquad
    J^\prime(\Gamma)(\bm{v}^\Gamma) = - \int_{\Gamma^\mathrm{d}} 
    \underbrace{\left( \mu \, \nabla \bm{u}\, \bm{n} \cdot \nabla \hat{\bm{u}} \, \bm{n} \right)}_{s} \, \bm{v}^\Gamma \cdot \bm{n} \, \mathrm{d} \Gamma
    \label{equ:zero_and_first_objec_derivative} \, ,
\end{align}
where $\bm{e}_1$ denotes the basis vector in the $x$-direction (the main flow direction), see \cite{kuehl2019decoupling} for a more detailed explanation.
Note that the objective is evaluated along the complete circular obstacle $\Gamma$, but its shape derivative is evaluated only along the section under design $\Gamma^\mathrm{d}$ as shown in Fig. \ref{fig:cylinder_2d_sketch}(a).
The decision of optimizing a section of the obstacle's shape instead of the complete shape is made to avoid trivial solutions such as, e.g. a singular point or a straight line without the need for applying additional geometric constraints.
The steady and laminar study is performed at $\mathrm{Re}_D = \rho \, U_\mathrm{in} \, D / \mu = \SI{20}{}$ based on the cylinder's diameter $D$ and the inflow velocity $U_\mathrm{in}$. The two-dimensional domain has a length and height of $40 \, D$ and $20 \, D$, respectively.
At the inlet, velocity values are prescribed, slip walls are used along the top as well as bottom boundaries and a pressure value is set along the outlet. 

To ensure the independence of the objective functional $J$ and its shape derivative $J'$ in Eq. (\ref{equ:zero_and_first_objec_derivative}) w.r.t. the spatial discretization, a grid study is first conducted, as presented in Tab. \ref{tab:mesh_sens}.
Since the monitored integral quantities do not show a significant change from refinement level 4 on, level 3 is employed for all following optimizations.
A detail of the utilized structured numerical grid 
is displayed in Fig. \ref{fig:cylinder_2d_sketch} (b) and
consists of approximately $\si{19000}$ control volumes. The cylinder is discretized with 200 surface patches along its circumference. 
\begin{table}
\caption{Cylinder ($\mathrm{Re}_\mathrm{D} = 20$): Results of the mesh dependence study. For illustrative purposes we denote here $\hat{J}'=\int_{\Gamma^{\mathrm{d}}} s \, d\Gamma$. Index $i$ refers to the mesh refinement level. Note $\rho = 20$ $\mathrm{kg/m^3}$, $\mu = 1$ $\mathrm{Pa \cdot s}$, $U_{in}=1$ $\mathrm{m/s}$ and $D= 1$ $\mathrm{m}$.}
\begin{tabular}[ht]{cccccc}
 \toprule
refinement level & number of FV &   $\frac{2J_i}{\rho U^2_{in} D^2}$ & $\frac{2\hat{J}'_i}{\rho U^2_{in} D}$ & $\frac{J_i-J_{i-1}}{J_{i-1}} (\%)$  & $\frac{\hat{J}'_i-\hat{J}'_{i-1}}{\hat{J}'_{i-1}} (\%)$ \vspace{0.9em}  \\  
\hline 
 M0	                 & 300  &   2.1197   & -3.325 & - & -   \\
\hline
 M1	                 & 1200  &  2.1433 &  -3.612 & 1.11 & -8.64   \\
\hline
 M2	                 & 4800  &   2.1356 &  -3.822 & -0.35 & -5.81  \\
\hline
 \textbf{M3}	                 & \textbf{19200}  &   \textbf{2.1334} &  \textbf{-3.937} & \textbf{-0.11} & \textbf{-3.01} \\
\hline
 M4	                 & 76800  &   2.1334 &  -3.932 & -0.003 & 0.14   \\
\hline
 M5	                 & 307200  &   2.1334 &  -3.936 & -0.001 & -0.11   \\
 \bottomrule
\end{tabular}
\centering
\centering
\label{tab:mesh_sens}
\end{table}
\par In contrast to the theoretical framework,
we now have to take into consideration further practical aspects in order to realize our numerical optimization process. A crucial aspect that needs to be taken into account in any CFD simulation is the quality of the employed numerical grid. As the optimization progresses, the grid is deformed on the fly rather than following a remeshing approach. Hence, we have to ensure that the quality of the mesh is preserved to such an extent that the numerical solution converges and produces reliable results. 
An intuitive method to ensure that grid quality is not heavily deteriorated is to restrict large deformations by using a small step size $\alpha$.

In the numerical investigations of the 2D case, the step size remains constant through the optimization process and is determined by prescribing the maximum displacement in the first iteration ($\theta^\mathrm{max}$) as described in Sec.~\ref{sec:mesh_morphing}.
We set it to two percent of the diameter of the cylinder, i.e. $\theta^\mathrm{max} = 0.02 \, D$, based on the experience of the authors on this particular case, cf. \cite{kuhl2022adjoint}. Further investigations in combination with the line search method are presented in Appendix A.

\subsubsection{Results}
\label{subsec:2d_results}
\begin{figure}
\centering
\includegraphics[width=\textwidth]{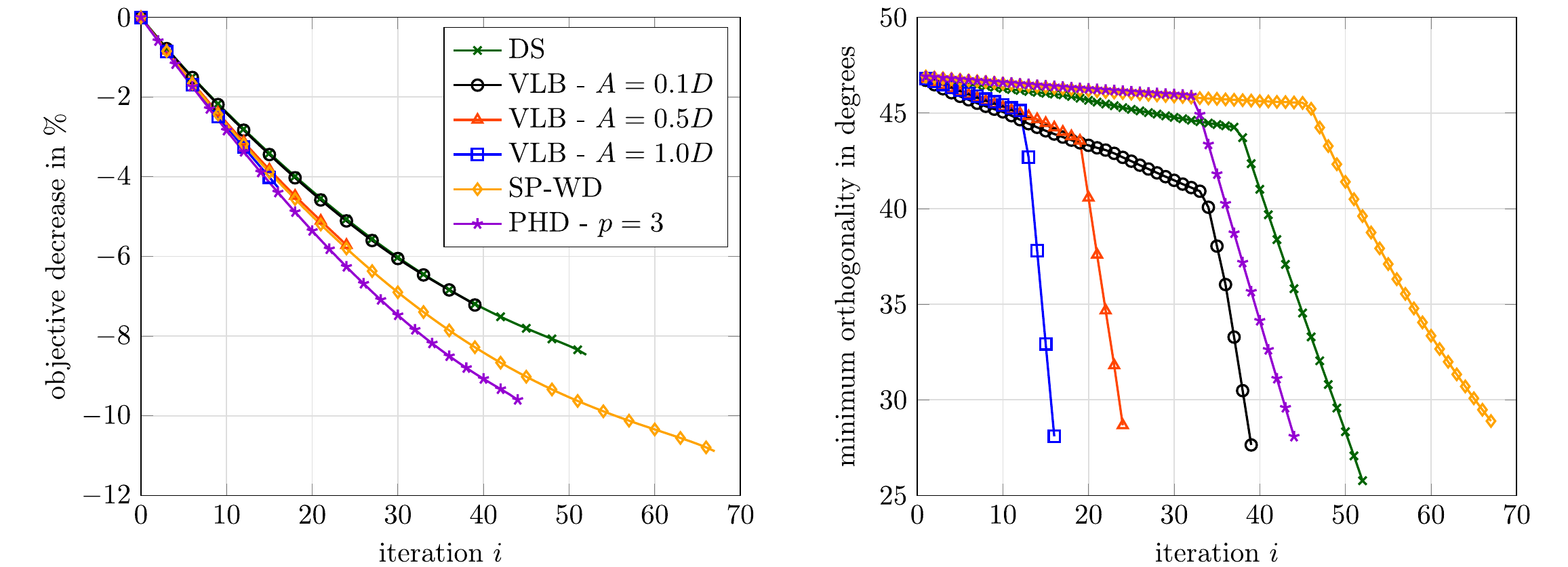}
\caption{Cylinder ($\mathrm{Re}_\mathrm{D} = 20$): (a) Relative decrease $(J_i- J_0)/J_0 \cdot 100\%$ of objective ($J(\Omega)$). (b) Minimum cell orthogonality of the computational meshes. Figures (a) and (b) share the same legend.}
\label{fig:cylinder_2d_convergence}
\end{figure}

\begin{figure}
\centering
\subfigure[]{\includegraphics[width=0.45\textwidth]{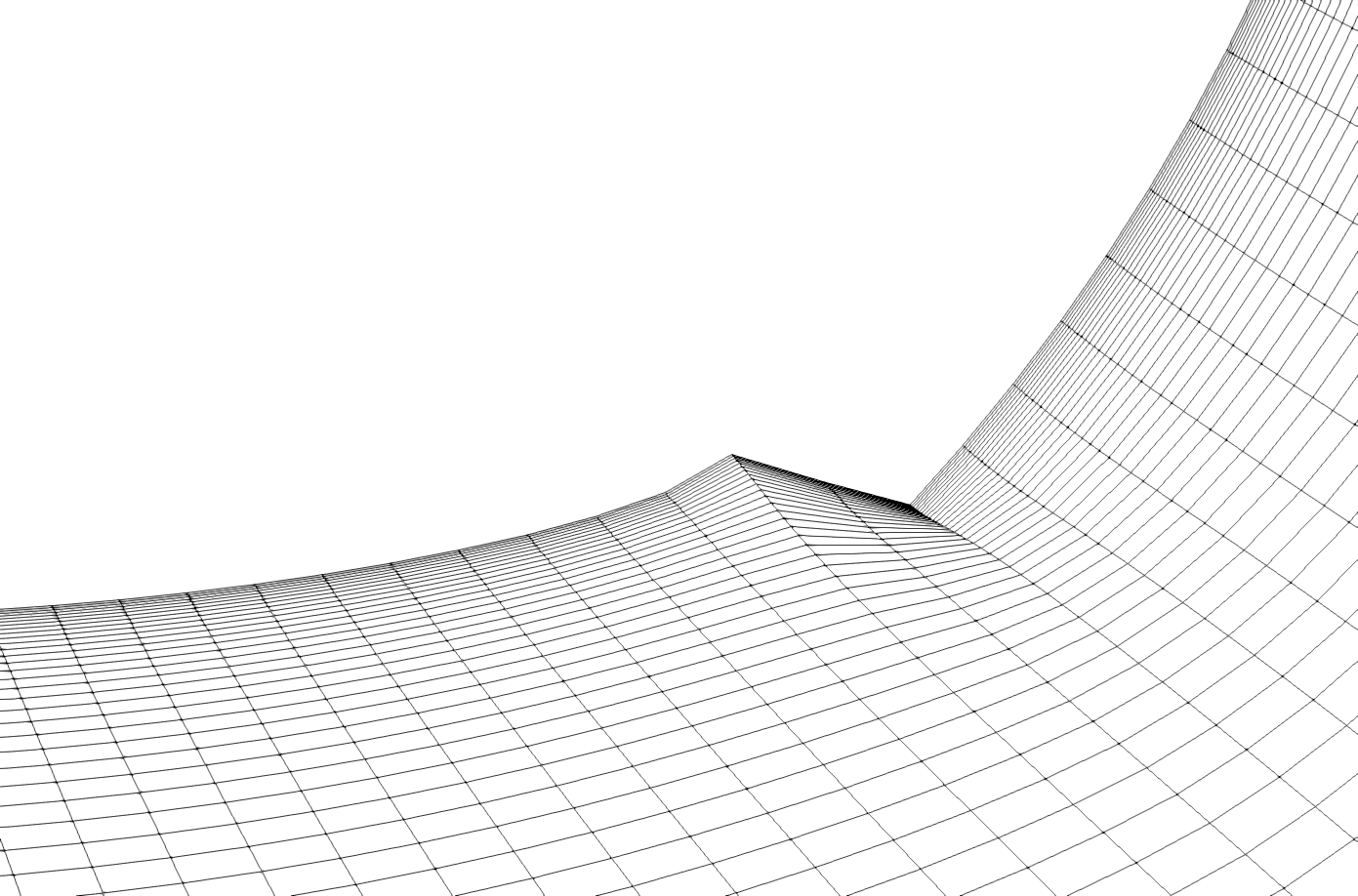}}
\subfigure[]{\includegraphics[width=0.45\textwidth]{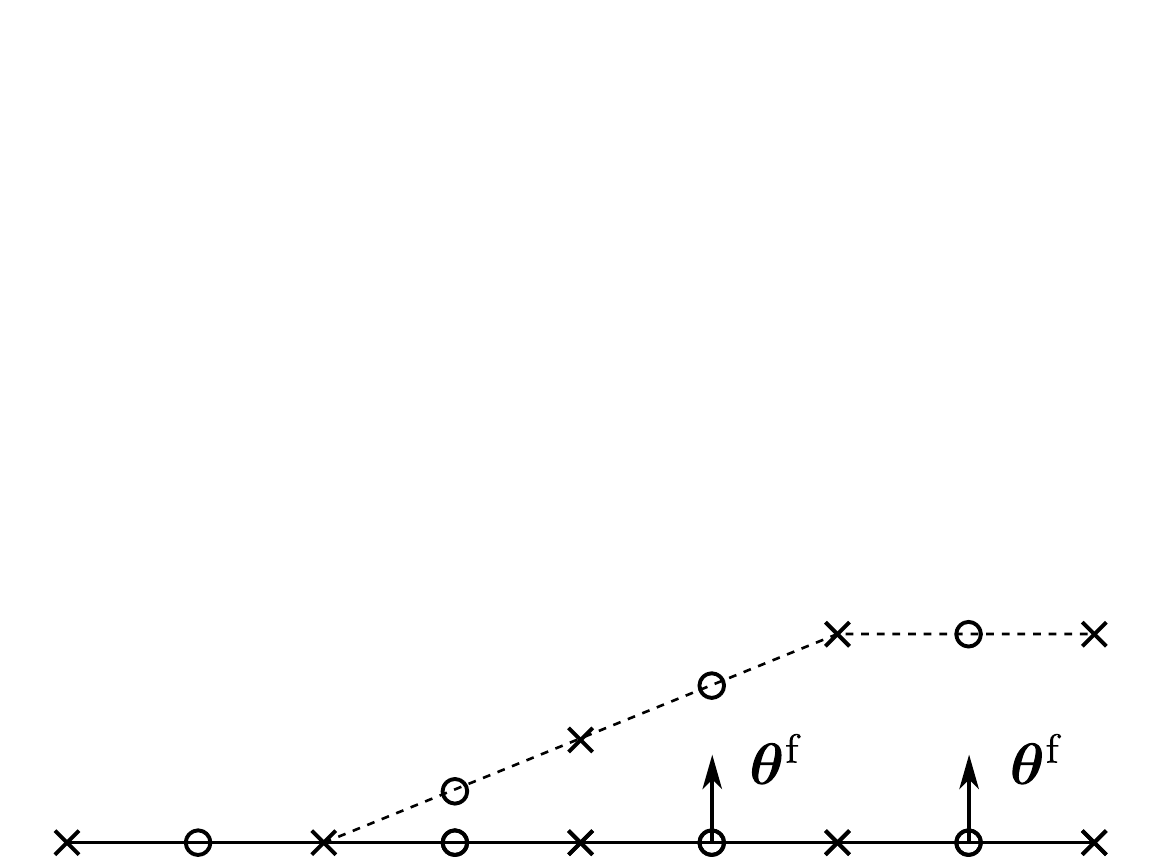}}
\caption{Cylinder ($\mathrm{Re}_\mathrm{D} = 20$): (a) Detail of the numerical grid at the rightmost connection point between $\Gamma \backslash \Gamma^\mathrm{d}$ and $\Gamma^\mathrm{d}$ in the last optimization iteration with the approach VLB - $A=0.1D$. (b) one-dimensional illustrative example for a mesh update (see Eq. (\ref{equ:mesh_update})). Face centers are shown with circles while vertices are displayed by $\times$ marks. The arrows denote the shape update direction, $\bm{\theta}^\mathrm{f}$, at the face centers. Solid line depicts the initial and dashed line the deformed discretized shape.}
\label{fig:cylinder_2d_connection}
\end{figure}

The investigated approaches are DS, VLB with $A=0.1D$, VLB with $A=0.5D$, VLB with $A=D$, SP-WD and PHD.
For all approaches that yield $\bm{\theta}^\Gamma$ only, the extension into the domain is done as described in Sec.~\ref{sec:mesh_morphing} (see Eq. (\ref{equ:domain_transport})) with a constitutive relation based on Eq. (\ref{eq:constituive_wall_distance}).
Figure \ref{fig:cylinder_2d_convergence}(a) shows the relative decrease of $J(\Gamma)$ w.r.t. the initial shape, for all aforementioned approaches. 
As it can be seen, the investigated domain expressions SP-WD \& PHD managed to reach a reduction greater than 9\% while the remaining boundary expressions fell shorter at a maximum reduction of 8.2\% by the DS approach. 
In the same figure one can notice, that none of the employed approaches managed to reach a converged state with its applied constant step size. 
The reason behind this shortcoming is shown in Fig. \ref{fig:cylinder_2d_convergence}(b) where the minimum orthogonality of the computational mesh is monitored during the optimization runs.
In all cases, mesh quality is heavily deteriorated during the final steps of the optimization algorithm leading to unusable computational meshes. 
This is partially attributed to the selected section of design ($\Gamma^\mathrm{d}$) and the mesh update approach, as described by Eq. (\ref{equ:mesh_update}). 
A natural question that one may ask by virtue of Eq. (\ref{equ:mesh_update}) is  \textit{what happens at nodes connecting a design and a non-design surface patch}. 
To this end, we present Fig. \ref{fig:cylinder_2d_connection}, in which we show the discretized rightmost connecting section of the cylinder between the aforementioned surfaces at the end of the optimization process of VLB - $A = 0.1D$. 
As can be seen, a sharp artificial kink appears at the connection between design and non-design surfaces. 
This is due to the displacement of the connecting vertex, which is computed based on contributions of all adjacent surface patches, as illustrated in Fig. \ref{fig:cylinder_2d_connection}(b). 
Therefore, if our auxiliary problem results in shape updates that do not smoothly fade out to zero at the connection between a design and non-design boundary, a kink is bound to appear. 
A resulting significant deterioration of the surrounding mesh leads to a premature termination of the computational study due to divergence of the primal or adjoint solver. 
This exact behavior, even though it is noticed for all shape updates, appears earlier or later w.r.t. the complete optimization run.

\begin{figure}
\centering
\subfigure[]{\includegraphics[trim={5cm 0 5cm 0},clip, width=0.3\textwidth]{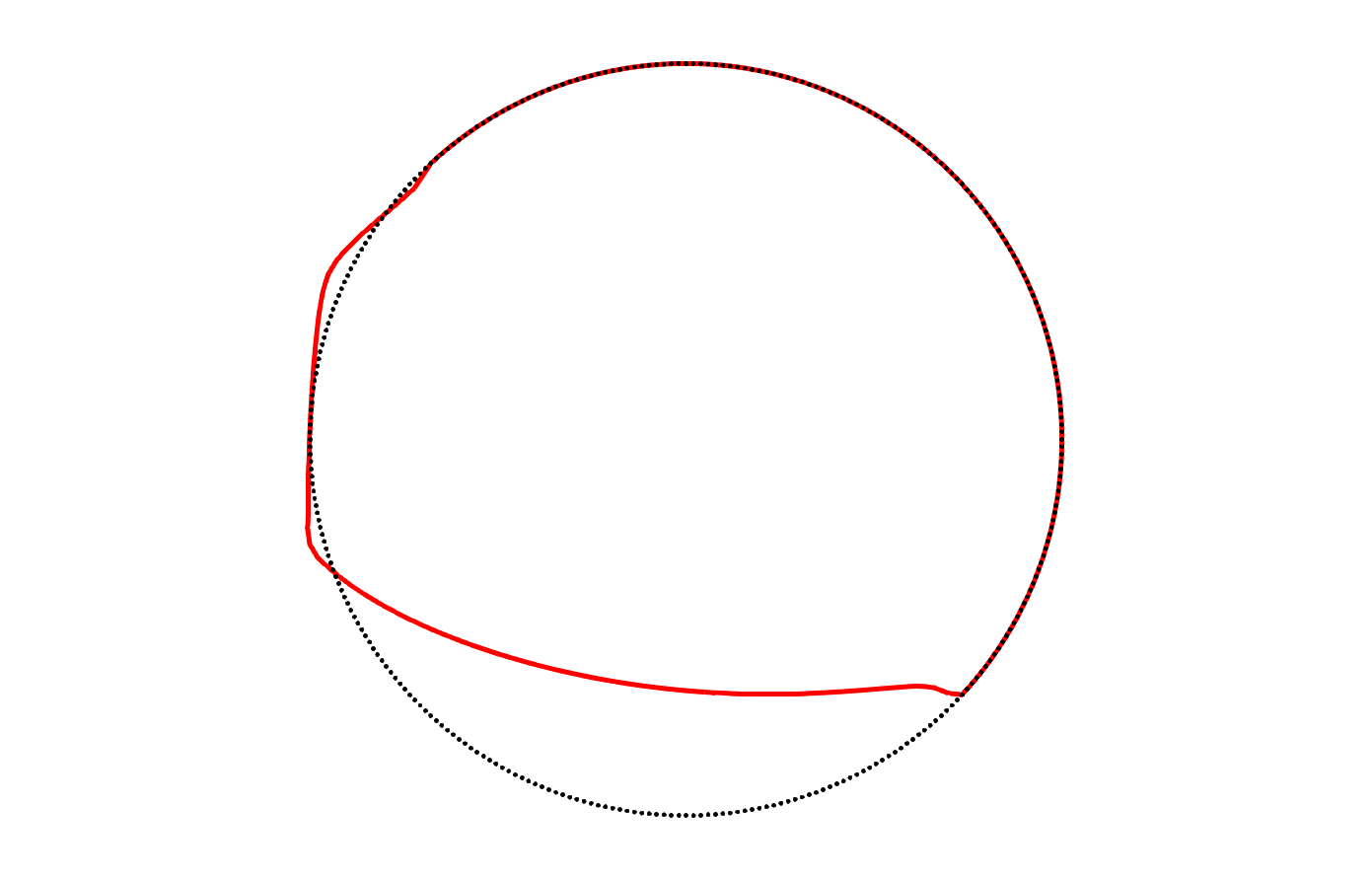}}
\subfigure[]{\includegraphics[trim={5cm 0 5cm 0},clip, width=0.3\textwidth]{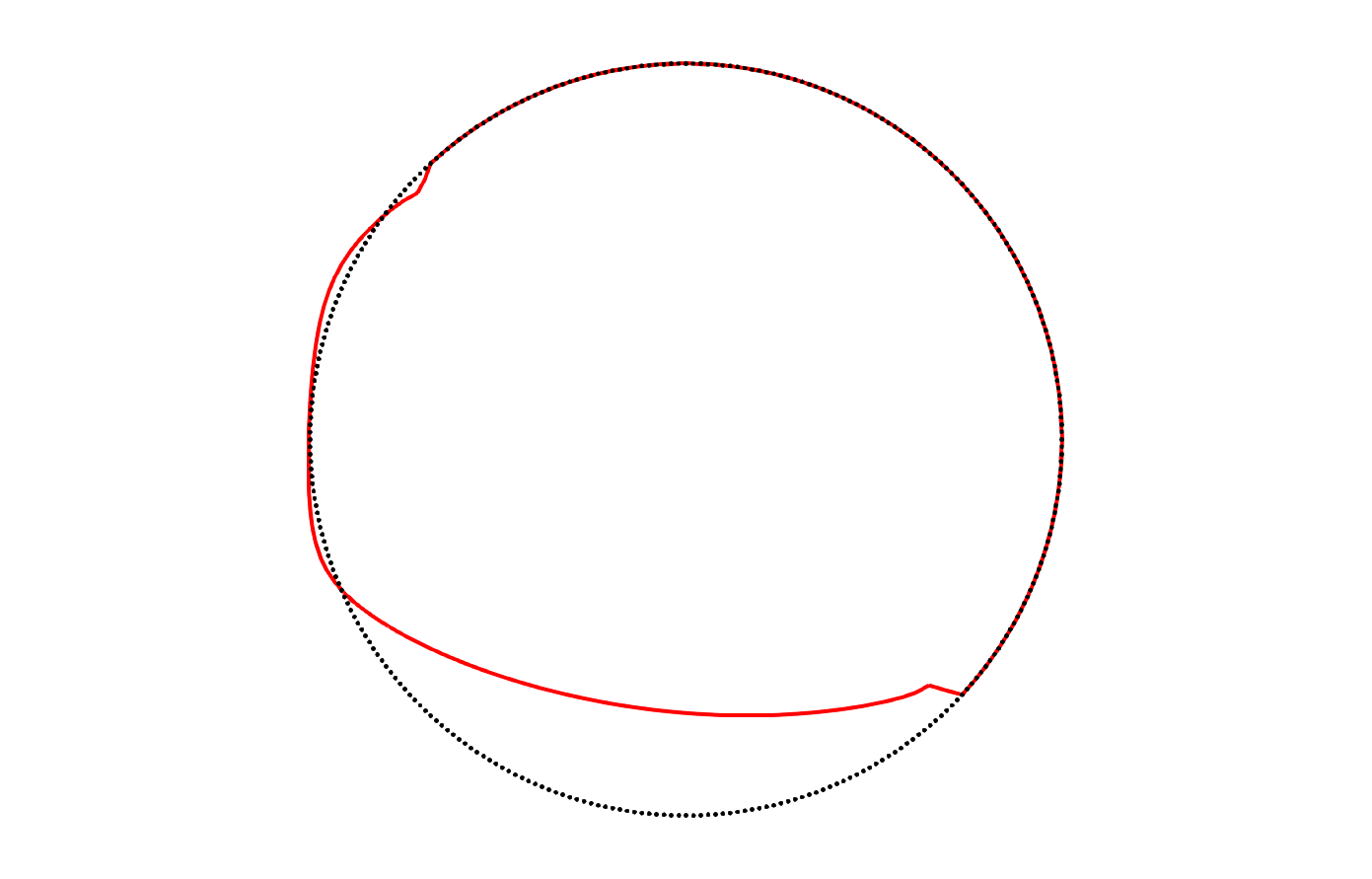}}
\subfigure[]{\includegraphics[trim={5cm 0 5cm 0},clip, width=0.3\textwidth]{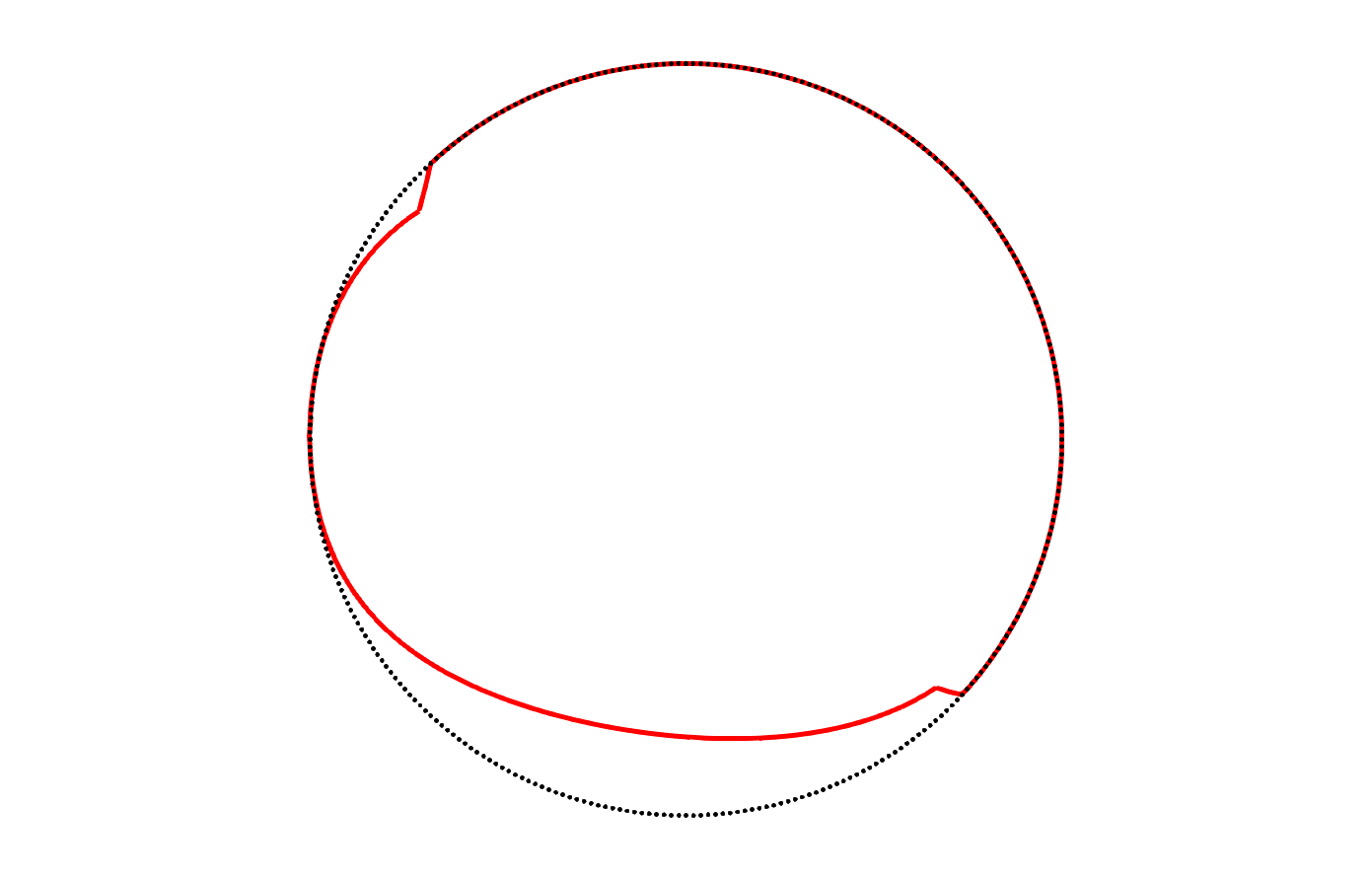}}
\subfigure[]{\includegraphics[trim={5cm 0 5cm 0},clip, width=0.3\textwidth]{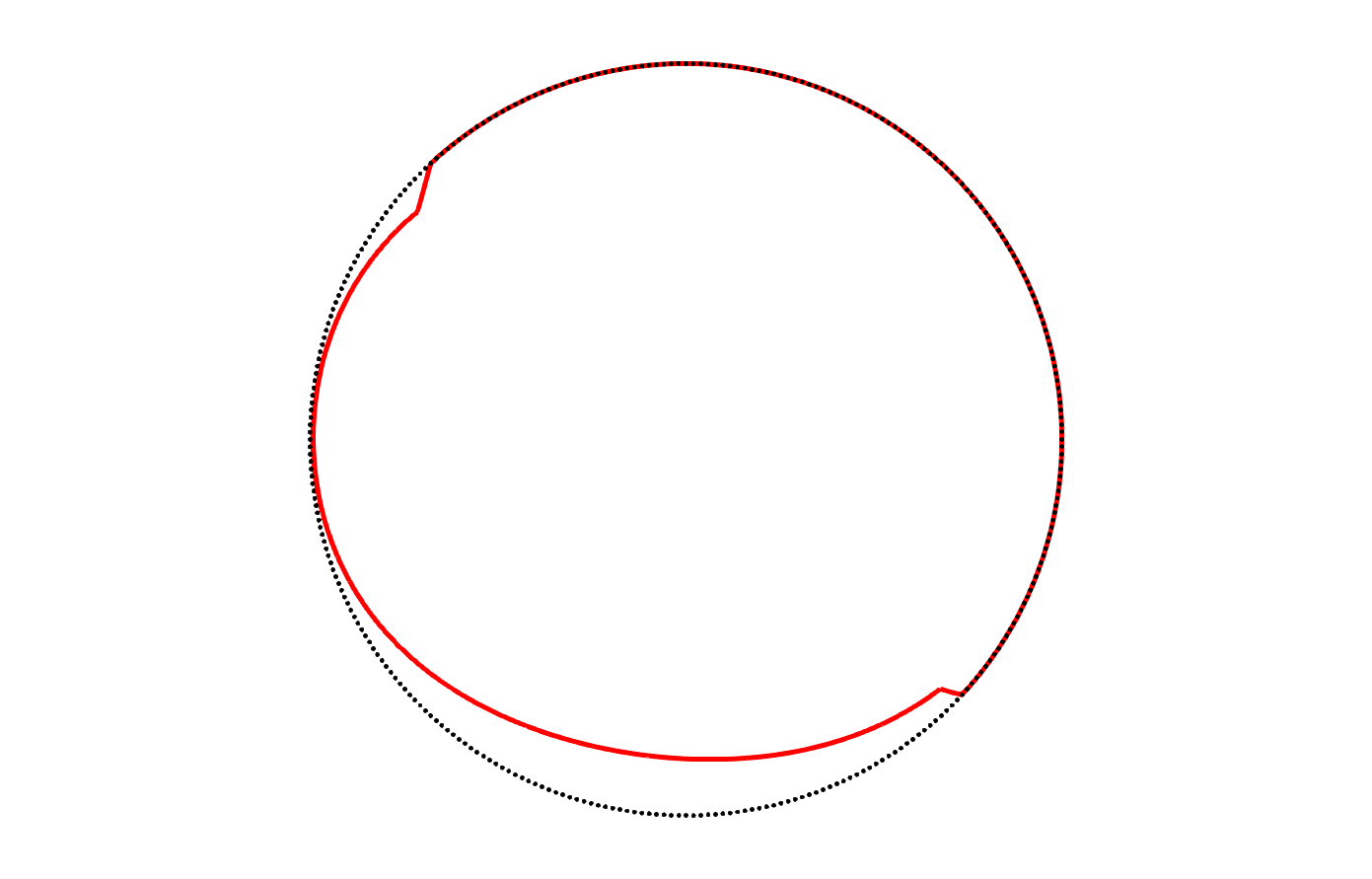}}
\subfigure[]{\includegraphics[trim={5cm 0 5cm 0},clip, width=0.3\textwidth]{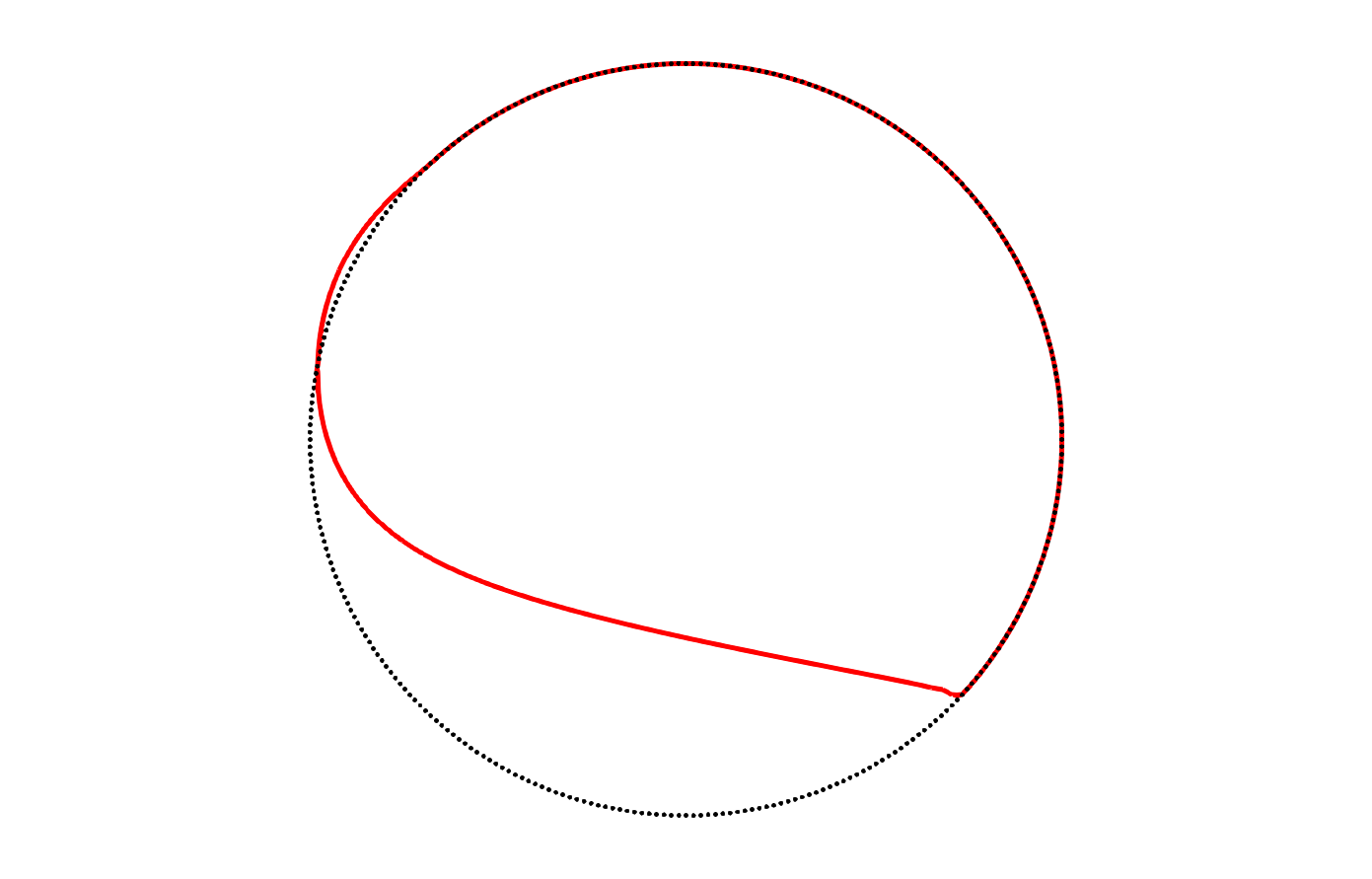}}
\subfigure[]{\includegraphics[trim={5cm 0 5cm 0},clip, width=0.3\textwidth]{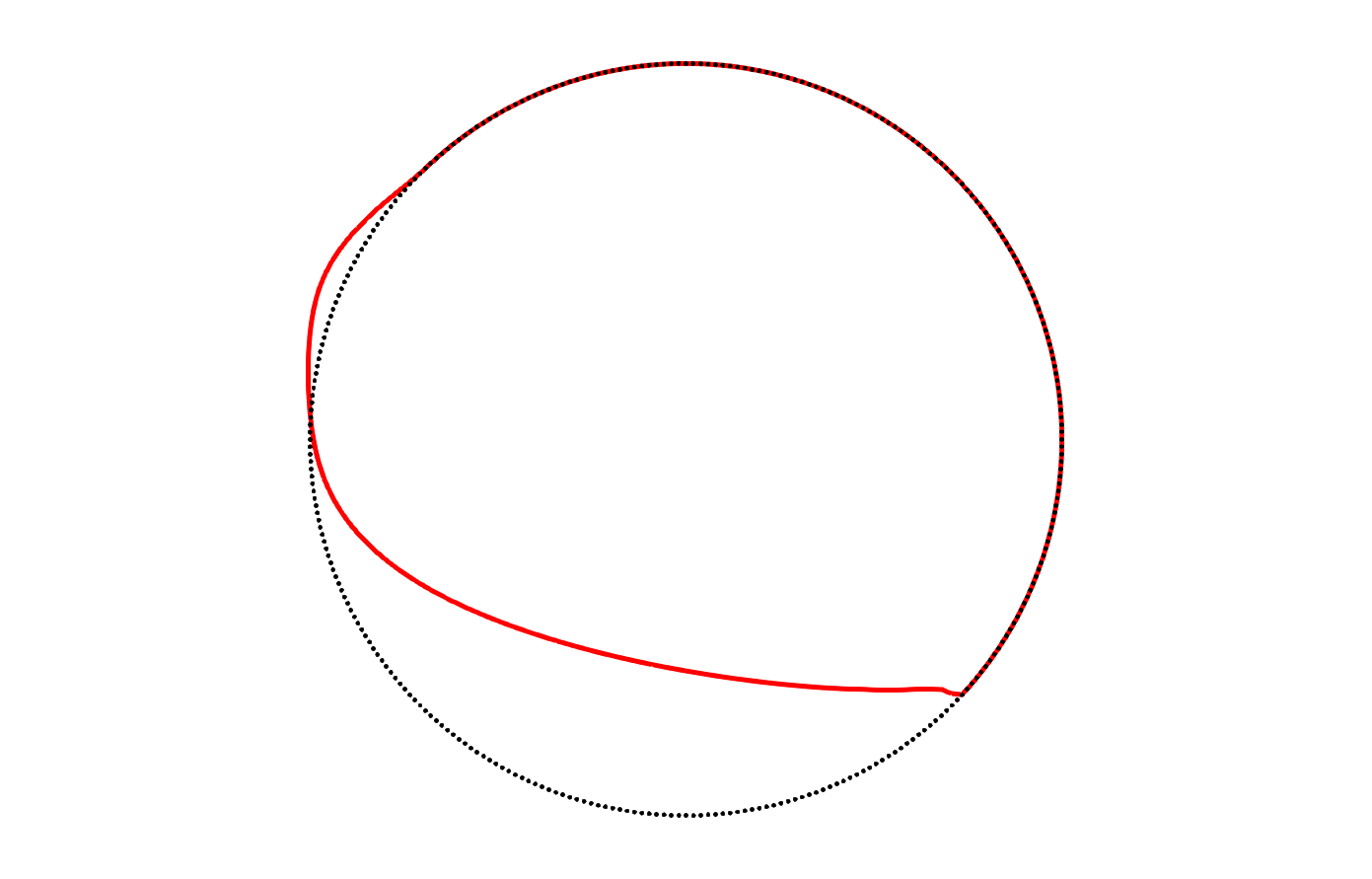}}
\caption{Cylinder ($\mathrm{Re}_\mathrm{D} = 20$): Outline of optimized (red) compared to initial (black) shapes. (a) DS, (b) VLB - $A=0.1D$, (c) VLB - $A=0.5D$, (d) VLB - $A=D$, (e) SP-WD and (f) PHD.}
\label{fig:cylinder_2d_shapes}
\end{figure}

Furthermore, it is interesting to note that the shapes found by each metric differ significantly and the paths towards them as well. 
This is shown in Fig. \ref{fig:cylinder_2d_shapes}. 
We note that SP-WD and PHD result in smoother solutions while shapes produced by the VLB approach become less and less smooth as $A$ decreases. 
Note that in the limit $A \xrightarrow{} 0$, VLB is equivalent to DS (see Eq. (\ref{eq:pde_lapalce_beltrami})).

\subsubsection{Step size control through line search}
\begin{figure}
\centering
\subfigure[]{\includegraphics[width=0.48\textwidth]{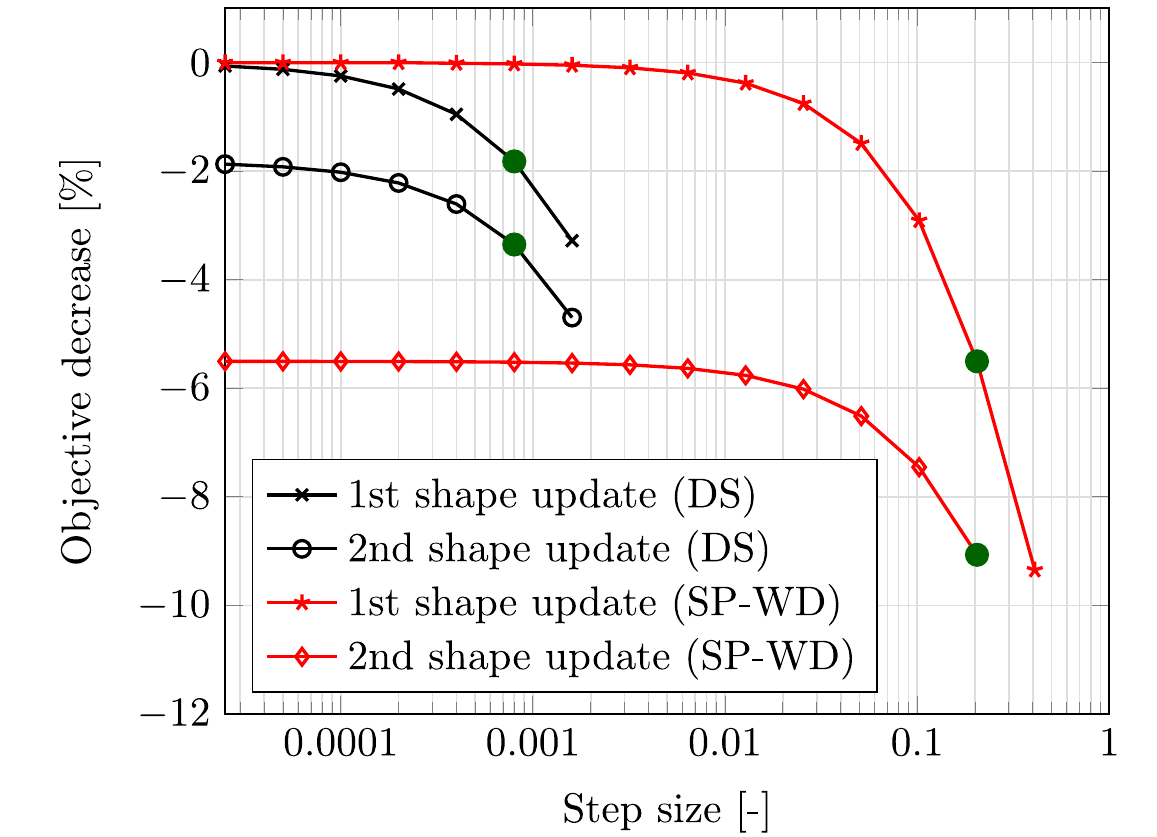}}
\hfill
\subfigure[]{\includegraphics[width=0.48\textwidth]{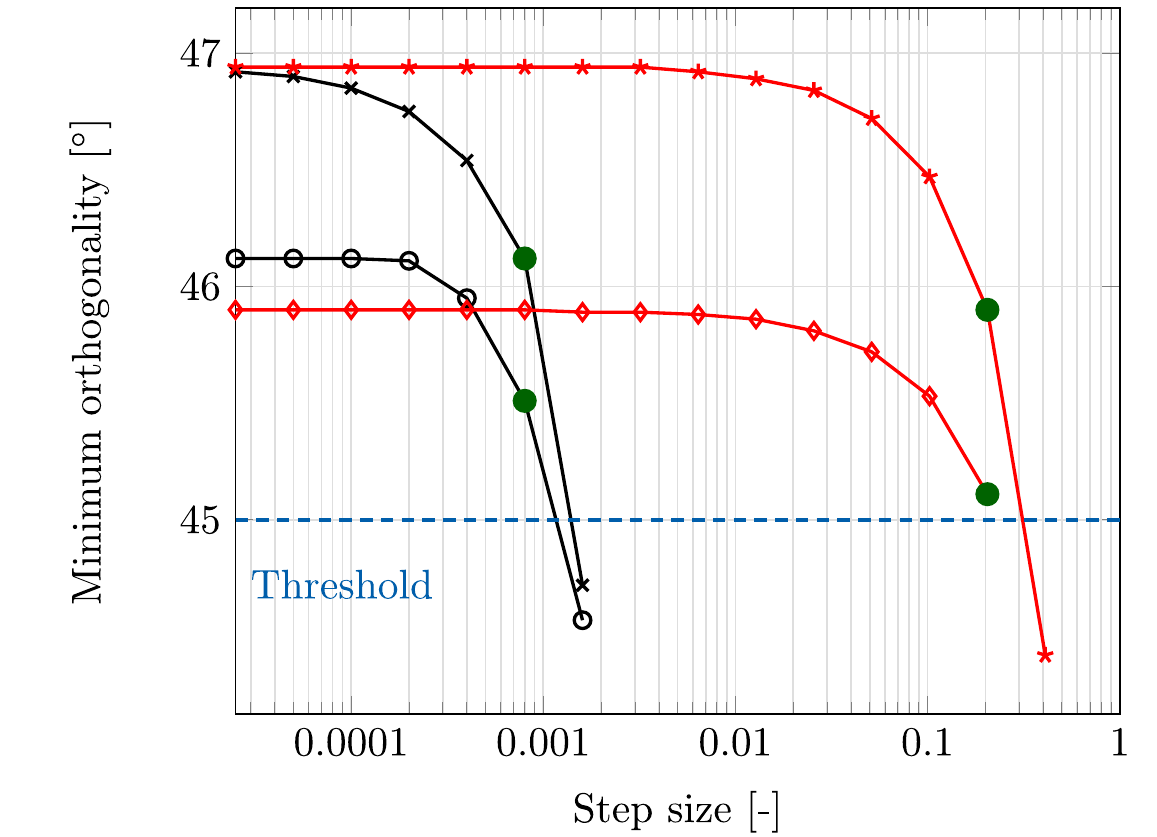}}
\caption{First two optimization iterations employing an optimal step size control based on a line search technique. Filled green circles denote results obtained for the optimally selected step size. (a) Relative decrease of objective. (b) Minimum cell orthogonality of employed computational grid. Figures (a)
and (b) share the same legend.}
\label{fig:step_size_control}
\end{figure}

Similar to the illustrative test case of Sec. \ref{sec:illustrative_example}, we apply the line search technique described in Sec.~\ref{sec:mesh_morphing} to find an optimal step size for the 2D cylinder application.
Due to significant numerical effort needed to test different step sizes, we restrict our investigations to the SP-WD and DS approach.
Figure~\ref{fig:step_size_control} (a) shows the dependence of the objective functional $J(\Gamma^{i+1})$ on the step size for the first two optimization iterations.
Contrary to the illustrative test case, we cannot reach a step size in which $J$ starts increasing. Instead, the line search ends early, due to a low mesh quality.
In particular, we monitor the minimum mesh orthogonality and quit at a threshold of $45^{\circ}$. 
This choice is confirmed by the results shown in Fig. \ref{fig:step_size_control} (b) where for most descent directions, a rapid deterioration of the mesh is noticed after $45^{\circ}$.
This study highlights the significant numerical restrictions that one may face when considering CFD-based shape optimization studies. While preferably, we would like to employ the optimal step size for each descent direction, we are inevitably restricted by the quality of the employed mesh. To this extent, one may pose the question of what the optimal balance between an extensive mesh refinement - which implies increased computational effort - and a straightforward, experienced-based choice of the step size is. An answer to such a question stems from the goal of the optimization at hand and the available computational resources of the user.

\subsection{Three-dimensional flow through a double-bent duct}
\label{subsec:3D_example}
\begin{figure}
\centering
\includegraphics[width=0.3\textwidth]{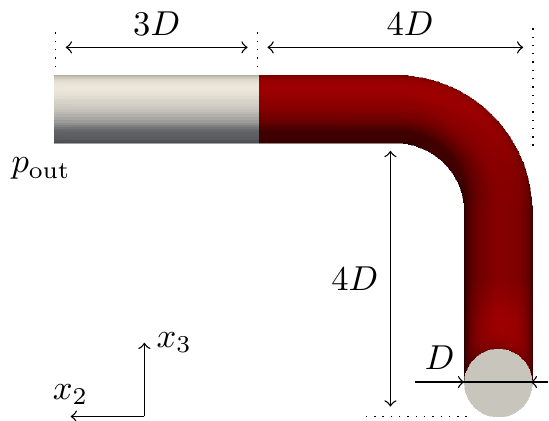}\hspace{2em}
\includegraphics[width=0.3\textwidth]{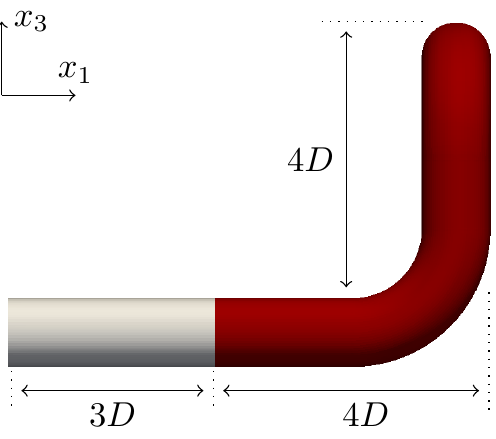}\hspace{2em}
\includegraphics[width=0.3\textwidth]{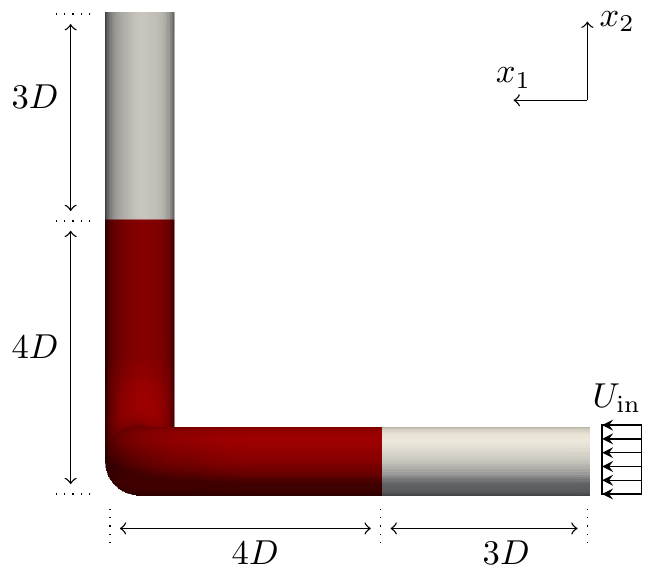}
\caption{Double-bent pipe ($\mathrm{Re}_\mathrm{D} = 500$): Several views on the initial geometry where red areas indicate the region free for design.}
\label{fig:duct_3d_sketch}
\end{figure}

The second test case examines a more involved, three-dimensional, double-bent duct as shown in Fig. \ref{fig:duct_3d_sketch}. The flow has a bulk Reynolds-number of $\mathrm{Re_D} = \rho U D / \mu = 500$ where $U$ and $D$ refer to the bulk velocity as well as the inlet diameter, respectively.
Along the inlet, a uniform velocity profile is imposed and a zero pressure value is prescribed at the outlet. 
The ducted geometry is optimized w.r.t. the total power loss, i.e.
\begin{align}
    J(\Gamma) = - \int_{\Gamma}
    \bm{n} \cdot \bm{u} \left( p + \frac{\rho}{2} \bm{u} \cdot \bm{u}\right) \, \mathrm{d} \Gamma,
    \label{equ:total_power_loss}
\end{align}
for which the corresponding shape derivative $J^\prime(\Gamma)(\bm{\theta})$ corresponds to that of 
the previous section, see Eq.~(\ref{equ:zero_and_first_objec_derivative}). A detailed explanation of the adjoint problem including boundary conditions is provided in \cite{stuck2013adjoint,kuhl2021adjoint_2}.

\begin{figure}
\centering
\subfigure[]{\centering \includegraphics[width=0.45\textwidth]{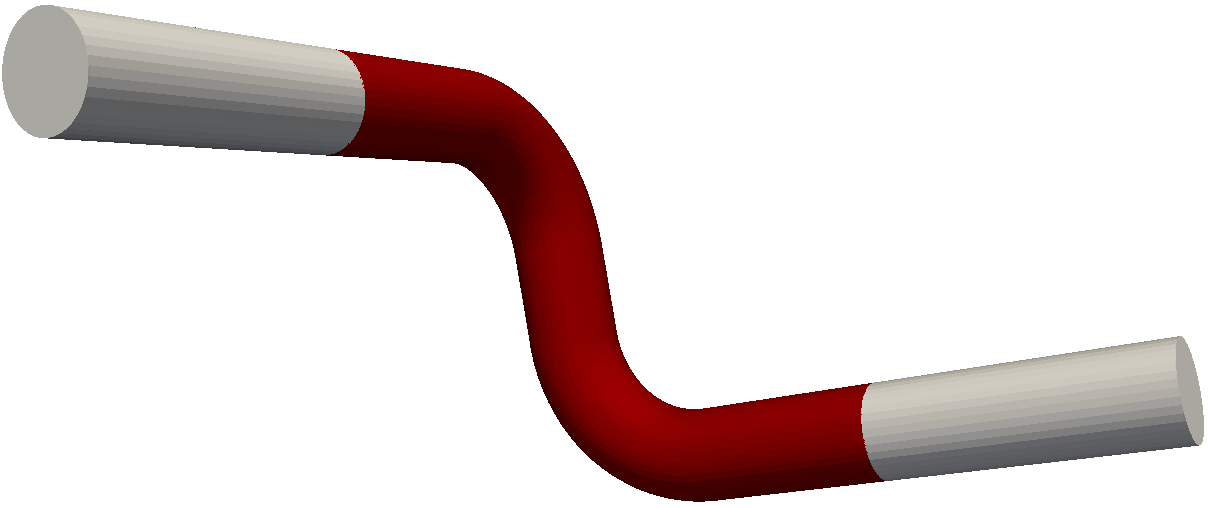} }
\hspace{1cm}
\subfigure[]{\centering \includegraphics[width=0.45\textwidth]{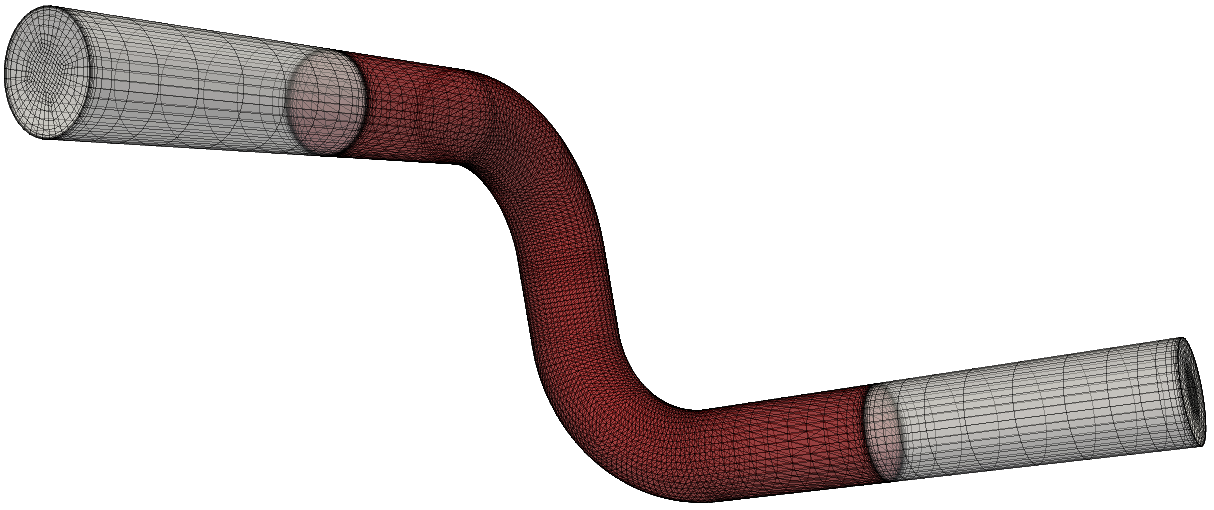} }
\caption{Double-bent pipe ($\mathrm{Re}_\mathrm{D} = 500$): Initial geometry (a) and employed numerical grid (b). Red areas indicate the design region.}
\label{fig:duct_3d_perspective}
\end{figure}
Like for the two-dimensional flow, a grid study is first conducted, as presented in Tab. \ref{tab:duct_3d_mesh_sensitivity}. 
In order to enable a computationally feasible study as well as ensure a reliable estimation of the objective, level 2 is employed for all cases presented hereafter. This corresponds to a structured numerical grid of 90000 control volumes.
Three diameters downstream of the inlet, the curved area is free for design and discretized with 5600 surface elements and the numerical grid is refined towards the transition region between design and non-design wall as depicted in Fig. \ref{fig:duct_3d_perspective}.
\begin{table}
\caption{Double-bent pipe ($\mathrm{Re}_\mathrm{D} = 500$): Results of the mesh dependence study. For illustrative purposes we denote here $\hat{J}'=\int_{\Gamma^{\mathrm{d}}} s \, d\Gamma$. Index $i$ refers to the mesh refinement level. Note $\rho = 500$ $\mathrm{kg/m^3}$, $\mu = 1$ $\mathrm{Pa \cdot s}$, $U = 1$ $\mathrm{m/s}$ and $D= 1$ $\mathrm{m}$.} 
\begin{tabular}[ht]{cccccc}
\toprule
refinement level & number of FV &  $\frac{2 \, J_i}{\rho U^3 D^2}$ & $\frac{2 \, \hat{J}'_i}{\rho U^3 D}$ & $\frac{J_i-J_{i-1}}{J_{i-1}} (\%)$  & $\frac{\hat{J}'_i-\hat{J}'_{i-1}}{\hat{J}'_{i-1}} (\%)$ \vspace{0.9em}  \\  
\hline 
 M0 & 11250 & 2.18 & -5.55 & - & - \\
\hline
 \textbf{M1} & \textbf{90000} & \textbf{3.091} & \textbf{-11.44} & \textbf{41.73} & \textbf{106.13 }  \\
\hline
 M2 & 720000 & 3.15 & -11.38 & 1.91 & -0.53 \\
\hline
 M3 & 5760000 & 3.17 & -11.38 & 0.41 & 0.0 \\
 \bottomrule
\end{tabular}
\centering
\centering
\label{tab:duct_3d_mesh_sensitivity}
\end{table}
During the optimization of the 3D case, the step size remains constant through the process and is determined by prescribing the maximum displacement in the first iteration ($\theta^\mathrm{max}$) as described in Sec.~\ref{sec:mesh_morphing}. We set it to one percent of the initial tube's diameter, i.e. $\theta^\mathrm{max} = 0.01 D$.
The investigated shape and domain updates are DS, SLB with $A / D = 1$, VLB with $A / D = 1$, SP-WD and PHD with $p=4$. Here $A$ is used in a similar context as in Sec.~\ref{sec:2d_cylinder}.
All investigated shape updates are extended into the domain as in the two-dimensional case.

\subsubsection{Results}
\begin{figure}
\centering
\subfigure[]{\includegraphics[width=0.48\textwidth]{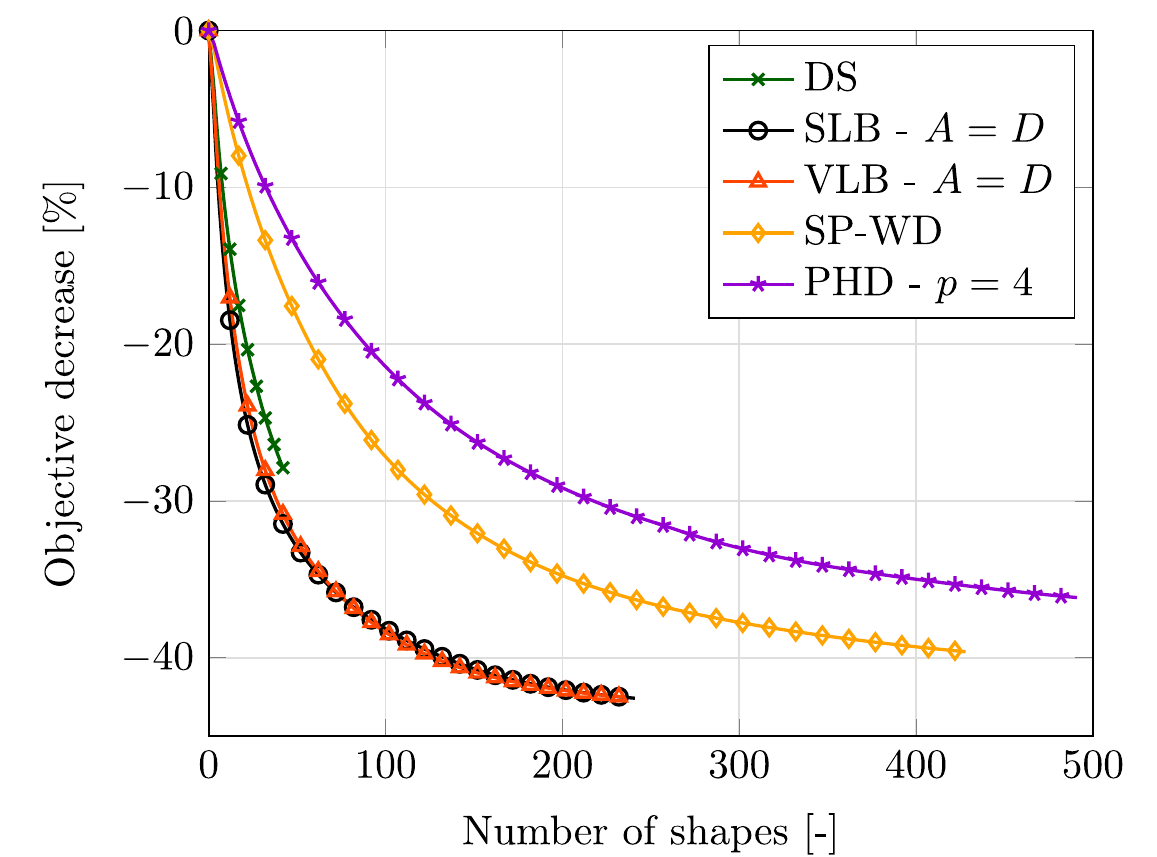}}
\subfigure[]{\includegraphics[width=0.48\textwidth]{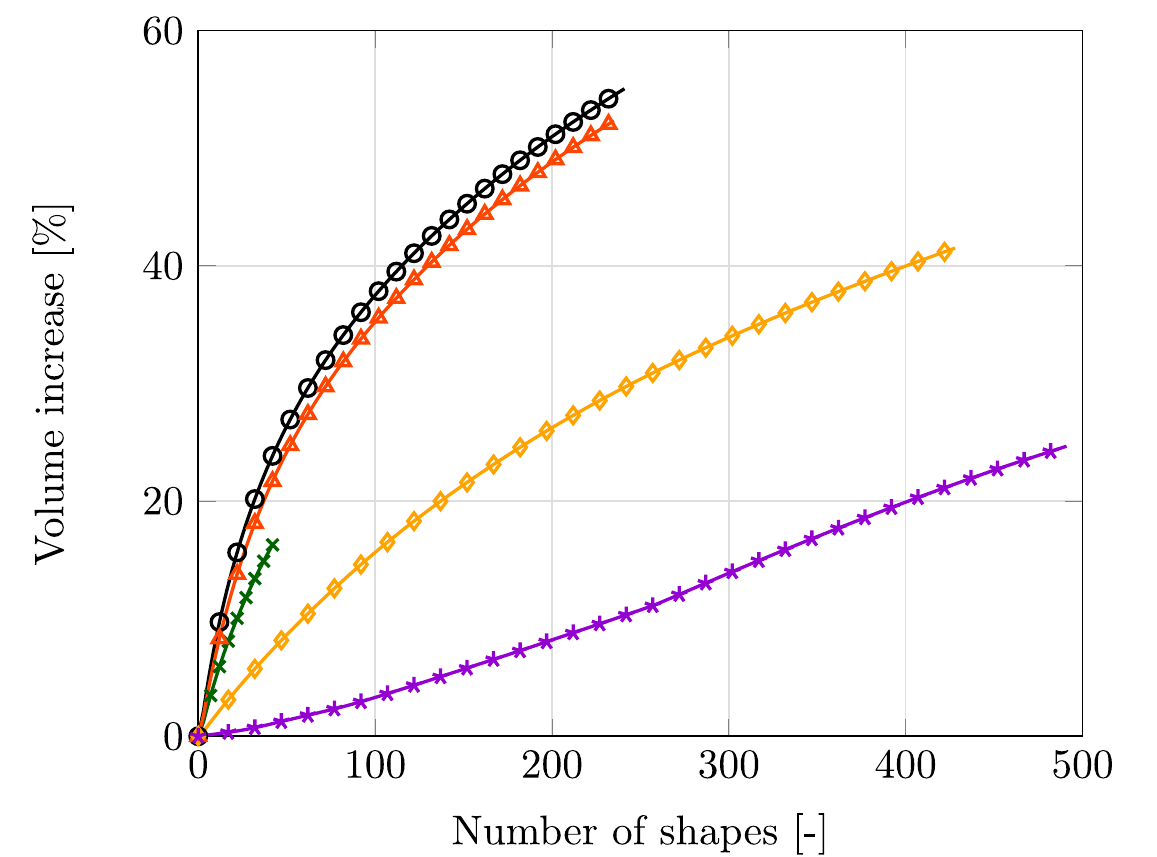}}
\caption{Double-bent pipe ($\mathrm{Re_D} = 500$): (a) Relative decrease of objective ($J(\Omega)$) during the optimization runs as in Fig. \ref{fig:cylinder_2d_convergence}. (b) Relative global volume $(V_i - V_0)/V_0 \cdot 100\%$ increase for each shape update. Figures (a) and (b) share the same legend.}
\label{fig:double_bent_duct_obj_conv}
\end{figure}

Figure \ref{fig:double_bent_duct_obj_conv} (a) shows the relative decrease of $J(\Omega)$ w.r.t. the initial shape. A stopping criterion of the optimization runs is fulfilled when the relative change of the objective functional between two domain updates falls below  $0.1\%$, i.e when $(J_i - J_{i-1})/J_{i-1} \cdot 100\% < 0.1\%$.
The investigated boundary-based approaches SLB \& VLB  managed to reach a reduction greater than 40\%, which corresponds to the SP-WD gain. The PHD approach minimizes the cost functional by $\approx 36 \%$ which is still 10\% more than the DS approach, that terminated due to divergence of the primal solver after 42 iterations.
The reason for termination is the divergence of the primal solver due to insufficient mesh quality, as already described in the previous section. Note that solver settings like relaxation parameters etc., are the same for all simulations during all optimizations.

\begin{figure}
\centering
\subfigure[]{
\centering
\includegraphics[scale=0.15]{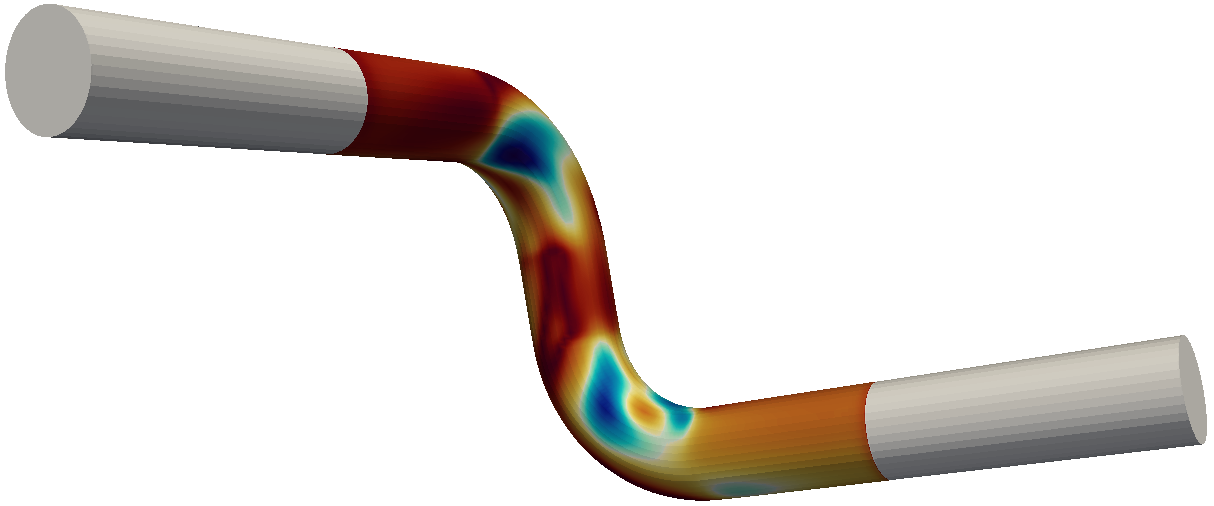}
}
\hspace{1cm}
\subfigure[]{
\centering
\includegraphics[scale=0.65]{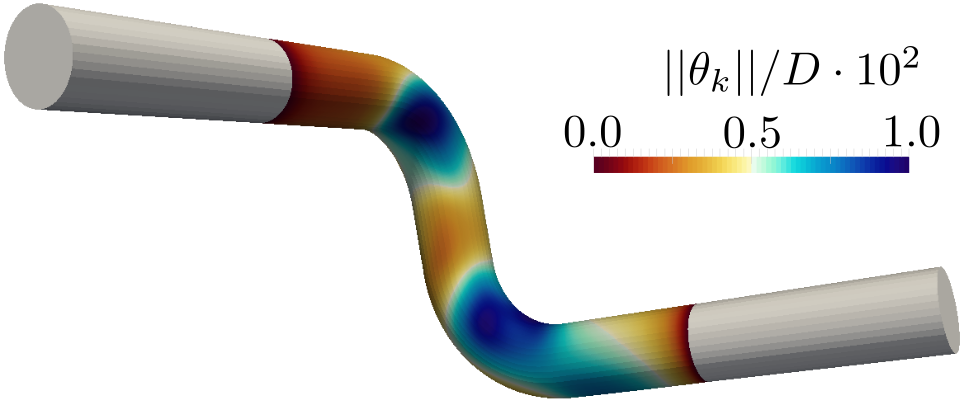}
}
\subfigure[]{
\centering
\includegraphics[scale=0.15]{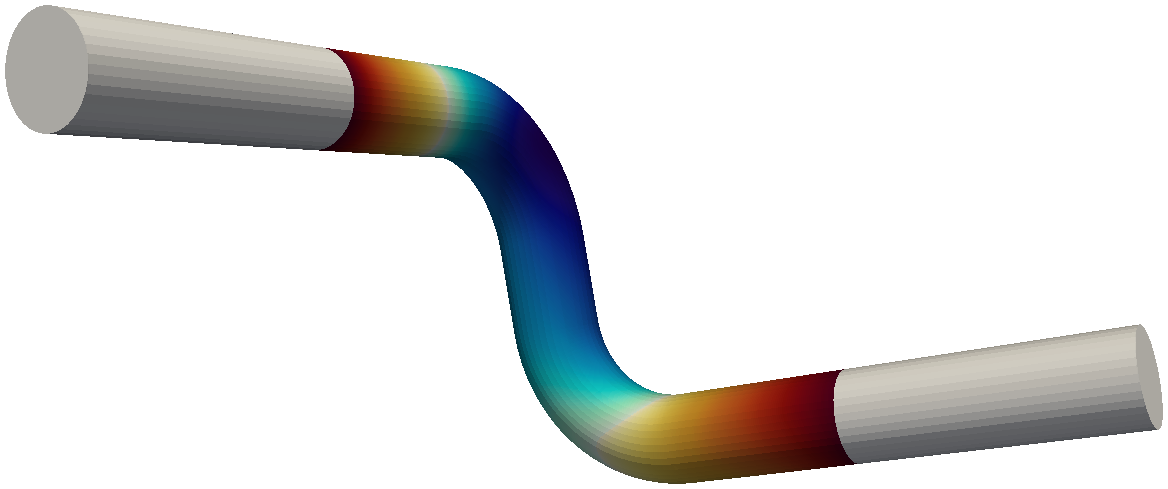}
}
\hspace{1cm}
\subfigure[]{
\centering
\includegraphics[scale=0.15]{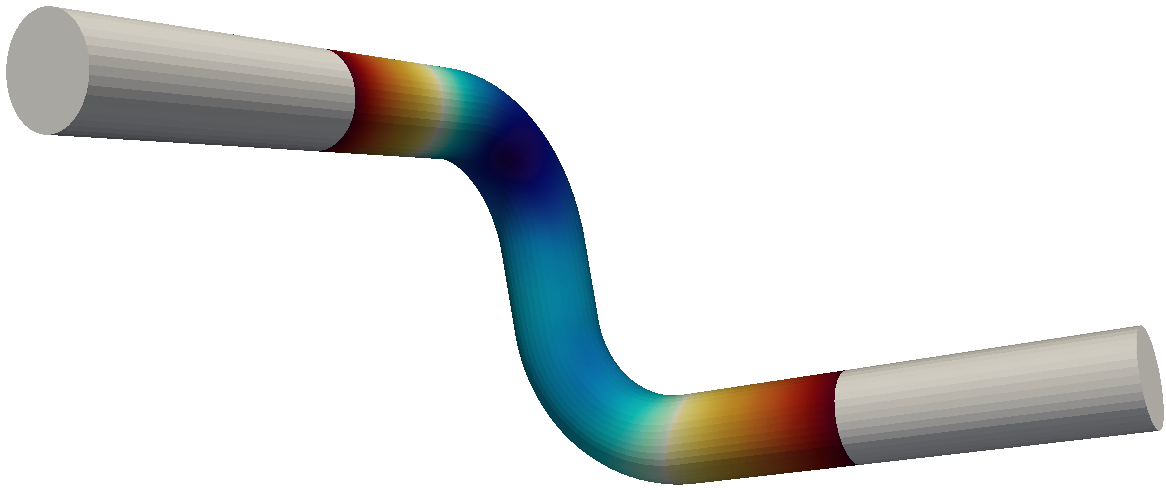}
}
\caption{Double-bent pipe ($\mathrm{Re}_\mathrm{D} = 500$): Normalized magnitude of displacement for the first shape update for the (a) DS, (b) SLB ($A = D$), (c) SP-WD, and (d) PHD ($p=4$) approaches along the design region.}
\label{fig:double_bent_pipe_initial_update}
\end{figure}

The degraded grid quality within the DS procedure can be anticipated from the representation of the shape update direction in Fig. \ref{fig:double_bent_pipe_initial_update} (a).
Compared to the shape updates in (b) SLB, (c) SP-WD, and (d) PHD with $p=4$, a rough shape update field is apparent for the DS approach, especially in the straight region between the two tube bends.
It is noted that the figure is based on the cell-centered finite-volume approximation, and the results have to be interpolated to the CV vertices using Eq. (\ref{equ:mesh_update}). 
This procedure results in a smoothing, which allows the numerical process to perform at least a few shape updates without immediate divergence of the solver.
Compared to the DS approach, the shape update is significantly smoother for the SLB approach with a filter width of $A / D = 1$, cf. Fig. \ref{fig:double_bent_pipe_initial_update} (b). Even smoother shape changes follow from the remaining approaches, with comparatively little difference in the respective deformation field between SP-WD and PHD in the region between the tube's bents.

\begin{figure}
\centering
\subfigure[]{
\centering
\includegraphics[scale=0.15]{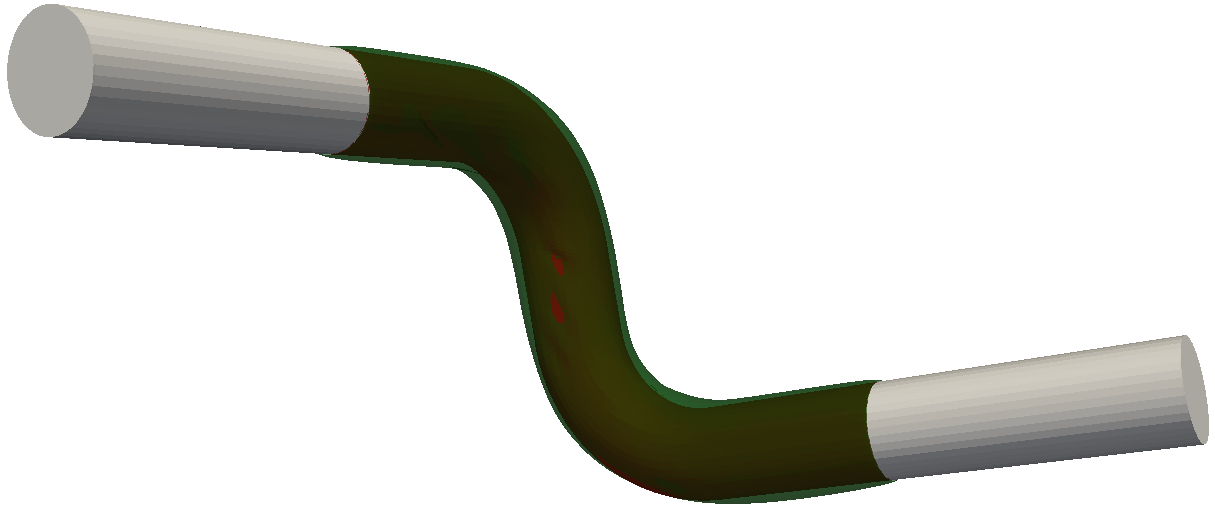}
}
\hspace{1cm}
\subfigure[]{
\centering
\includegraphics[scale=0.15]{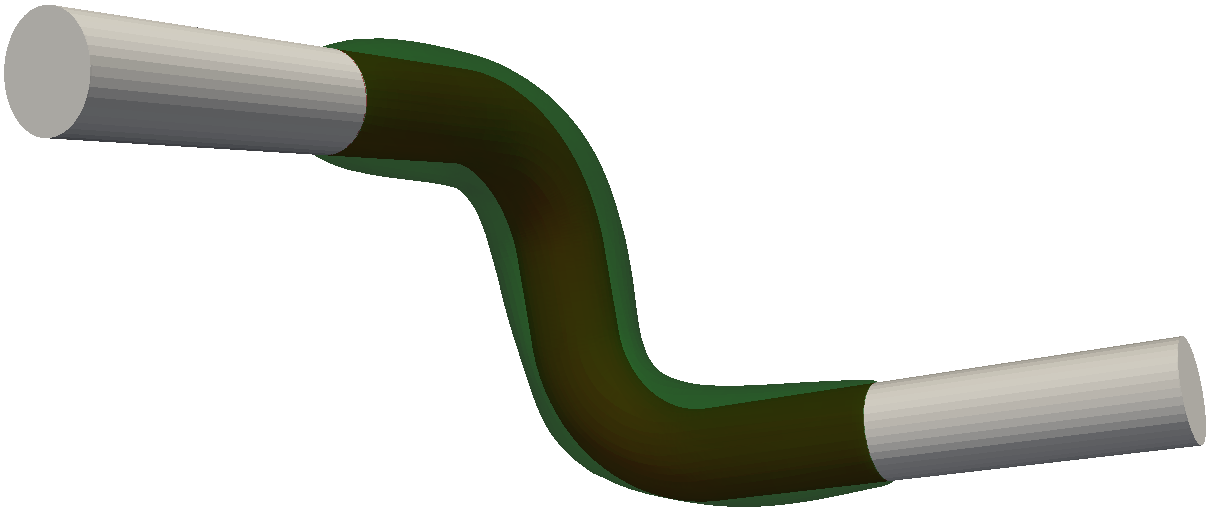}
}
\subfigure[]{
\centering
\includegraphics[scale=0.15]{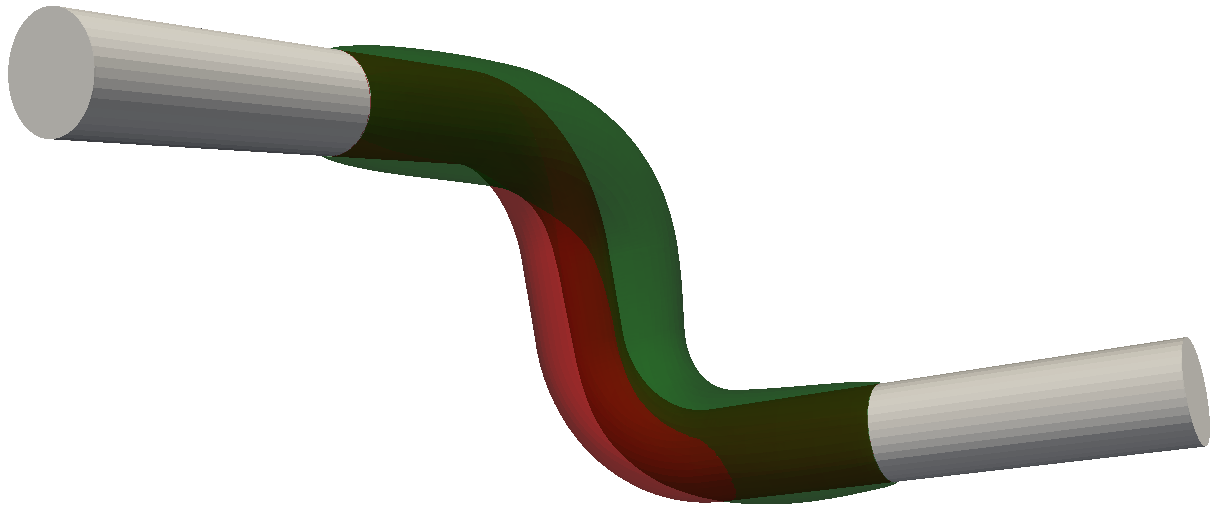}
}
\hspace{1cm}
\subfigure[]{
\centering
\includegraphics[scale=0.15]{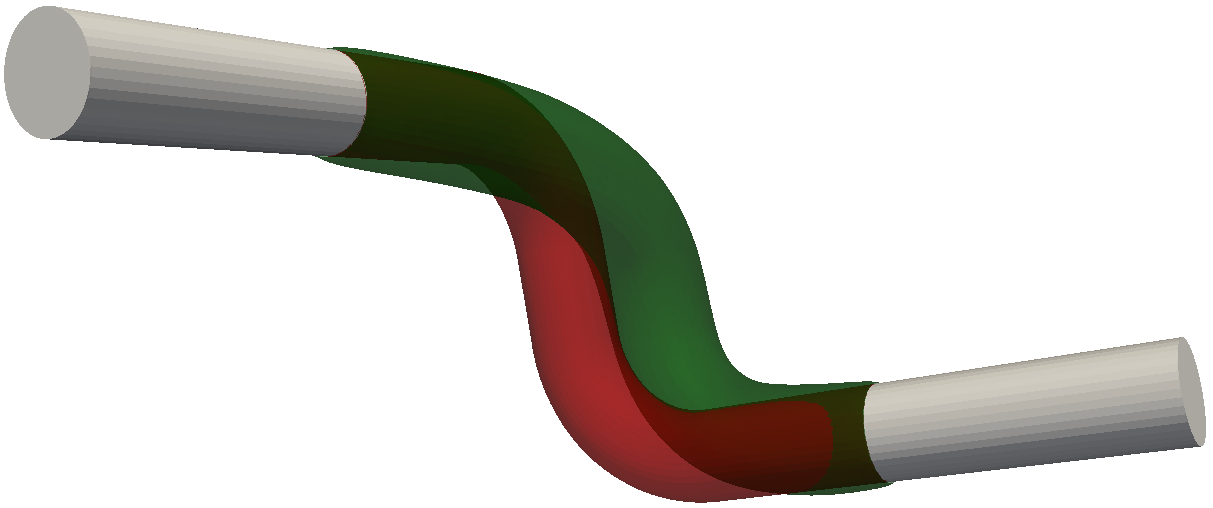}
}
\caption{Double-bent pipe ($\mathrm{Re}_\mathrm{D} = 500$): Initial (red) and optimized (green) shapes based on the (a) DS, (b) SLB ($A = D$), (c) SP-WD, and (d) PHD ($p=4$) approach.}
\label{fig:double_bent_pipe_opt_shapes}
\end{figure}

Perspective views of the final shapes obtained with the four different approaches are shown in Fig. \ref{fig:double_bent_pipe_opt_shapes}.
Again, it can be seen that the DS approach (a) results in local dents in the region between the bends, which is ultimately the reason for the divergence of the SIMPLE solver after a few iterations.
On the other hand, shape updates of the SLB, SP-WD, and PHD approaches are all smooth but still noticeably different. 

The results in Fig. \ref{fig:double_bent_pipe_opt_shapes} are consistent with the expectation that an increased volume should accompany a reduction in pressure drop. 
The fact that the different shape update approaches yield different final shapes can be alternatively observed by tracking the pipe's volume.
For this purpose, Fig. \ref{fig:double_bent_duct_obj_conv} (b) is presented, in which the relative volume changes (i.e., the sum of all FVs) over the number of shape changes are depicted for all approaches. 
The LB-based methods require about 55\% relative volume increase to achieve roughly 43\% relative cost functional reduction.
On the other hand, the SP-WD approach converts relative volume change of approximately 40\% almost directly into a relative objective decrease of also 40\%. 
Only the PHD and DS approaches reduce the cost functional significantly more than the volume increase. 
Thus, the PHD [DS] approach gained about 36\% [26\%] relative objective decrease with about 25\% [17\%] relative volume increase. 

Due to the increased computational effort required for this study compared to the two-dimensional example shown previously, it is interesting to compare the methods with respect to computation time. Such a comparison is given in Tab. \ref{tab:3D_comp_time}, distinguishing between mean primal and mean adjoint computation time. 
For the underlying process, the mesh is adjusted before each primal simulation and thus, the averaged primal time consists of the time required to compute the shape update and the solution to the primal Navier-Stokes system (\ref{eq:primal_continuity})-(\ref{eq:primal_momentum}).
In all cases, the average adjoint simulation time is in the range of 0.1 CPUh. Interestingly, the values of the optimizations based on the Laplace-Beltrami approach are slightly below while all others slightly above this value. 
Starting from an approximately similar simulation time of all primal NS approximations, a significant increase in computation time can be seen for the volume-based methods. Therein, the PHD approach is particularly costly, since the nonlinear equation character in (\ref{eq:p_laplacian_pde}) is elaborately iterated in terms of Picard linearization, which drastically increases the total simulation time.
\begin{table}
\caption{Double-bent pipe ($\mathrm{Re}_\mathrm{D} = 500$): Measured computation time CPUh ($n^\mathrm{opt} \cdot \overline{t^\mathrm{wc}} \cdot  n^\mathrm{CPU}$) for all five optimization studies, where $\overline{t^\mathrm{wc}}$ refers to the mean wall clock time per primal/adjoint run and $n^\mathrm{opt}$ as well as $n^\mathrm{CPU}$ denote the number performed optimization steps as well as employed CPU cores.}
\begin{tabular}[ht]{ccccc}
\toprule
approach & $n^\mathrm{opt}$ [-] & primal $\overline{t^\mathrm{wc}}$ [h] & adjoint $\overline{t^\mathrm{wc}}$ [h] & total CPUh [h] \vspace{0.9em}  \\  
\hline 
DS & 42 & 0.1325 & 0.1176 & 10.5042 \\
\hline
SLB ($A / D = 1$) & 241 & 0.1005 & 0.0994 & 48.1759 \\
\hline
VLB ($A / D = 1$) & 235 & 0.0991 & 0.0981 & 46.342 \\
\hline
SP-WD & 441 & 0.1255 & 0.1109 & 86.9652 \\
\hline
PHD ($p = 4$) & 491 & 0.1914 & 0.1070 & 146.5144 \\
\bottomrule
\end{tabular}
\centering
\label{tab:3D_comp_time}
\end{table}

\subsection{Discussion}
Overall, the numerical studies shown herein highlight how different shape updates on the same CFD-based optimization problem impact not only the steepness of the objective reduction curve but also the final shape. 
From a practical point of view, we identified mesh quality preservation to be the bottleneck of the applied approaches. 
Indeed, one can sustain a better mesh quality or even progress the optimization of non-converged runs by auxiliary techniques, such as remeshing or additional artificial smoothing, however, this goes beyond the scope of the paper. 
Furthermore, it is of interest to note that the computational cost for each shape update is not the same but rather increases when the complexity of the utilized shape update increases as well. 
Finally, based on the presented results, we would like to emphasize that the intention is not to enable a direct comparison of different shape updates with regard to performance in general.
Rather we would like to show how a range of practical shape updates may result in different shapes because typically, the optimization runs have to be stopped before an optimal shape is reached due to mesh distortion issues.
Which shape update yields the largest reduction until the mesh becomes heavily deteriorated depends on the application. 
For example, by comparing the applications presented herein, one can notice that VLB performs much better in the double-bent pipe than in the cylinder case. 

\FloatBarrier

\section{Summary and conclusion}
We have explained six approaches to compute a shape update based on a given sensitivity distribution in the scope of an iterative optimization algorithm.
To this end, we elaborated on the theory of shape spaces and Riemannian shape gradients from a mathematical perspective, before introducing the approaches from an engineering perspective.
We included two variants of the well known Hilbertian approaches based on the Laplace-Beltrami operator that yield first order Sobolev gradients (SLB and VLB).
For comparison, a discrete filtering technique and a direct application of the sensitivity was considered as well (FS and DS). 
Further, two alternative approaches that have not yet been extensively used for engineering applications were investigated (SP and PHD).
They directly yield the domain update direction, such that an extra step that extends the shape update direction into the domain can be avoided.

Based on an illustrative example, the characteristic behavior of the approaches was shown. 
While the FS and the DS approach manage to find the optimal shape even in regions where it is not smooth or has a high curvature, the SP approaches yields shapes which differ in these regions.
For the PHD, VLB and SLB approach, the parameters $p$ and $A$ can be used to regulate the smoothness of the obtained shape.
Due to the possibility of remeshing for the comparably simple problem, mesh quality was not an issue.

Regarding the simulations of the CFD problems, for which remeshing was not realized, the decrease in mesh quality became a severe issue preventing the optimization algorithm from convergence.
For the two-dimensional case, the PHD approach yielded the steepest decrease in the objective functional, however, the smallest objective functional value was obtained using the SP method, which managed to preserve a reasonable mesh quality for more iterations than all other approaches.
For the three-dimensional case, the VLB and SLB approaches outperformed all other approaches in terms of steepest decrease of the objective functional as well as the smallest value that could be achieved before the mesh quality became critical.

Concluding, we have observed that the behavior of the approaches is strongly connected to the considered problem.
We suggest to use the SP as a first choice, as it is computationally less involved than the PHD approach and does not require an extension of the shape update into the domain in a second step like the SLB and the VLB approach.
The performance of the latter shall still be compared for a given application scenario -- despite the extension in a separate step, the overall computational cost may still be reduced compared to the SP approach due to a steeper descent.
Finally, we suggest not to use the DS approach, since it was weaker than all other approaches in terms of mesh quality, irrespective of the problem.

\section*{Acknowledgment}
The current work is a part of the research training group 'Simulation-Based Design Optimization of Dynamic Systems Under Uncertainties' (SENSUS) funded by the state of Hamburg within the Landesforschungsf\"orderung under project number LFF-GK11. 

The authors gratefully acknowledge the computing time made available to them on the high-performance computers Lise and Emmy at the NHR centers ZIB and G\"ottingen. These centers are  jointly supported by the Federal Ministry of Education and Research and the  state governments  participating  in the NHR  (www.nhr-verein.de/unsere-partner).

\section*{Contribution}

\textbf{Lars Radtke:} Conceptualization, Software, Formal analysis, Investigation, Writing -- Original Draft  (Sec. 1, 3, 4, 6), Writing -- Review \& Editing, Visualization, Project Administration.
\textbf{Georgios Bletsos, Niklas K\"uhl:} Conceptualization, Software, Formal analysis, Investigation, Writing -- Original Draft (Sec.~5), Writing -- Review \& Editing, Visualization.
\textbf{Tim Suchan:} Conceptualization, Formal analysis, Writing -- Original Draft (Sec.~2), Writing -- Review \& Editing
\textbf{Kathrin Welker:} Conceptualization, Formal analysis, Writing -- Review \& Editing, Supervision, Project administration, Funding acquisition.
\textbf{Thomas Rung, Alexander D\"uster:} Writing -- Review \& Editing, Supervision, Project administration, Funding acquisition.

\bibliographystyle{abbrv}
\bibliography{literature}

\end{document}